\documentclass[%
 reprint,
 amsmath,amssymb,
 aps,
onecolumn]{revtex4-2}

\usepackage{graphicx}
\usepackage{dcolumn}
\usepackage{bm}
\usepackage{natbib}
\usepackage[dvipsnames]{xcolor}
\definecolor{mycolor}{rgb}{0,0.286275, 0.658824}
\usepackage{natbib}
\graphicspath{{Images/}}
\usepackage{dsfont}
\usepackage{amsmath}

\begin{document}
\let\WriteBookmarks\relax
\def\floatpagepagefraction{1}
\def\textpagefraction{.001}

\title [mode = title]{Quantum non-Gaussianity of light and atoms}  

\author{Lukáš Lachman}
\email{lachman@optics.upol.cz}
\author{Radim Filip}%
 \email{filip@optics.upol.cz}
\affiliation{%
 Department of Optics, Faculty of Science, Palack\' y University,\\
17. listopadu 1192/12,  771~46 Olomouc, \\
Czech Republic 
}%

\begin{abstract}
Quantum non-Gaussian states of photons and phonons are conclusive and direct witnesses of higher-than-quadratic nonlinearities in optical and mechanical processes. Moreover, they are proven resources for quantum sensing, communication and error correction with diverse continuous-variable systems. This review introduces theoretical analyses of nonclassical and quantum non-Gaussian states of photons and phonons. It recapitulates approaches used to derive operational criteria for photons tolerant to optical losses, their application in experiments and their nowadays extension to quantum non-Gaussian photon coincidences. It extends to a recent comparison of quantum non-Gaussianity, including robustness to thermal noise, and sensing capability for high-quality phononic Fock states of single trapped cooled ions. The review can stimulate further development in the criteria of quantum non-Gaussian states and experimental effort to prepare and detect such useful features, navigating the community to advanced quantum physics and technology.  
\end{abstract}


\maketitle

\section{Introduction}\label{introduction}

Quantum physics of the bosonic states of individual photons emitted by atoms and phonons of atomic vibrations are the core of modern quantum technology. Since photons and phonons are bosons, many of them can have identical properties, and therefore they can be treated collectively by a single wave function in an optical or mechanical mode. This collective behaviour is so significant for the light that classical optics investigates only the wave features of mode occupations and ignores the discrete corpuscular aspects completely. We will, therefore, start first reviewing progress with bosonic states of photons and subsequently move to bosonic states of single-atom motion.  

A fundamental property of the waves that classical optics examines is the first-order coherence \cite{Loudon2000}. A series of experiments, including the famous Young double-slit interference experiment \cite{D.F.Walls2008} or Arago’s investigation of white spot in the center of a shadow, thrown by a circular object \cite{Harvey1984}, explored this basic coherence in many details. Despite ongoing investigations, the physics developed in the twentieth century showed that the wave description of light is incomplete. The first witnesses of this appeared in statistical physics and thermodynamics with an idea to quantize the electromagnetic field enabling clarification of the spectral properties of the thermal radiation \cite{Born1999}. It opened a wide path leading to an explanation of corpuscular aspects of light, bringing hidden and unexplored laws of nature.
 
The corpuscular aspects say that the energy of the light field possesses an integer multiple of a unit of energy that corresponds to the non-vanishing energy of a single photon. The first-order coherent aspects from classical optics are preserved even when light is so weak that it comprises only a single photon. Thus, all the interference experiments from the classical optics are reproducible for individual photons, and interference fringes remain visible for many independent repetitions \cite{D.F.Walls2008}. It immediately brings statistical aspects to the description of such interference experiments. It strikes our intuitive concept of nature where objects are either particles or waves, which can always split and spread to the whole space by propagation, diffraction and interference. Photons pick their behaviour according to the experimental setup and detection. Importantly, all the first-order coherence effects are independent of the field intensity in a single run of experiments. That is reflected later in their measurement and evaluation.  

A theory explaining the classical coherence by terms of quantum theory was established by R. Glauber and E. C. G. Sudarshan in 1963. They identified quantum states of light – coherent states that behave like an ideal classical wave when a detector measures the intensity of these states \cite{Glauber1963b, Sudarshan1963, Glauber1963}. They also generalized classical correlation functions to quantum correlation functions to predict classical coherence effects with quantum theory. Such coherent states are simple results of coherent classical driving of linear quantum harmonic oscillators. Their photon statistics is Poissonian, and they do not exhibit correlations on linear optical elements, like beam splitter. Therefore, they are statistically analogous to discrete shot noise of independent electrons.     

Simultaneously, their approach allows the differentiation of attractive quantum states going beyond classical coherence effects in statistical optics. It was the birth of quantum optics with nonclassical light. Nonclassicality means here incompatibility with the classical coherence theory of oscillations and waves. Operationally, it proves that the generation process went beyond just classical linear driving of linear oscillator. Such nonclassical states were achieved experimentally almost eight decades after explaining the black-body radiation provided by Max Planck in the year 1900. The concept of the quantized electromagnetic field had been only a subject to theoretical considerations for those eight decades. Nonclassical states justified the principal need for a quantum theory of light. 

The first experiment verifying the incompleteness of the classical wave optics was realized in the year 1977 by Kimble, Dagenais and Mandel \cite{Kimble1977}, see \cite{Davidovich1996} for a review. They statistically observed the antibunching phenomena of a single photon on a beam splitter using the second-order correlation function. It conclusively proved the indivisibility of photons by linear optics.  Such effects are impossible by classical optical fields or equivalently quantum coherent states found by Glauber and Sudarshan. Moreover, single-photon states approach eigenstates of field energy and, as such, they reduce energy uncertainty below the level of any coherent states. Such light can demonstrate sub-Poissonian (sub-shot-noise) behaviour. 


Experimental demonstration of these different nonclassical aspects, reducing quantum fluctuations in discrete and continuous quantum variables of field, showed a diversity of nonclassical states, their features and applications.  Since then, theoretical analysis and experimental tests of nonclassical states have been extensively ongoing. It is increasingly stimulated by new and better sources, detectors and applications of nonclassical light.              
Later, in 1985, Slusher, Hollberg, Yurke, Mertz, and Valley won a race for the first generation of nonclassical squeezed states \cite{Slusher1985}, see \cite{Andersen2016} for a review. Squeezed states are nonclassical because they reduce the uncertainty in a field amplitude below the level of any coherent state. As field amplitude is time (phase) sensitive, this nonclassicality varies as the field oscillates. It is different from that phase-insensitive nonclassicality of single-photon states. As a consequence of the uncertainty principle, when the optical field suppresses fluctuations below coherent states periodically, field noise has to increase adequately at some other times of field oscillations. Interestingly, such squeezed states can still have Gaussian statistics of field variables similarly to the coherent states representing classical fields.  

Nowadays, modern experiments focus on the manipulation and detection of individual photons since they represent a valuable resource in already proposed applications such as, for example, quantum cryptography \cite{Bennett2014}, quantum metrology \cite{Boto2000}, or quantum computing \cite{Knill2001}. Simultaneously, theory development aims to recognize novel quantum properties of nonclassical states motivated by the currently progressing experiments. A violation of the classical theory of coherence has stimulated many investigations since the 1970s \cite{Mandel1986}. As a response to rapid progress in engineering quantum states during the last decade, quantum non-Gaussianity (QNG) of single-photons emerged as a more demanding benchmark for quantum aspects \cite{Filip2011}. Quantum non-Gaussian single-photon states form a narrow subset of nonclassical states. This subset includes the ideal single-photon eigenstate of field energy - Fock state; therefore, their detection proves that single-photon states are of higher quality. 

Importantly, all statistical mixtures of squeezed coherent states are outside this narrow subset. Thus, the states remaining inside the subset are quantum non-Gaussian. The adjective quantum has been used to distinguish it from classical non-Gaussianity that appears simply by a mixture of fields in classical optics. QNG states require a generation process that involves nonlinear dynamics beyond up-to-quadratic Hamiltonians. Therefore, their detection is inconclusive prove of genuine nonlinear nature of the generation. Moreover, QNG single-photon states were recognized as a valuable indicator for applications of quantum optics \cite{Lasota2017a} and useful probe of storage of light in the matter \cite{Rakhubovsky2017}. 

The quantum non-Gaussianity directly inspects whether light overcomes linearized dynamics in the quantum optics, which is sufficient to generate the squeezed states of light \cite{Yuen1976}. Coherently displacing such Gaussian squeezed states allows arbitrary energy (intensity) noise suppression below the Poissonian limit of coherent states. Quantum non-Gaussianity, therefore conclusively distinguishes imperfect Fock states from such purely Gaussian effects.  Thus, the single-photon quantum non-Gaussianity considerations currently aim at three main points. Firstly, it represents a strict reference challenging the single-photon experiments where nonclassical light is already routinely produced. Secondly, it reveals that the quantum nonlinearity beyond the second-order \cite{Boyd2008} has been used to emit the light. Thirdly, it indicates relevant features of quantum states of light for applications, where only nonclassical light is insufficient. In this way, analysis of quantum non-Gaussian features of light advances fundamental analysis of nonclassical phenomena. Simultaneously, it leaves roots to quantum coherence theory based on correlation function in terms of normally-ordered moments of field variables.

This Review focuses on a comprehensive analysis and comparison of the nonclassicality and the quantum non-Gaussianity in the context of currently developing quantum technologies. Chapter 2 provides methods for the subsequent Chapters. In the beginning, it mentions a  path leading to a quantized electromagnetic field. Further, it introduces formal representations of light and gives a concept of the nonclassicality in quantum optics. The chapter also describes basic Gaussian operations in quantum optics: coherent displacement, linear optical beam-splitter and linearized single-mode squeezing together with a description of the Gaussian states stemming from them. Finally, it proposes calculations predicting how quantum states propagate through linear optical networks.

Chapter 3 introduces the nonclassicality manifested in the traditional Hanbury Brown and Twiss layout to interconnect the concepts. Firstly, the chapter outlines a historical background together with contemporary research related to nonclassicality. Further, it describes an ab-initio approach proposed in \cite{Filip2011}, which is applied for derivation of a reliable criterion of the nonclassicality. The usefulness of this benchmark is analysed for a model of light source that is relevant for the current experimental sources of quantum light. It is proved there that the nonclassical light is not restricted only to weak light. Finally, the chapter describes an experiment supporting the theory. 


Chapter 4 inspects in detail quantum non-Gaussianity representing a fundamental property needed for challenging current quantum technologies. In the beginning, the quantum non-Gaussianity is related to contemporary research. The presented criteria are derived analogously to the ones introduced in Ref.\cite{Filip2011}. Extending \cite{Filip2011}, the criteria form a sequence of conditions, which can be exploited for detection of the quantum non-Gaussianity even on multi-photon light sources. The chapter analyses such applicability for realistic states of light. The theory is supported by an experimental realization of quantum non-Gaussian light.

Chapter 5 extends the certification of the quantum non-Gaussianity to photonic states in two modes. It allows for a detection scheme exploited for exposing the nonclassical coincidences \cite{Friberg1985} and provides a criterion of quantum non-Gaussian coincidences relying on response of either single-photon avalanche photo-diodes or photon number resolving detectors. It discusses capability of a realistic photon pair to obey this criterion.

Chapter 6 proposes a natural extension of the quantum non-Gaussianity. It introduces a hierarchy of quantum features denoted as genuine $n$-photon quantum non-Gaussianity. They identify quantum aspects that are possessed by the Fock state $\vert n \rangle$ but not possessed by the lower Fock states. The recognition of such properties exploits the methodology from \cite{Filip2011} as well. The chapter provides criteria for these quantum features and analysis for relevant quantum states of light. Finally, it describes a realized experiment where these quantum features were observed. 

Chapter 7 deals with evaluation of quantum aspects manifested by phononic states of an ion. It allows for the strictest criteria of genuine $n$-phonon quantum non-Gaussianity and describes their feasibility for realistic phononic states. Simultaneously, it explores the capability of phononic states to provide an advantage in quantum sensing. The chapter analyses quality of realistic states according to the strict criteria of genuine $n$-phonon quantum non-Gaussianity and their application in quantum metrology.

\section{Basics of nonclassical light}

\subsection{Quantized light field and Fock states}
The quantum theory of light assumes an operator form of linear Maxwell's equations derived for classical electromagnetic field in vacuum where the operators of the field are measurable Hermitian operators. A formal solution of the operator Maxwell equations for a situation when the quantum field is confined in a virtual box is determined from behaving of a Hermitian operator $\mathbf{A}$ representing the vector potential \cite{D.F.Walls2008}. The temporal and spatial evolution of $\mathbf{A}$ works out to be
\begin{equation}
\mathbf{A}(\mathbf{r},t)=\sum_{\mathbf{k}}\left(\frac{\hbar}{2 \omega_\mathbf{k} \epsilon_0 V}\right)^{1/2}\mathbf{u}_{\mathbf{k}} \left[a_{\mathbf{k}} e^{i \mathbf{k}\cdot \mathbf{r}-i \omega_{\mathbf{k}} t}+a_{\mathbf{k}}^{\dagger} e^{-i \mathbf{k}\cdot \mathbf{r}+i \omega_{\mathbf{k}} t} \right],
\label{vecPot}
\end{equation}
where the wave vector $\mathbf{k}$ obtains only discrete values depending on size of the box, the frequency $\omega_{\mathbf{k}}$ satisfies the dispersion relation with the wave vector $\boldsymbol{k}$, $V$ is the volume of the box, $\mathbf{u}_{\mathbf{k}}$ is a unit vector identifying polarization, $\hbar$ is the reduced Planck constant and $\epsilon_0$ is the permitivity of the vacuum. Finally, $a_{\mathbf{k}}$ and $a_{\mathbf{k}}^{\dagger}$ are fundamental annihilation and creation operators obeying the commutation relation
\begin{equation}
    \left[a_{\mathbf{k}},a_{\mathbf{k'}}^{\dagger}\right]=\delta_{\mathbf{k},\mathbf{k'}}.
    \label{comRel}
\end{equation}
The annihilation and the creation operators are not Hermitian operators, and therefore they do not correspond to any measurement. Their introduction is useful for building the quantum theory of light, and their actions on the quantum state can be conditionally approached in a laboratory. Their formal role in the theory stems from the Hamiltonian of an electromagnetic field
\begin{equation}
    H=\frac{1}{2}\int (\epsilon_0 \mathbf{E}\cdot \mathbf{E}+\mu_0 \mathbf{B}\cdot \mathbf{B})\mathrm{d}\mathbf{r},
\label{Hammiltonian}
\end{equation}
which is formulated in terms of the operators of electric intensity $\mathbf{E}=-\partial_t \mathbf{A}$ and the  magnetic induction $\mathbf{B}=\nabla \times \mathbf{A}$. Substituting the solution of the Maxwell equations (\ref{vecPot}) into the Hamiltonian (\ref{Hammiltonian}) gives \cite{D.F.Walls2008}
\begin{equation}
    H=\sum_{\mathbf{k}}\omega_{\mathbf{k}}\left(a_{\mathbf{k}}^{\dagger}a_{\mathbf{k}}+\frac{1}{2}, \right)
    \label{ham}
\end{equation}
which is expressed by those fundamental operators. 
It is useful to define the canonical coordinate $X_k$ and the canonical momentum  $P_k$ observables according to
\begin{eqnarray}
   X_k&=&\frac{1}{\sqrt{\omega_{{\mathbf{k}}}}}\left(a_k+a_k^{\dagger}\right)\nonumber \\
   P_k&=&i \sqrt{\omega_{\mathbf{k}}} (a_k-a_k^{\dagger}).
   \label{XP}
\end{eqnarray}
Using these directly measurable observables of light detectable by homodyne detection, the Hamiltonian reads
\begin{equation}
    H=\sum_{\mathbf{k}}\left(P_k^2+ \omega_{\mathbf{k}}^2X_k^2 \right),
    \label{Hammiltonian1}
\end{equation}
which resembles the Hamiltonian of independent linear harmonic oscillators. It shows the light mode behaves formally as an oscillator of the electric field. So far, formalism has been introduced for light in the virtual box with a finite volume. Description of light propagating in the free space before and after the interaction is determined from (\ref{vecPot}), (\ref{Hammiltonian}) and (\ref{Hammiltonian1}) in a limit of the box with the infinite size. As a consequence, the wave vector $\mathbf{k}$ can obtain arbitrary value and the summation over the modes in (\ref{vecPot}), (\ref{Hammiltonian}) and (\ref{Hammiltonian1}) is replaced by integration.

Let us further investigate the light occupying only a single mode with a given wave vector, consider the frequency $\omega_{\mathbf{k}}=1$ and avoid the dependence of the operators on the wave vector in the notation for simplicity.
The impacts of the creation and annihilation operators can be inspected on the eigenstates of the Hamiltonian, which are identified by
\begin{equation}
    H \vert n \rangle = E_{n} \vert n \rangle.
    \label{eigH}
\end{equation}
Employing the commutation relation (\ref{comRel}) reveals \cite{JohnDepartmentofPhysics2008}
\begin{eqnarray}
    H a^{\dagger} \vert n \rangle &=&a^{\dagger}(E_{n}+1)\vert n \rangle \nonumber \\
    H a \vert n \rangle &=& a(E_{n}-1)\vert n \rangle. 
\end{eqnarray}
It entails the creation operator increases the energy by a single unit, whereas the annihilation operator decreases the energy by a single unit. It means the single-mode light occupies the Hilbert space spanned by the eigenstates $\vert n \rangle$ where $n$ is an integer. The states $\vert n \rangle$ are called Fock states. The vacuum $\vert 0 \rangle$ is defined as a state whose energy cannot be diminished, i. e. $a \vert 0 \rangle=0$. Any Fock state $\vert n \rangle$ is produced by a sequential application of the creation operator on the vacuum
\begin{equation}
    \vert n \rangle=Z_n \left(a^{\dagger} \right)^n \vert 0 \rangle,
\end{equation}
where $Z_n$ is a normalization constant determined from
\begin{equation}
    a^{\dagger}a\vert n \rangle=n \vert n \rangle.
\end{equation}
It leads to $Z_n=1/\sqrt{n!}$ and to identities
\begin{eqnarray}
   a^{\dagger} \vert n \rangle &=& \sqrt{n+1} \vert n + 1 \rangle \nonumber \\
   a \vert n \rangle &=& \sqrt{n} \vert n - 1 \rangle.
   \label{cran}
\end{eqnarray}
The relations (\ref{cran}) show that creation and annihilation operators change the superpositions in the Fock state basis. Although the annihilation and the creation operators are not observables, their acting on a quantum state was approached probabilistically in an experiment \cite{Parigi2007}. The Fock states have simultaneously represented primary target states and resources for application in quantum optics since they were predicted theoretically.

\subsection{Wigner representation of the quantized light}

Light description in the Fock state basis is practically convenient when the number of involved photons is limited and relatively small. For multi-photon nonclassical light, it is better to use a continuous variable representation. 
Such continuous-variable representation is based on a displacement unitary operator $ e^{\beta a^{\dagger}-\beta^* a}$, analogous to the Fock state representation using the anihination and creation operators.  Average of such displacement operator defines a characteristic function given by
\begin{equation}
    \chi_s(\beta, \beta^*)=\mbox{Tr} \left[ \rho  e^{\beta a^{\dagger}-\beta^* a}\right],
    \label{xis}
\end{equation}
Where subscript s stands for symmetric ordering of annihilation and creation operators. Employing the integral transformation defines the central continuous representation in quantum optics - Wigner function
\begin{equation}
    W(\alpha,\alpha^*)=\frac{1}{4\pi^2}\int \chi_s(\beta, \beta^*)e^{-\beta \alpha^*+\beta^* \alpha}\mathrm{d}^2\beta.
    \label{wRep}
\end{equation}
Since the eigenstates of the symmetrically ordered operators are the eigenstates of the rotated canonical coordinate $X_{\phi}=a e^{i\phi}+a^{\dagger}e^{-i\phi}$, it is convenient to express the arguments of the Wigner function in terms of real quantities
\begin{eqnarray}
    x_{\phi}&=& \alpha e^{i\phi}+\alpha^* e^{-i\phi} \nonumber \\
    p_{\phi}&=&i\left[\alpha e^{i\phi}-\alpha^* e^{-i\phi}\right].
\end{eqnarray}
Then, the Wigner function obeys \cite{Wigner1932}
\begin{equation}
    \langle x_{\phi} \vert \rho \vert x_{\phi} \rangle = \int W(x_{\phi},p_{\phi})\mathrm{d}p_{\phi}
    \label{wigInt}
\end{equation}
with $X_{\phi}|x_{\phi}\rangle=x_{\phi}|x_{\phi}\rangle$, which represents probability density function of the rotated coordinate. However, the Wigner function itself does not always corresponds to a probability density function since it can obtain negative values \cite{Kenfack2004}. Moreover, the uncertainty principle does not allow the Wigner function to become the Dirac delta function. On other hand, it can have a structures below standard deviation used in uncertianty principles. Due to negative values, Wigner function is therefore a quasi-probability function meaning that only the integration (\ref{wigInt}) always leads to a probability density function for any observable. 
Any squeezed coherent state will have a positive semidefinite Wigner function. Their classical mixtures can form complex non-Gaussian shapes of the Wigner function; however, positivity remains. Therefore, the negativity of the Wigner function is a particular aspect of the quantum non-Gaussianity. Precisely, it is a sufficient condition for quantum non-Gaussianity. Until 2011, it was the only criterion of quantum non-Gaussianity. 
The Wigner function can be reconstructed from the homodyne measurement directly detecting statistics of the continuous operator $X_{\phi}$. Scanning the homodyne measurement for different $\phi$ enables tomographic reconstruction of the Wigner function \cite{Lvovsky2009}. The Wigner function $W(\alpha)$ can also be detected directly in any point of its phase space by measuring the mean value of the parity operator $\left(-1\right)^{a^{\dagger} a}$ after the proper displacement by $e^{\alpha a^{\dagger}-\alpha^* a}$ since
\begin{equation}
    W(0)=\frac{1}{2\pi}\langle \left(-1\right)^{a^{\dagger} a}\rangle. 
    \label{parWig}
\end{equation}
Measurement of parity can be substituted by photon counting, as these observables commute. This enables detection of the Wigner function in a single point belonging to the origin. It provides an operationalistic definition of the Wigner function. The displacement operator purely transforms the arguments of the Wigner function \cite{Royer1977}. 

\subsection{Classical theory of coherence and coherent states}

The classical theory of the coherence investigates impacts of stochastic processes on the coherence of waves in optics. A fundamental object in this theory is a stochastic coherent wave with a fluctuating amplitude $\alpha(t)$. Detectors do not measure this amplitude, but the integrated intensity $W(t)$ given by \cite{Loudon2000}
\begin{equation}
    W(t_1,\Delta t)=\frac{1}{\Delta t}\int_{t_1}^{t_1+\Delta t} \alpha^*(t)\alpha(t)\mathrm{d}t,
    \label{intInt}
\end{equation}
where the continuous measurement is carried out between time $t_1$ and $t_1+\Delta t$. Because the amplitude $\alpha(t)$ can be influenced by random processes of both source and detector, measurement results are quantified by the statistical moments of the integrated intensity $\langle W(t)^n \rangle =\langle \left(\alpha^* \alpha \right)^n \rangle $ averaged over several realizations of the measurement. A broad class of physical situations is described sufficiently with the assumptions that the $\langle W(t)^n \rangle$ are stationary , i.e. independent of time \cite{Loudon2000}. 

Normalized correlation functions describe influence of the stochastic processes on the coherence. They statistically quantify random changes of the amplitude $\alpha(t)$. The correlation functions are ordered according to the largest moment of the intensity that involve. For stationary processes, the correlation function of the first-order $g^{(1)}(\tau)=\langle \alpha^*(\tau) \alpha(0) \rangle/\langle (\alpha^*(0) \alpha(0)) \rangle$ quantifies the visibility in the Mach-Zehnder or Michelson interferometer and as such, it is sensitive to the multimode structure of the light \cite{Loudon2000}. For a single mode of radiation, all photons are completely indistinguishable and $ g^{(1)}(\tau) =e^{i \omega \tau}$ with $\omega$ being the angular frequency of the mode. The second-order correlation function $g^{(2)}(\tau)=\langle \alpha^*(\tau) \alpha^*(0) \alpha(\tau) \alpha(0) \rangle/\langle (\alpha^*(0) \alpha(0)) \rangle^2$ describes coherence aspects of intensity fluctuations for the stationary light. It is measured in a Hanbury Brown and Twiss experiment layout where light is split on a beam-splitter and measured by two detectors responding to the integrated intensity \cite{BROWN1956}. Since the functions $g^{(1)}(\tau)$ and $g^{(2)}(\tau)$ do not provide complete information about the coherence, one can introduce the $n$th-order correlation function \cite{Loudon2000}
\begin{eqnarray}
&\ &g^{(n)}(\tau_1,..,\tau_{n-1})=\nonumber \\
&\ &\frac{\langle  \alpha^*(\tau_{n-1})\times ... \times \alpha^*(\tau_1)\times \alpha^*(0)  \alpha(\tau_{n-1})\times ... \times\alpha(\tau_1)\times \alpha(0) \rangle}{\langle \alpha^*(0) \alpha(0) \rangle^n}.
\label{gntau}
\end{eqnarray}
Ideal classical coherent wave obeys $g^{(n)}(0)=1$ for every order $n$ and any times $\tau_1$, ..., $\tau_{n-1}$. R. Glauber provided a revolutionary explanation of this classical theory in terms of quantum optics. It stems from a substitution of the classical amplitudes by normally ordered combinations of the annihilation and creation operators, i. e. \cite{Glauber1963a}
\begin{eqnarray}
&\ &g^{(n)}(\tau_1,..,\tau_{n-1})\rightarrow \nonumber \\
&\ &\frac{\langle  a^{\dagger}(\tau_{n-1})\times ... \times a^{\dagger}(\tau_1)\times a^{\dagger}(0)  a(\tau_{n-1})\times ... \times a(\tau_1)\times a(0) \rangle}{\langle a^{\dagger}(0) a(0) \rangle^n}, 
\label{gnQuantum}
\end{eqnarray}
where the times $\tau_i$ are ordered increasingly, i. e. $\tau_i \geq \tau_{i-1}$. The normal ordering is chosen because the historically first available optical detectors used the photon absorption for the detection, and therefore they measured the operators in this ordering \cite{Glauber1963}. The relations $g^{(n)}(\tau_1,..,\tau_{n-1})=1$ specify unambiguously a pure quantum state acting as a coherent classical wave \cite{Glauber1963a}. Because of the normal ordering in the definition, the state is determined from $a \vert \alpha \rangle = \alpha \vert \alpha \rangle$. The stochastic processes on coherent states $\vert \alpha \rangle$ establish a class of states corresponding to classical waves
\begin{equation}
    \rho=\int P(\alpha)\vert \alpha \rangle \langle \alpha \vert \mathrm{d}^2\alpha,
    \label{classicalStates}
\end{equation}
where the function $P(\alpha)$ is the probability density function. Although all states can be expressed formally in form (\ref{classicalStates}), only the states with $P(\alpha)$ being the probability density function behave like classical noisy waves in the detection based on the photon counting. On the other hand, any state of light is nonclassical, i.e. beyond classical wave description in coherence theory, when the function $P(\alpha)$ is negative or more singular than the Dirac delta function \cite{Sudarshan1963}. The mixtures of coherent states constitute a very broadly accepted definition of classical states in the optics if the exploited classical detectors measure  the integrated intensity (\ref{intInt}) \cite{Mandel1986}. The boundary of the classical states is considered as a reference for detection of quantum phenomena beyond the classical wave description. Nowadays, the nonclassical states can be extended from quantum optics to any bosonic system. Their use can go beyond quantum coherence motivation based on operational definition of coherent states rising from linear external drive of linear harmonic oscillators. Such operational definition does not depend on the detection only on the state preparation.   

\subsection{Nonclassical features of quantum states}

The states beyond the semi-positive regular $P$ function are nonclassical.  The $P$ function can be reconstructed from homodyne tomography \cite{Kiesel2008}. However, it requires a lot of different measurements and the deconvolution procedure is not stable. It increases bias and statistical errors and limits the reliability of nonclassicality detection. It is necessary to develop criteria recognizing the nonclassicality from available direct measurements. 

The correlation functions (\ref{gnQuantum}) can be expressed directly by the function $P$ due to the normal ordering, and therefore they can expose some cases when the function $P$ represents nonclassical states. The simplest correlation function enabling that is the second-order correlation function
\begin{equation}
    g^{(2)}(\tau)=\frac{\langle a^{\dagger}(\tau)a^{\dagger}(0) a(\tau) a(0) \rangle}{\langle a^{\dagger}(0)a(0) \rangle \langle a^{\dagger}(0)a(0) \rangle},
    \label{g2tau}
\end{equation}
where $\tau$ is delay time involved for measurement of time correlations.

Due to the Cauchy-Schwarz inequality, only the nonclassical states can gain \cite{Volovich2016}
\begin{equation}
    g^{(2)}(\tau)<1
    \label{g2tauCond}
\end{equation}
for some delay time $\tau$. When measurement results obey the condition for $\tau=0$, it is said that light manifests sub-Poissonian photon statistics \cite{Mandel1986}. The inverse situation when $g^{(2)}(0)>1$ corresponds to super-Poissonian photon statistics. Both classical and nonclassical states can exhibit the super-Poissonian photon statistics, but only nonclassical states can be sub-Poissonian \cite{Mandel1986}.  
The sub-Poissonian photon statistics can also be formulated to suppress energy fluctuations below the Poissonian (shot-noise) limit of coherent states. It is quantified by the Fano factor \cite{Mandel1979}
\begin{equation}
    F=\frac{\langle n^2 \rangle -\langle n \rangle^2 }{\langle n \rangle}
    \label{fano}
\end{equation}
representing a ratio between the variance of the intensity and its mean value. The commutation relation (\ref{comRel}) together with the definition of the function $g^{(2)}(0)$ leads to an identity
\begin{equation}
    F=\langle n \rangle \left[g^{(2)}(0)-1\right]+1,
\end{equation}
according to which the sub-Poissonian light exhibits $F<1$. However, Fano factor and $g^{(2)}(0)$ can change differently with statistics of photons. While $g^{(2)}(0)$ for an attenuated single-photon is always strictly vanishing, Fano factor $F=1-\langle n \rangle=1-\eta$ varies with the attenuation factor $\eta$. Therefore, no fluctuations in number of photons (energy) are achieved only for $\eta=1$. In general, the ideal Fock states $\vert n \rangle$ defined in (\ref{eigH}) yield
\begin{eqnarray}
    g^{(2)}(0)&=&1-\frac{1}{n}\nonumber \\
    F&=&0
    \label{g2Fano}
\end{eqnarray}
showing the sub-Poissonian statistics for $n$. In the quantum non-Gaussian Wigner function, such possibility is accompanied by negative values forming rings rising with $n$. Note that any Fano factor value can be asymptotically obtained by displacing Gaussian squeezed states of light. Their recognition is one of many reasons why to introduce directly measurable criteria of quantum non-Gaussianity. 

For $\tau>0$, nonclassicality evaluation from the correlation function (\ref{g2tau}) involves temporal effects and relations (\ref{g2Fano}) do not hold anymore. In that case, time correlation $g^{(2)}(\tau)$ can identify different aspect of nonclassicality employing the criterion
\begin{equation}
    g^{(2)}(\tau_2)>g^{(2)}(\tau_1),
    \label{antibunch}
\end{equation}
where $\tau_2>\tau_1>0$, as can be proved using the Cauchy-Schwarz inequality \cite{Volovich2016}. The nonclassical phenomenon when $g^{(2)}(\tau)$ grows with the positive delay time $\tau$ is called antibunching. It means, that photon correlation is weaker for smaller time delay, the opposite of the case for thermal or even Poissonian radiation. Although the antibunching is often associated with the sub-Poissonian statistics, the conditions (\ref{g2tauCond}) and (\ref{antibunch}) are not equal in general \cite{Mandel1986}. It can be detected for light radiated from a couple of single-photon emitters where detection does not distinguish which emitter radiates the light. In this case, the light manifests antibunching but super-Poissonian statistics \cite{Skornia2001}.

Both sub-Poissonian light and antibunching represent only sufficient conditions of nonclassicality. There are many nonclassical states that obey neither (\ref{g2tauCond}) or (\ref{antibunch}). Then, one can use the photon correlation function of a higher order than two and, generally, a function of the creation and annihilation operators set in the normal order \cite{Shchukin2005}. It can recognize the nonclassicality in cases when (\ref{g2tauCond}) or (\ref{antibunch}) fails. However, a necessary condition of nonclassicality requires fulfilling all possible conditions that can be formulated. That can never be verified for general states. Therefore, continuous development of nonclassical criteria with different strengths and suitable for new light sources or matter effects detected by light statistics is necessary. 

All the criteria mentioned above are defined using moments of the creation and annihilation operators. These moments can be computed from the detection of the integrated intensity (\ref{intInt}) of strong beams of light in classical optics. However, those intensity detectors cannot measure weak light manifesting the nonclassicality due to their low detection efficiency and noise \cite{Eisaman2011}. Detection of such weak light by developing single-photon detectors gradually involved a phototube \cite{Kimble1977}, a single-photon avalanche diode \cite{Achilles2003} and,  recently, a transition edge sensor \cite{Hoepker2019}. These detectors are very sensitive to the individual photons, but they do not allow measurement or estimation of the integrated intensity \cite{Sperling2012}, as they are not sufficiently photon-number-resolving. Therefore, the nonclassicality was revealed only when the intensity moments were approximately calculable from the detector outcomes. Accurate and reliable nonclassicality detection requires criteria incorporating measurement layout and probabilities of direct responses of the employed detectors. 

\subsection{Basics of experimental detection of nonclassical light}

\begin{figure}
    \centering
    \includegraphics[width=0.35 \linewidth]{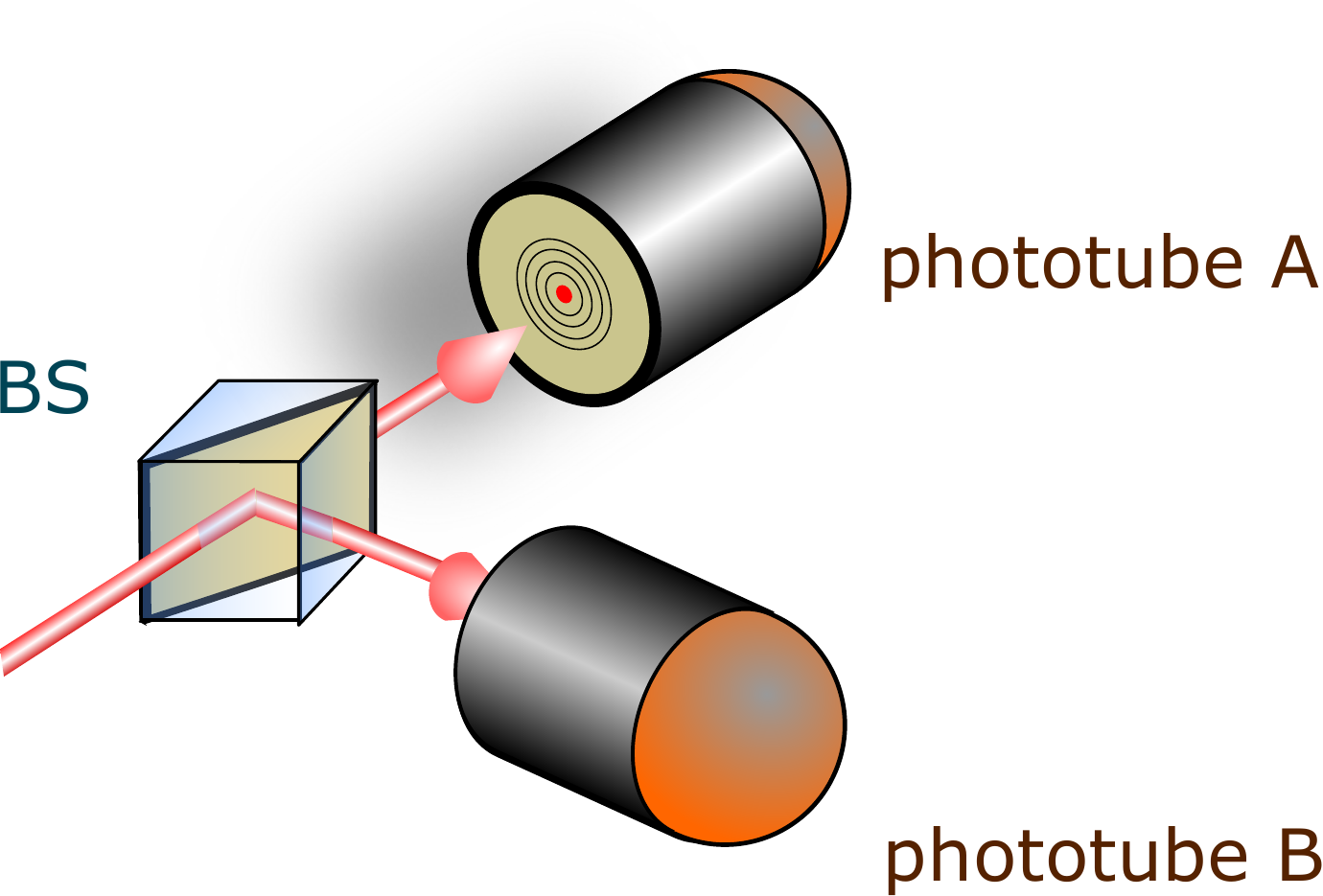}
    \caption{A layout providing detection of nonclassicality in the first experiments \cite{Kimble1977,Short1983, Grangier1986}. An impinging signal was split by a beam splitter and directed to two phototubes.}
    \label{fig:oldExp}
\end{figure}

The nonclassicality was revealed in an experimental by observation of the antibunching \cite{Kimble1977}. It was followed by the detection of the sub-Poissonian light \cite{Short1983, Grangier1986}. In all these experiments, an atom scattered light resonantly, the scattered light impinged on a beam-splitter and was measured by two phototubes as depicted in Fig.~\ref{fig:oldExp}. The phototube is a detector that converts incoming light to an electric current. The electric current signals incoming photons without distinguishing their number. In modern experiments, single-photon avalanche diodes have a binary response as well but operate with higher quantum efficiency compared to phototubes.

For very weak emitted single mode light, density matrix can be approximated by
\begin{equation}
    \rho \approx (1-\eta_1-\eta_2)\vert 0 \rangle \langle 0 \vert +\eta_1 \vert 1 \rangle \langle 1 \vert + \eta_2 \vert 2 \rangle \langle 2 \vert,
    \label{approxSt}
\end{equation}
where contributions of three and more photons are neglected and $\eta_1 \gg \eta_2$. In that case, probabilities $\eta_1$ and $\eta_2$ can be approximated by the normally-ordered moments of the annihilation and the creation operators $\eta_1 \approx \langle a^{\dagger} a \rangle$ and $\eta_2 \approx \langle \left(a^{\dagger})^2 a^2 \right\rangle /2$. The detection events can be quantified by probabilities
\begin{eqnarray}
   P_a &\approx& T q  \langle a^{\dagger} a \rangle \nonumber \\
     P_b &\approx& (1-T) q \langle a^{\dagger} a \rangle \nonumber \\
    P_{ab} &\approx& T(1-T) q^2 \langle \left(a^{\dagger})^2 a^2 \right\rangle
    \label{photocurrent}
\end{eqnarray}
where $P_{a}$ ($P_b$) is a probability that the detector $A$ ($B$) registers photons, $P_{ab}$ denotes the probability of simultaneous registering by both detectors,  $q$ is the quantum efficiency of the detectors and $T$ is the transmission of the beam-splitter. A comparison with the definition of the second-order correlation function in (\ref{g2tau}) leads to
\begin{equation}
    g^{(2)}(0) = \frac{\langle a^{\dagger 2} a^2\rangle}{\langle a^{\dagger} a\rangle^2} \approx \frac{P_{ab}}{P_a P_b}
    \label{approxg2}
\end{equation}
valid in the limit (\ref{approxSt}). In this approximation, the sub-Poissonian light is recognized from such detection layout when 
\begin{equation}
    \frac{P_{ab}}{P_a P_b}<1.
\end{equation}
The antibunching can be also detected in the approximation if the phototube $B$ registers signal with a time delay $\tau$ against the phototube $A$. In this case, the measurement leads to
\begin{equation}
     g^{(2)}(\tau)\approx \frac{P_{ab}(\tau)}{P_a P_b},
     \label{approxg2tau}
\end{equation}
where $P_{ab}(\tau)$ quantifies the time correlation of clicks of both phototubes. Here, the probabilities $P_{a,b}$ depends only on the time delay $\tau$ for the stationary emission of light. The denominator normalizing probablity $P_{ab}(\tau)$ of detector coincidence fulfils the identity
\begin{equation}
    P_a P_b=\lim_{\tau \rightarrow \infty} P_{ab}(\tau)
\end{equation}
simply using that very delayed signals are uncorrelated from any source. This normalization gives a physical meaning to the expression in (\ref{approxg2}) as a ratio comparing the probability $P_{ab}(\tau)$ with itself measured for very large $\tau$.

In all these cases, the detection of nonclassical light using the basic HBT layout depends on the accuracy of the approximation in (\ref{approxSt}). The light sources in the first experiments \cite{Kimble1977, Short1983, Grangier1986} emitted weak light and approached well the density matrix of the form (\ref{approxSt}). The current sources of nonclassical light can exceed this limit when the single-photon rate and multi-photon rate increase. Although attenuating any state can prepare the state in this approximation, it slows the experiment, makes it sensitive to instabilities and increases the experimental error bars. It can cause nonclassicality detection to fail due to poor statistics of clicks. Therefore, ab initio criteria derived for a general detector layout without such approximations of unknown light statistics are required.

\subsection{Gaussian states at a beam splitter}

The starting example considers classical linear driving of an optical mode, which corresponds to a unitary operator defined as
\begin{equation}
    D(\alpha)=e^{\alpha^* a-\alpha a^{\dagger}}.
\end{equation}
The operator is called a displacement operator because it transforms the annihilation and creations operator according to
\begin{eqnarray}
   D(-\alpha)a D(\alpha)&=&a+\alpha \nonumber \\
   D(-\alpha)a^{\dagger}D(\alpha)&=&a^{\dagger}+\alpha^*
\end{eqnarray}
by linear adding coherent classical energy to the optical mode. The former relation implies an identity
\begin{equation}
    a D(\alpha)\vert 0 \rangle=D(\alpha)D(-\alpha)a D(\alpha)\vert 0 \rangle=D(\alpha)(a+\alpha)\vert 0 \rangle =\alpha D(\alpha)\vert 0 \rangle.
\end{equation}
Since by the definition, the coherent state are an eigenstate of the annihilation operator, the coherent state can be expressed as
\begin{equation}
    \vert \alpha \rangle = D(\alpha)\vert 0 \rangle.
    \label{cohDef2}
\end{equation}
It represents an equivalent and more operational definition of the coherent state independent of the photodetection. According to that definition, a coherent state is a result of dynamics driven by an interaction Hamiltonian $H=i(\alpha a^{\dagger}-\alpha^* a)$ linear in the annihilation and creation operators. Nonclassical states of bosons, therefore, require dynamics beyond such linear interaction Hamiltonians.   

An example of such operations is amplification/squeezing operator
\begin{equation}
S(\xi)=e^{\xi \left(a^{\dagger}\right)^2-\xi^* a^{2}}
\label{sqOp}
\end{equation}
generated by a quadratic interaction Hamiltonian. In quantum optics, it rises from a trilinear Hamiltonian $H=i\left[g\left(a^{\dagger}\right)^2 b-g^*a^2 b^{\dagger}\right]$, where $b$ ($b^{\dagger}$) represents the annihilation (creation) operator of pumping light. When the pumping is by a strong classical beam undepleted by such interaction, the operators $b$ and $b^{\dagger}$ can be substituted by the amplitudes $\beta$ and $\beta^*$ and the Hamiltonian driving the evolution gets a quadratic form (\ref{sqOp}) \cite{D.F.Walls2008}. The parameter $\xi$ in (\ref{sqOp}) is given by the product of $g \beta$ and time of the evolution.
The squeezing operator transforms the annihilation and the creation operators according to
\begin{eqnarray}
   S(-\xi)a S(\xi)&=&a \cosh{2\vert \xi\vert}+a^{\dagger}e^{i\phi}\sinh{2\vert \xi\vert}\nonumber \\
   S(-\xi)a^{\dagger} S(\xi)&=&a e^{-i\phi}\sinh{2\vert \xi\vert}+a^{\dagger} \cosh{2\vert \xi\vert},
\end{eqnarray}
where $\phi$ is determined from $\xi=\vert \xi \vert e^{i \phi}$. Substituting the transformations into the canonical coordinate $X=(a+a^{\dagger})$ and the canonical momentum $P=i\left(a-a^{\dagger}\right)$ illustrates a role of the squeezing operator in measurement of these continuous observables. It results in
\begin{eqnarray}
    S(\xi)X S(-\xi)&=&X \left(\cosh{2\vert \xi\vert}+\cos{\phi}\sinh{2\vert \xi\vert}\right)\nonumber \\
    &-&P\sin{\phi}\sinh{2\vert \xi\vert} \nonumber \\
     S(\xi)P S(-\xi)&=&P \left(\cosh{2\vert \xi\vert}-\cos{\phi}\sinh{2\vert \xi\vert}\right)\nonumber \\
    &+&X\sin{\phi}\sinh{2\vert \xi\vert}.
    \label{squeezing}
\end{eqnarray}
For $\phi=0$ the squeezing operator deamplifies the $P$ operator, i. e. $\langle P \rangle$ decreases.  The variance $\langle P^2 \rangle-\langle P \rangle^2$ squeezes as well, potentially, below the variance of classical coherent states. Then, it directly demonstrates nonclassical phenomena and generates squeezed states. However, both mean value $\langle X \rangle$ and the variance $\langle X^2 \rangle-\langle X \rangle^2$ has to amplify due the necessary preservation of commutation relation between $X$ and $P$. 
 
Generalizing the definition of the coherent states, one can introduce a squeezed coherent (displaced) state \cite{Stoler1970,Yuen1976}
\begin{equation}
  \vert \alpha,\xi \rangle= S(-\xi)D(\alpha)\vert 0 \rangle
  \label{gaussSt}
\end{equation}
simply by squeezing a coherent state $\vert \alpha \rangle$. Formally, the state $\vert \alpha,\xi \rangle$ is eigenstate of complex superposition of annihilation and creation operators \cite{Yuen1976}
\begin{equation}
    (a \mu+a^{\dagger} \nu)\vert \alpha,\xi \rangle=(\alpha \mu+\alpha^* \nu)\vert \alpha,\xi \rangle,
\end{equation}
where $\mu=\cosh{2\vert \xi \vert}$ and $\nu=e^{i\phi} \sinh{2\vert \xi \vert}$.
The states $\vert \alpha,\xi \rangle$ are the only states that saturate the uncertainty principle \cite{Stoler1970}
\begin{equation}
    \mbox{var} (X) \mbox{var}(P) \geq 1
\end{equation}
stemming from the commutation relation $\left[ X, P \right]=2i$. Since the coherent states represent a special case of the state (\ref{gaussSt}), they saturates the uncertainty inequality as well but their $    \mbox{var} (X)=\mbox{var}(P)$. When the rotated canonical coordinate $X_{\phi}$ is measured on the state $\vert  \alpha,\xi \rangle$ the quantum noise can go below the vacuum fluctuation, i. e.
\begin{equation}
    \langle \alpha,\xi \vert X_{\phi}^2 \vert \alpha,\xi \rangle- \langle \alpha,\xi \vert X_{\phi} \vert \alpha,\xi\rangle^2<1,
    \label{supprVacPhi}
\end{equation}
for $\phi$ given by $\xi=\vert \xi \vert e^{2i\phi}$. It is nonclassical aspect because mixtures of the coherent states never reach (\ref{supprVacPhi}). The states with $\alpha=0$ are called squeezed vacuum states. The displacement operator acting on the vacuum squeezed states can produce sub-Poissonian statistics although the squeezed vacuum does not exhibit that \cite{Lemonde2014}. An expansion of the squeezed states in the Fock state basis results in \cite{Yuen1976}
\begin{equation}
    \langle n \vert \alpha,\xi \rangle = \frac{1}{\sqrt{n! \mu}} \left(\frac{\nu}{2\mu}\right)^{n/2}H_n\left(\frac{\beta}{\sqrt{2\nu \mu}}\right)e^{-\frac{1}{2}\vert \beta \vert^2+\frac{\nu^*}{2\mu}\beta^2},
    \label{gaussPn}
\end{equation}
where $\beta=\alpha \mu+\alpha^* \nu$ and $H_n$ is the Hermite polynomial of order $n$. It is a direct transfer of squeezing of the vaccum to squeezing in the number of quanta.  
In the Wigner representation, described in Section 2.8, the state $\vert \alpha,\xi \rangle$ obtains a Gaussian form. Therefore, the first and all second moments of $X$ and $P$ identify the states $\vert \alpha,\xi \rangle$ unambiguously. It is convenient to organize the second moments to a covariance matrix for detailed calculations. In Chapter 5, such Gaussian states will be used to define a threshold for the quantum non-Gaussianity beyond the negative values of the Wigner function.

The squeezed coherent state $\vert \alpha,\xi \rangle$ can be further generalized beyond the Gaussian states by applying the displacement and squeezing operators sequentially to the Fock states, giving 
\begin{equation}
    \vert \alpha,\xi, n \rangle = S(-\xi)D(\alpha)\vert n \rangle
    \label{betan}
\end{equation}
called a squeezed-displaced Fock state straightforwardly. The states with $n>0$ are not Gaussian states anymore as their Wigner functions exhibit negative values. In Chapter 6, the states $\vert \alpha,\xi, n \rangle$ will be exploited to build a hierarchy of the non-Gaussian quantum aspects of the bosons. An extensive description of the features of these states is presented in \cite{Kral1990}.

A last introduced operator describing mixing of two distinct light modes obtains a form
\begin{equation}
    U(\kappa)=e^{\kappa a_1 a_2^{\dagger}-\kappa^* a_1^* a_2},
\end{equation}
where the subscripts of the annihilation and creation operators denote the respective modes. The operator describes interference occurring in a beam-splitter (BS) in the free-space optics and, simultaneously, it corresponds to interference in linear optical couplers in the fiber optics. The operator acts on the annihilation operator as
\begin{eqnarray}
    U_{BS}(-\kappa)a_1 U_{BS}(\kappa)&=&a_1 \cos \vert \kappa \vert -e^{-i\phi} a_2\sin \vert \kappa \vert \nonumber \\
    U_{BS}(-\kappa)a_2 U_{BS}(\kappa)&=&e^{i\phi} a_1 \sin \vert \kappa \vert + a_2\cos \vert \kappa \vert
    \label{ubs}
\end{eqnarray}
with $\kappa=\vert \kappa \vert e^{i \phi}$. Let us set $\phi=0$ for simplicity and inspect how the operator influences the displacement operator. We arrive at an identity
\begin{equation}
    U_{BS}(-\kappa)D_1(\alpha)D_2(\beta)U_{BS}(\kappa)=D_1(\sqrt{T}\alpha+\sqrt{1-T}\beta)D_2(-\sqrt{1-T}\alpha+\sqrt{T}\beta),
\end{equation}
where $T=\cos^2 \vert \kappa \vert$ is the transmission of a BS and subscripts distinguish the modes. It holds due to the Baker – Campbell – Hausdorf theorem \cite{Scully1997}. Because the coherent state is defined as a result of acting of the displacement operator on the vacuum, they are affected by
\begin{equation}
    U_{BS}(T)\vert \alpha \rangle \vert \beta \rangle =\vert \sqrt{T}\alpha+\sqrt{1-T}\beta\rangle \vert -\sqrt{1-T}\alpha+\sqrt{T}\beta\rangle.
\end{equation}
The amplitudes $\alpha$ and $\beta$ transform identically with the amplitudes of classical coherent waves that are split on a BS with the transmission $T$. The relations (\ref{ubs}) establish also how the Fock states interfere on a BS. Since any Fock state can be expressed as $\vert n \rangle = (a^{\dagger})^n/\sqrt{n!}\vert 0 \rangle$, the interference on a BS leads to
\begin{eqnarray}
    U_{BS}(T)\vert m\rangle_1 \vert n \rangle_2 &=& \frac{1}{\sqrt{m!n!}}\left(\sqrt{T}a_1^{\dagger} +\sqrt{1-T}a_2^{\dagger}\right)^m \nonumber \\
    &\times& \left(-\sqrt{1-T}a_1^{\dagger} +\sqrt{T}a_2^{\dagger}\right)^n \vert 0 \rangle_1 \vert 0 \rangle_2.
\end{eqnarray}
As a particular case, let us consider the second mode is occupied by the vacuum. It follows in
\begin{equation}
U_{BS}(T)\vert m\rangle_1 \vert 0 \rangle_2=\sum_{k=0}^{m} \sqrt{\binom{n}{k}} (-1)^{m-k}T^{k/2}(1-T)^{(m-k)/2}\vert k \rangle_1 \vert m-k \rangle_{2}.
\end{equation}
Apparently, the state $\vert m \rangle$ behaves as $m$ classical particles that are transmitted according to the binomial distribution.
When more BSs are used the light is split among several modes. Since such networks are used commonly for increasing the knowledge about a distribution of photons in measured light, we focus on these networks in more details.    

\subsection{Multi-photon detection}

\begin{figure}
\centering
\includegraphics[width=0.9\linewidth]{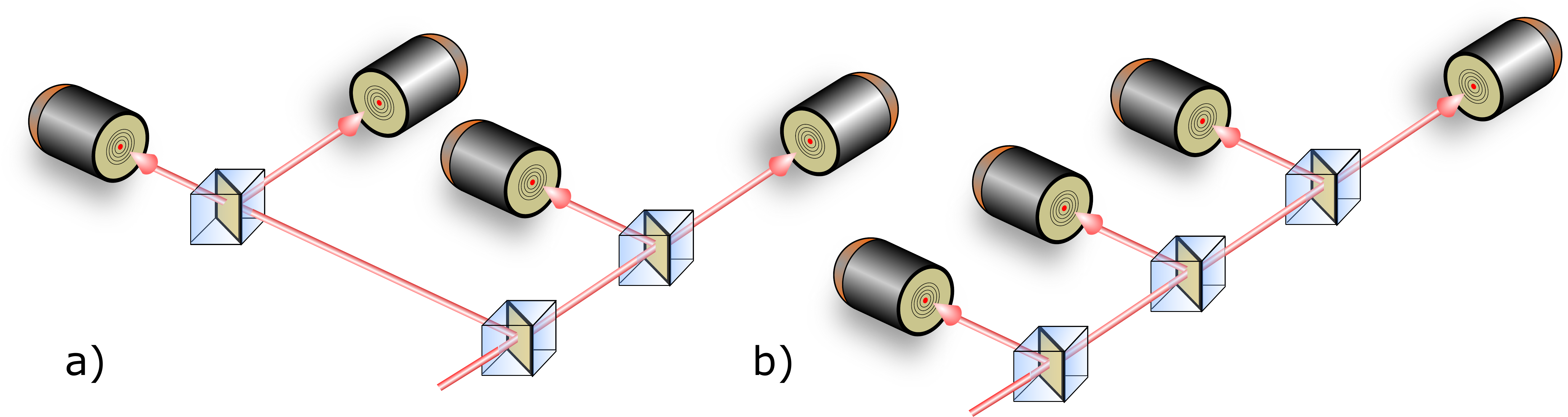}
\caption{Examples of networks splitting a state of light among several emerging modes detected by SPAD. The network can have a tree structure (\emph{a}) or split the light successively by a series of BSs (\emph{b}).}
\label{fig:splitting}
\end{figure}

A multiphoton detector can approximately measure statistics of many arriving photons by only splitting into many modes with single-photon detectors \cite{Hlousek2019}. Still, such detection has systematic errors when two photons do not split. To avoid risks that this will compromise conclusive and faithful detection of nonclassical phenomena, we resign to rely on photon number statistics or even modify criteria for photo-click statistics. We instead use the ab initio approach and derive the criteria for the well-characterized detector used in the laboratory. 
The multiphoton detector consists of $N-1$ BSs guiding the light towards $N$ single-photon avalanche diodes (SPAD) as depicted in Fig.~\ref{fig:splitting}. After the splitting, when any mode is occupied by at least one photon, the SPAD placed in that mode provides a detection event - a click. The click statistics approach photon statistics only when $N$ is sufficiently large. However, it is challenging to specify the systematic error even in that case. Assuming detector inefficiency and noise conservatively to be projected to measured statistics, we describe each SPAD by a positive-operator valued measure (POVM) having two-components $\mathbb{1}_i-\vert 0 \rangle_i\langle 0 \vert$ and  $\vert 0 \rangle_i\langle 0 \vert $, where the subscript $i$ denotes individual emerging modes. For example, we further consider a balanced network splitting the light equally among $N$ SPADs individually in $N$ modes.  We denote the respective probability of $n\leq N$ simultaneous clicks by
\begin{equation}
    P_n=1+\sum_{k=0}^n (-1)^k \binom{n}{k} P_0(k/N),
    \label{splittingFormula}
\end{equation}
where for coherent state $\vert \alpha \rangle$, $P_0(\tau)=\exp(-\vert \alpha \vert^2 \tau)$ leads to
\begin{equation}
    P_n=\left(1-e^{-\vert \alpha \vert^2/N}\right)^n
\end{equation}
which will be useful for ab initio nonclassicality criteria. For the Gaussian states we derive the click statistics from the Wigner representation.

According to the relation (\ref{parWig}), the Wigner function in the origin of the phase space represents a mean value of the parity operator $\Pi=\frac{1}{2\pi}\int \vert - x \rangle\langle x \vert \mathrm{d}x$. A value in any point of phase-space reads 
\begin{equation}
    W(\alpha,\alpha^*)=\frac{1}{2\pi}\int \langle x \vert D(\alpha) \rho D^{\dagger}(\alpha) \vert - x \rangle \mathrm{d}x,
\end{equation}
where $D(\alpha)=\exp(\alpha a^{\dagger}-\alpha^* a)$ is a displacement operator. Substituting the annihilation and creation operators by the canonical coordinate and momentum (\ref{XP}), using the Baker-Haussdorf formula and letting them act on the state $\vert x \rangle$ we obtain \cite{D.F.Walls2008} 
\begin{equation}
    W(x,p)=\frac{1}{2\pi}\int\langle x+\beta_2\vert \rho \vert x-\beta_2\rangle e^{i p \beta_2}\mathrm{d}\beta_2,
    \label{waveW}
\end{equation}
where $x=\alpha+\alpha^*$ and $p=i (\alpha-\alpha^*)$. Thus, the Wigner function is a Fourier transformation of off-diagonal elements of the density matrix expressed in eigenstates of the canonical coordinate. It follows, the vacuum with the wave function
\begin{equation}
    \vert 0 \rangle=\frac{1}{(2\pi)^{1/4}}\int e^{-\frac{x^2}{4}}\vert x \rangle \mathrm{d}x
\end{equation}
possesses the Wigner function with a Gaussian form
\begin{equation}
    W_0(x,p)=\frac{1}{2\pi}e^{-\frac{x^2+p^2}{2}},
    \label{vacW}
\end{equation}
saturating the uncertainty principle. Wigner functions of other states stemming from unitary linear  dynamics (described by Hamiltonians at most quadratic in position and momentum variables) of the vacuum state are given by transformation of the arguments in the Wigner function \cite{Scully1997}. Let us assume an initial state possesses $W(x,p)$. The unitary operator $D(\alpha)$ effects
\begin{equation}
   W(x,p)\rightarrow W(x+\alpha+\alpha^*,p+i(\alpha-\alpha^*)).
   \label{displW}
\end{equation}
The squeezing operator (\ref{sqOp}) with $\xi=\vert \xi \vert e^{i\phi}$, scaling the rotated coordinate and the rotated momentum, transforms the arguments according to
\begin{equation}
   W(x,p)\rightarrow W \left( x' e^{2\vert \xi\vert},p' e^{-2\vert \xi\vert}\right),
   \label{sqW}
\end{equation}
where $x'=\cos (\phi/2)x+\sin(\phi/2)p$ and $p'=-\sin(\phi/2)x+\cos(\phi/2)p$. Moreover, the beam splitter transforms arguments of the Wigner functions $W_1(x_1,p_1)$ and $W_2(x_2,p_2)$ as
\begin{eqnarray}
   W_1(x_1,p_1)W_2(x_2,p_2)&\rightarrow & W_1(\sqrt{T}x_1+\sqrt{1-T}x_2,\sqrt{T}p_1+\sqrt{1-T}p_2)\nonumber \\
    \times W_2(-\sqrt{1-T}x_1&+&\sqrt{T}x_2,-\sqrt{1-T}p_1+\sqrt{T}p_2),
    \label{BSW}
\end{eqnarray}
where $T$ is the transmission. Finally, the results of applying POVM on a state with the Wigner function $W(x,p)$ corresponds to an overlap \cite{Nehra2019}
\begin{equation}
    \int W_Q(x,p)W(x,p)\mathrm{d}x\mathrm{d}p,
\end{equation}
with the Wigner function $W_Q(x,p)$ of the POVM that can be obtain analogous using (\ref{waveW}) \cite{Nehra2019}. As an example, we mention a POVM corresponding to a click response of a SPAD
\begin{equation}
    W_{c}(x,p)=1-2 e^{-\frac{x^2+p^2}{2}}
    \label{povmCl}
\end{equation}
which will be used later for deriving the click statistics.

The transformations (\ref{displW})-(\ref{BSW}) applied on the Wigner function of the vacuum (\ref{vacW}) allow us to establish the Wigner function of any Gaussian state that is split among $M$ modes through a network of BSs in a form
\begin{equation}
    W(\boldsymbol{x})=\frac{1}{\pi^M \det \boldsymbol{\sigma^{-1}}}\exp\left[-\frac{1}{2}(\boldsymbol{x}-\boldsymbol{r}) \boldsymbol{\sigma} (\boldsymbol{x}-\boldsymbol{r})^T)\right],
    \label{wigG}
\end{equation}
where $\boldsymbol{x}=(x_1,p_1,...,x_M,p_M)$ is a vector with $x_i$ being the coordinate of the $i$th mode and $p_i$ being its momentum in the Wigner representation, $\boldsymbol{r}$ represents a vector of first moments of the coordinates and the momenta and $\boldsymbol{\sigma}$ is a covariance matrix. Explicitly, the elements of the vector $\boldsymbol{r}$ are given by $r_{2i+1}=\langle X_i \rangle$ and $r_{2i}=\langle P_i \rangle$ where $i$ distinguishes the modes. The covariance matrix $\boldsymbol{\sigma}$ has elements
\begin{eqnarray}
   \sigma_{2i+1,2j+1}&=&\frac{1}{2}(\langle X_i X_j \rangle+\langle X_j X_i \rangle)-\langle X_i \rangle \langle X_j \rangle \nonumber \\
    \sigma_{2i+1,2j}&=&\frac{1}{2}(\langle X_i P_j \rangle+\langle P_j X_i \rangle)-\langle X_i \rangle \langle P_j \rangle\nonumber \\
     \sigma_{2i,2j+1}&=&\frac{1}{2}(\langle P_i X_j \rangle+\langle X_j P_i \rangle)-\langle P_i \rangle \langle X_j \rangle\nonumber \\
      \sigma_{2i,2j}&=&\frac{1}{2}(\langle P_i P_j \rangle+\langle P_j P_i \rangle)-\langle P_i \rangle \langle P_j \rangle.
\end{eqnarray}
They represent symmetrically ordered moments fully specifying quantum noise in the Gaussian states. Transformations (\ref{displW})-(\ref{BSW}) change $\boldsymbol{r}$ and $\boldsymbol{\sigma}$ but preserve the Gaussian form. We can directly use this formalism for the calculation of the vacuum statistics 
\begin{equation}
    P_0(T)=2\frac{e^{-\frac{\vert \beta \vert^2 \tau}{2}\left[\frac{\cos^2 \phi}{\gamma(1/V,T)}+\frac{\sin^2 \phi}{\gamma(V,T)}\right]}}{\sqrt{\gamma(V,T)\gamma(1/V,T)}}
    \label{p0GaussM}
\end{equation}
after a beam splitter, where $\beta=\vert \beta \vert e^{i\phi}$ and $\gamma(V,\tau)=2 V+T \left(1-V\right)$, where $V$ is the minimal variance of the canonical coordinate, i. e. $V=e^{-2\vert \xi \vert}$. The click statistics is expressed by inserting (\ref{p0GaussM}) into  (\ref{splittingFormula}). It was achieved because all the considered transformations preserved the Gaussian shape of the Wigner function. Expressing the probability $P_0(T)$ of more complex states such as the displaced squeezed Fock states $\vert \alpha,\xi, n \rangle$ in (\ref{betan}) still only requires the integration of a Gaussian function multiplied by a polynomial. Alternatively, it can be also convenient to get the click statistics from a convolution of the photon distribution with a response of a detector on $n$ incoming photons.

\section{Nonclassical light}

Historically, optical detectors were not sensitive to single photons, and they only measured a time-average stream of photons producing an integrated continuously fluctuating intensity. In quantum optics, the output of such detection had been described by statistical moments of the normally-ordered annihilation and creation operators \cite{Glauber1963}. In the normal ordering, creation operators always stand left from annihilation ones. Such normally-order moments can be substituted for a mixture of coherent states by a classical averaging of a complex stochastic amplitude \cite{Glauber1963}. This classical theory exploits first-order (amplitude) and second-order (intensity) autocorrelation. The amplitude autocorrelation characterizes primary phase-sensitive interference effects. The intensity fluctuations of classical waves were investigated firstly by Hanbury Brown and Twiss (HBT) in their ground-breaking experiment \cite{BROWN1956}. They split the incoming light by a beam-splitter (BS) towards two intensity detectors giving the first and the second moment of the integrated intensity $W$. The breakthrough was that thermal light radiated by a star obeyed $\langle W^2 \rangle > \langle W \rangle^2$, which proved the light fluctuations were correlated. 

In quantum optics, a similar layout is used to detect light at a single photon level; however, single-photon avalanche diodes (SPADs) need to replace the detectors measuring the integrated intensity. SPAD sensitively converts one or more photons to an electronic pulse – a click. The SPAD produces a different electric signal than intensity as every photon can initiate an amplified output that does not linearly correspond to the number of arriving photons. Therefore, the SPAD produces binary events and indicates the presence of the photons or the vacuum. The pioneering experiments exploited the HBT setup to measure light scattered on atoms by two phototubes. When the pumping light was resonant with some addressed transition in the atoms, the scattered light exhibited antibunching \cite{Kimble1977} and sub-Poissonian statistics \cite{Short1983,Grangier1986}. Explanation of these phenomena by coherent states and their mixtures was insufficient, which proved in an experiment that quantum optics is a more general theory than the classical theory of light.

 Quantum technology has developed since these early experiments. For the last three decades, the workhorse for the generation of nonclassical light has been parametric processes in a nonlinear crystal \cite{Eisaman2011}. It allowed generation of heralded sub-Poissonian light \cite{Mosley2008}. Also, four-wave mixing in a fiber \cite{Fulconis2007} or in atomic vapour \cite{Mika2018} produces the sub-Poissonian light after heralding. Currently, platforms exploiting ions, molecules, or solid-state sources are being developed extensively \cite{Aharonovich2016}. They use discrete transitions emitting, in principle, precisely a single-photon. However, background noise often deteriorates photon statistics and also, the collection and the detection efficiency are low in many experiments \cite{Aharonovich2016}. Moreover, the solid-state emitters are often fabricated in a cluster behaving as several independent emitters with different emission characteristics  \cite{Qi2018}.

Nonclassical light exhibits different statistics than classical waves arising from a mixture of external coherent drives of light modes. As quantum optics theory grew from classical coherence theory, the correlation function $g^{(2)}(0)$ was used for introducing the first criteria of nonclassicality. To relate the HBT setup equipped with SPADs to the definition of $g^{(2)}(0)$, a conjecture that the source emits a weak light was used. Let us illustrate that by introducing the probability of just a single SPAD click $P_1$ and both SPADs clicking $P_2$. It can be tempting to approximate the moments as accurately as possible by
\begin{eqnarray}
    \langle a^{\dagger}a \rangle & \approx & \langle 1 \vert \rho \vert 1 \rangle+ 2\langle 2 \vert \rho \vert 2 \rangle \approx P_1+4P_2 \nonumber \\
    \langle \left(a^{\dagger}\right)^2 a^2 \rangle & \approx & 2\langle 2 \vert \rho \vert 2 \rangle \approx 4 P_2.
\end{eqnarray}
However, evaluation of the function $g^{(2)}$ for coherent states $\vert \alpha \rangle$ using such approximation yields \cite{Sperling2012}
\begin{equation}
    g^{(2)}(0)\approx 1-\vert \alpha \vert^2
\end{equation}
manifesting fake nonclassicality. The nonclassical criteria that are reliable can be manipulated only with the click statistics of the SPADs. The splitting networks depicted in Fig.~\ref{fig:splitting} can be employed to formulate \cite{Sperling2012a}
\begin{equation}
    F_B=N\frac{\langle c^2 \rangle - \langle c \rangle^2}{\langle c \rangle (N-\langle c \rangle)},
    \label{FB}
\end{equation}
where $N$ is the number of SPADs in the splitting network, and $c$ denotes the number of simultaneous clicks. The nonclassicality is recognized reliably when $F_B<1$. Moreover, the parameter $F_B$ converges to the Fano factor (\ref{fano}) for large $N$. This approach is already based on directly measurable click statistics. However, it corrects only already existing characteristics of light like $g^{(2)}$ and Fano factor and interpretation of (\ref{FB}) is not clear for small $N$.  A viable approach is to consider only probabilities of a few specific events, specifically, click of $n$ SPADs (successful event) and click of $n+1$ SPADs (error event), and derive a criterion involving only the probability quantifying these two events \cite{Filip2013}. When a detector at least partially distinguishes the number of photons $n-1$, $n$ and $n+1$ \cite{Hoepker2019}, the nonclassicality is detected when \cite{Klyshko1996}
\begin{equation}
    \frac{(n+1)p_{n-1}p_{n+1}}{n p_n^2}<1
    \label{klCrit}
\end{equation}
with $p_m=\langle m \vert \rho \vert m \rangle$. For example, these criteria (\ref{klCrit}) are useful for recognition of the nonclassicality occurring during Jaynes-Cummings interaction of a two-level system in the ground state with classical light \cite{Marek2016}. However, there was not any systematic approach to derive nonclassicality criteria for the given detection layout yet. Finally, the nonclassicality can be also directly observed from the homodyne measurement giving the probability density function for the rotated canonical coordinate $P(x,\phi)$.  One can determine a function $G(k,\phi)$ through the Fourier transformation
\begin{equation}
    P(x,\phi)=\frac{1}{2\pi}\int e^{-i k x} G(k,\phi)\mathrm{d}k.
\end{equation}
The nonclassicality occurs if the function exceeds its value for the vacuum state, i.e. \cite{Vogel2000}
\begin{equation}
    \vert G(k,\phi)\vert>e^{-k^2/2}.
    \label{VogelG}
\end{equation}
However, the homodyne measurement requires a local oscillator to interfere with the signal. It is challenging for many atomic and solid-state experiments. All the introduced criteria (\ref{FB}), (\ref{klCrit}) and (\ref{VogelG}) provide only sufficient conditions of the nonclassicality for diverse direct detection methods. A necessary condition of nonclassicality would require satisfying an infinite number of conditions \cite{Richter2002}. Because real experiments cannot verify all the conditions, it raises the practical question: which criteria are appropriate for a specific experimental realization. We omitted here a discussion of indirect measurement of nonclassicality combining various detectors and ultimate tomographic reconstruction of density operator.

\subsection{Ab-initio nonclassical criteria for measurement layout}
\begin{figure}[t]
\centering
\includegraphics[width=0.45 \linewidth]{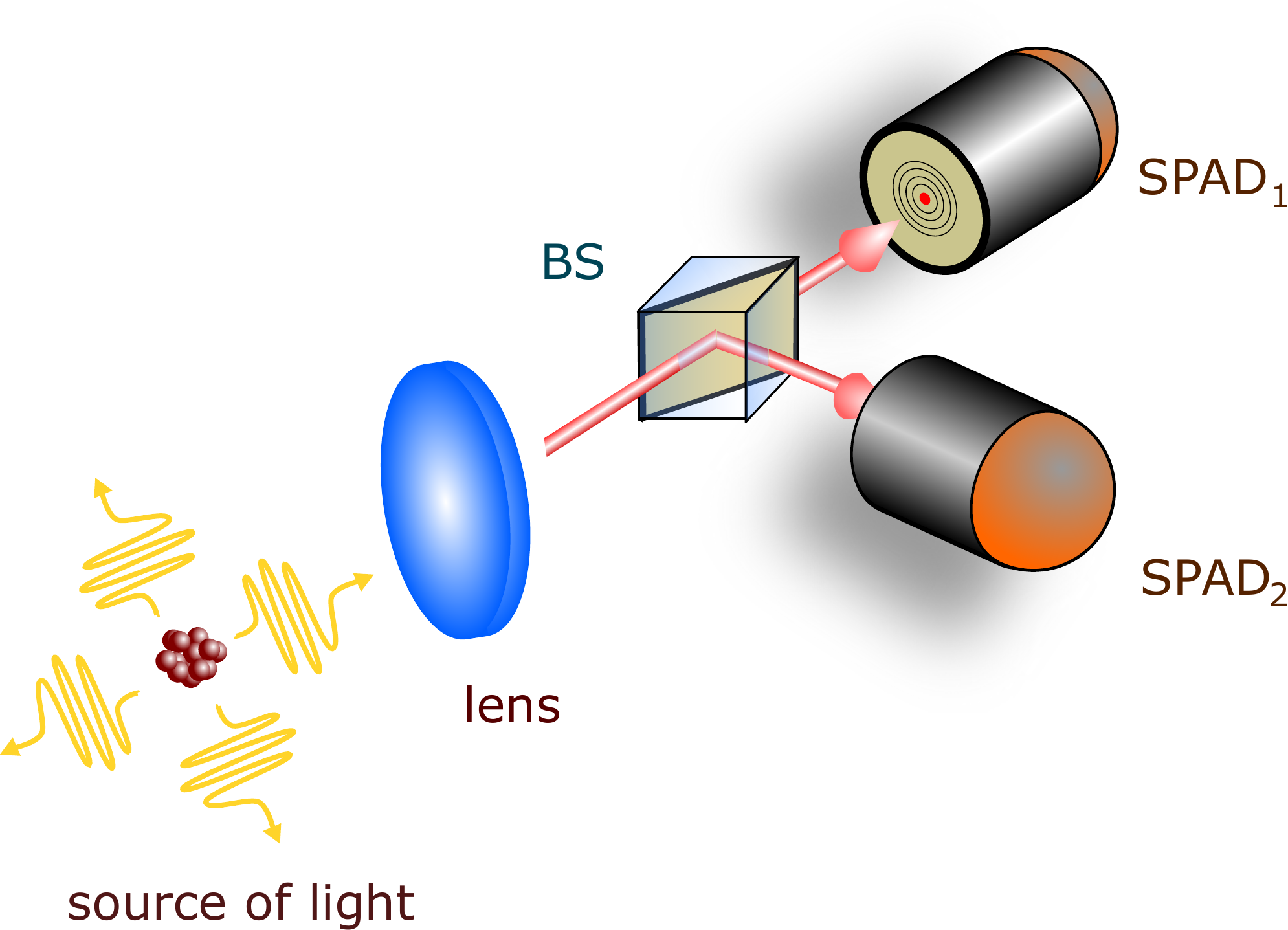}
\caption{The detection layout for measurement nonclassical light radiated from an ensemble of single-photon emitters. The light is collected by a lens and then directed toward single-photon avalanche diodes SPAD$_1$ and SPAD$_2$ through a BS.}
\label{fig:HBT}
\end{figure}

The Hanbury-Brown and Twiss layout is the simplest detection scheme to witness nonclassical light. It consists of a single beam splitter BS with the transmissivity $T$ and two SPADs, as depicted in Fig.~\ref{fig:HBT}. The measurement produces three different detection events: a click of SPAD$_1$, a click of SPAD$_2$ and, finally, a simultaneous click of both detectors SPAD$_1$ and SPAD$_2$. In all other cases, no detection event is registered. A nonclassical condition on probabilities quantifying those events can be obtained from the Cauchy-Schwarz inequality \cite{Grangier1986}. However, the same condition can be derived exploiting the systematic methodology in the Ref. \cite{Filip2011}  which do not require the Cauchy - Schwarz inequality. It is convenient to employ no-click events instead of click events to simplify calculations. Let $P_{0;1}$, $P_{0;2}$ and $P_{0;12}$ denote probabilities quantifying successively the no-click event in SPAD$_1$, the no-click event in SPAD$_2$ and, finally, the no-click event observed when neither SPAD$_1$ or SPAD$_2$ click. Note, $P_{0;1}$ and $P_{0;2}$ are not generally equal since the used BS with the transmission $T$ can be unbalanced or the employed detectors can have different quantum efficiency. The criterion can be systematically derived from a linear combination of the probabilities
\begin{equation}
F_{a;i} (\rho)=P_{0;i}+a P_{0;12},
\label{linF}
\end{equation}
where $a$ is a free parameter and $i=1,2$  distinguishes two functions $F_{a;i}$ according to a choice between probabilities $P_{0;1}$ or $P_{0;2}$. To obtain criterion, we optimize (\ref{linF}) over mixtures of the coherent states. Importantly, both functions $F_{a,i} (\rho)$ are linear in a state $\rho$, which means
\begin{equation}
    F_{a;i}(\sum_j p_j \vert \alpha_j \rangle \langle \alpha_j \vert )=\sum_j p_j F_{a;i}(\vert \alpha_j \rangle \langle \alpha_j \vert).
    \label{famixture}
\end{equation}
When the functions $F_{a;i}$ are optimized over mixtures of coherent states, the optimal amplitudes in the right side of (\ref{famixture}) are the same for each $j$. Because $\sum_j p_j=1$, the optimum is determined from optimizing over a coherent state, i. e.
\begin{equation}
    F_{a;i}=\max_{\alpha_j,p_j} F_{a;i}(\sum_j p_j \vert \alpha_j \rangle \langle \alpha_j \vert )=F_{a;i}(\alpha_{0,i})
\end{equation}
with $\alpha_{0,i}$ representing the optimal amplitude. It leads to \begin{eqnarray}
F_{a;1}&=&-\left(-\frac{T}{a}\right)^{\frac{1}{1-T}}\frac{1-T}{T}a \nonumber \\
F_{a;2}&=&-\left(-\frac{1-T}{a}\right)^{\frac{1}{T}}\frac{T}{1-T}a.
\label{Fia}
\end{eqnarray}
Both functions $F_{a;1}$ and $F_{a;2}$ are results of optimizing over all mixtures of coherent states in a single mode. An optimum over classical states occupying several modes is always identical to $F_{a;1,2}$. The reason is that the optima are achieved by pure coherent states, which exhibit Poissonian distribution of photons independently of a number of modes that they occupy. A sufficient condition of nonclassicality reads $\exists a: P_{0;i}+ a P_{0;12}>F_{i;a}$, which can be formulated equivalently by $P_{0;i}>\min_a \left[F_{i;a}-a P_{0;12}\right]$. The optimal parameters fulfill
\begin{eqnarray}
    a_{0;1}=-\frac{T}{P_{00}^{1-T}}\nonumber \\
    a_{0;2}=-\frac{1-T}{P_{00}^{T}}
\end{eqnarray}
Inserting it to (\ref{Fia}) results in requirements $P_{0;1}^T>P_{0;12}$ or $P_{0;2}^{1-T}>P_{0;12}$. Their combination yields the final condition
\begin{equation}
\frac {P_{0;12}}{P_{0;1}P_{0;2}}<1,
\label{finalCond}
\end{equation}
which does not depend on the transmission $T$. Criterion (\ref{finalCond}) can be reformulated in terms of click probabilities 
\begin{eqnarray}
P_{s,1}&=&1-P_{0,1} \nonumber \\
P_{s,2}&=&1-P_{0,2} \nonumber \\
P_c&=&1-P_{0,1}-P_{0,2}+P_{00}
\label{p0pnRelation}
\end{eqnarray}
referring to the probability of click of SPAD$_1$, to the probability of click SPAD$_2$ and to a simultaneous click of both SPAD$_1$ and SPAD$_2$. Inverting the relations (\ref{p0pnRelation}) and inserting it to inequality (\ref{finalCond}), leads finally to a nonclassicality criteria
\begin{equation}
\frac {P_c}{P_{s,1} P_{s,2}}<1. 
\label{g2}
\end{equation}
This systematic approach substitutes more ad hoc derivation using the Cauchy - Schwarz inequality \cite{Grangier1986}. Moreover, the left side of (\ref{g2}) converges to $g^{(2)}(0)$ for weak states, and therefore it is independent of losses in the approximation of the weak states. Although both conditions (\ref{finalCond}) and (\ref{g2}) are equivalent, their left sides represent two different parameters, which can become helpful for an analysis of the nonclassical light.

 \begin{figure}[t]
     \centering
     \includegraphics[width=0.5 \linewidth]{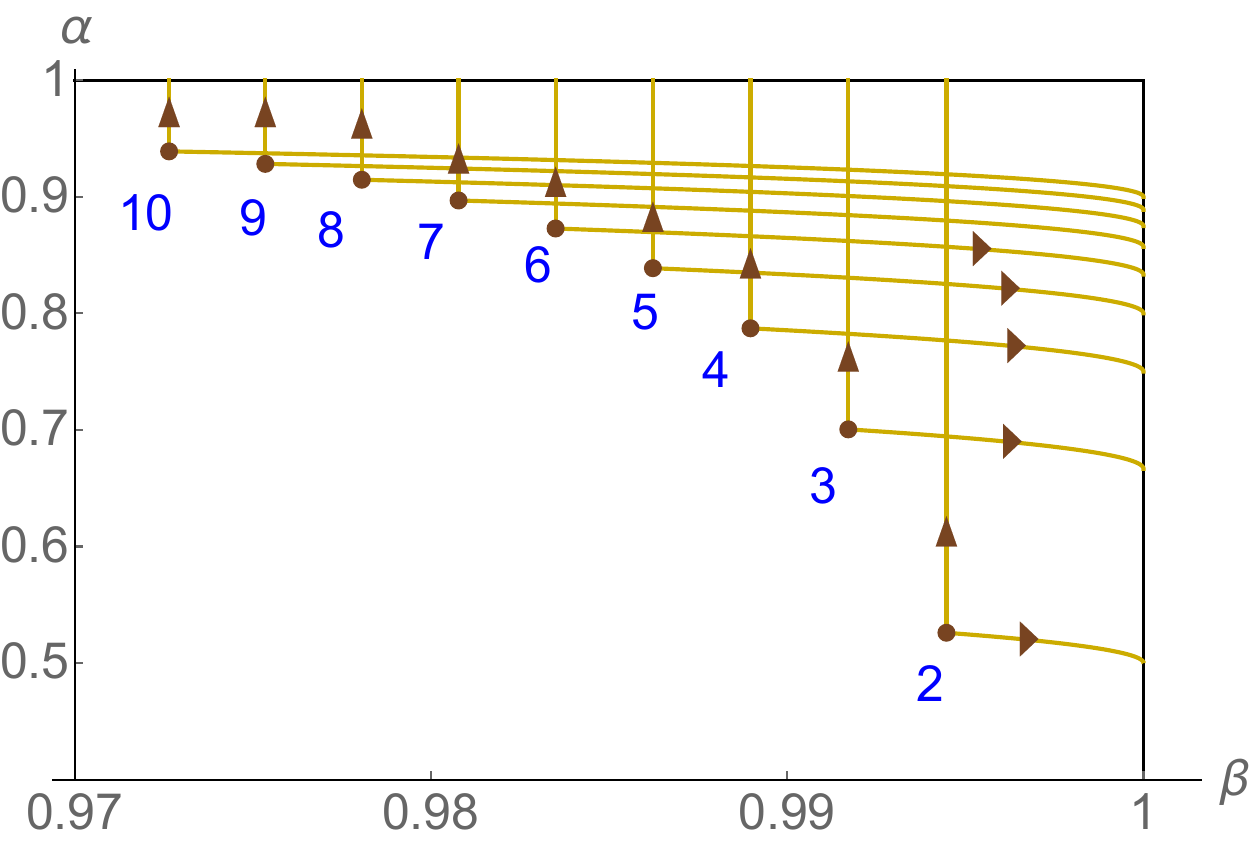}
     \caption{Comparison of parameters $\alpha=P_c/(P_{s,1}P_{s,2})$ and $\beta=P_{00}/(P_{0,1}P_{0,2})$ revealing nonclassicality. The blue points represent values of these parameters for model state (\ref{state}) with $\eta=0.1$ and the attached numbers correspond to the number of contributing single-photon emitters. Number of emitters goes from two to ten and grows from left to right. The slightly declining horizontal lines shows shifting of the states during attenuation and the vertical lines exhibit impacts of the Poissonian background noise. In both cases, a direction of the shifting of states in the plot is shown by the arrows.}
     \label{fig:comparison}
 \end{figure}

Atomic and solid-state sources can be also fabricated as clusters of single-photon emitters, which can radiate multiphoton light. Before such light is efficiently emitted to a single optical mode to form Fock states for the applications, their independent emission needs to be investigated. A density matrix of the emitted light approaches 
\begin{equation}
\rho=\left[ (1-\eta)\vert 0 \rangle	 \langle 0 \vert + \eta \vert 1 \rangle \langle  1 \vert \right]^{\otimes N} \otimes \rho_{\bar{n}},
\label{state}
\end{equation}
where $\eta$ is an efficiency of photon emission from a single emitter, $N$ is a number of emitters presented in the radiating cluster and $\rho_{\bar{n}}$ is background noise that has Poissonian statistics with a mean number of photons $\bar{n}$, i. e.
\begin{equation}
    \rho_{\bar{n}}=e^{-\bar{n}}\sum_{n=0}^{\infty}\frac{\bar{n}^n}{n!}\vert n \rangle \langle n \vert.
    \label{rhoModel}
\end{equation}
In a different possible model, the background noise occupies more modes depending on the number of contributing emitters. Since the multimode noise preserves the Poissonian distribution of the photons, it can be described effectively by the density matrix with the form (\ref{rhoModel}) where the mean number of photons of the noise grows with the number of the contributing emitters.
In a case of a balanced HBT detection, the detector response on the state follows
\begin{eqnarray}
P_0&=&(1-\eta/2)^N e^{-\bar{n}/2} \nonumber \\
P_{00}&=&(1-\eta)^N e^{-\bar{n}},
\label{noClicks}
\end{eqnarray}
where $P_{0}=P_{0,1}=P_{0,2}$. Inserting these quantities into the condition (\ref{finalCond}) recognizes the nonclassicality of the state $\rho$ for any number of emitters $N$ if $\eta>0$. Moreover, the nonclassicality remains observable if the state is deteriorated by Poissonian background noise with arbitrarily large mean number of photons $\bar{n}$. Also, the nonclassicality tolerates losses since losses only decrease parameters $\eta$ and $\bar{n}$ but preserve the form of density matrix (\ref{state}). Although both the losses and the background Poissonian noise do not break nonclassical nature of the state (\ref{state}) they have a different impacts on parameters
\begin{eqnarray}
\alpha &=& P_c/P_s^2, \nonumber \\
\beta &=& P_{00}/P_0^2 
\label{probingPar}
\end{eqnarray}
revealing the nonclassicality in conditions (\ref{finalCond}) and (\ref{g2}). Fig.~\ref{fig:comparison} depicts how these parameters are changing during losses and increasing background noise. The parameter $\alpha$ is slightly affected by loss, but it grows for more single-photon emitters and larger Poissonian background noise. These two contributions are, therefore, hard to distinguish. On the other hand, the parameter $\beta$ is independent of the background noise with the Poissonian statistics. Simultaneously, the parameter $\beta$ decreases with the number of single-photon emitters in an ensemble. Using both parameters, we gain a deeper insight into the nonclassical aspects of the multiphoton light distinguishing growth of the number of emitters from increasing contributions of the background Poissonian noise.

\begin{figure}[t]
\centering
\includegraphics[width=0.5 \linewidth]{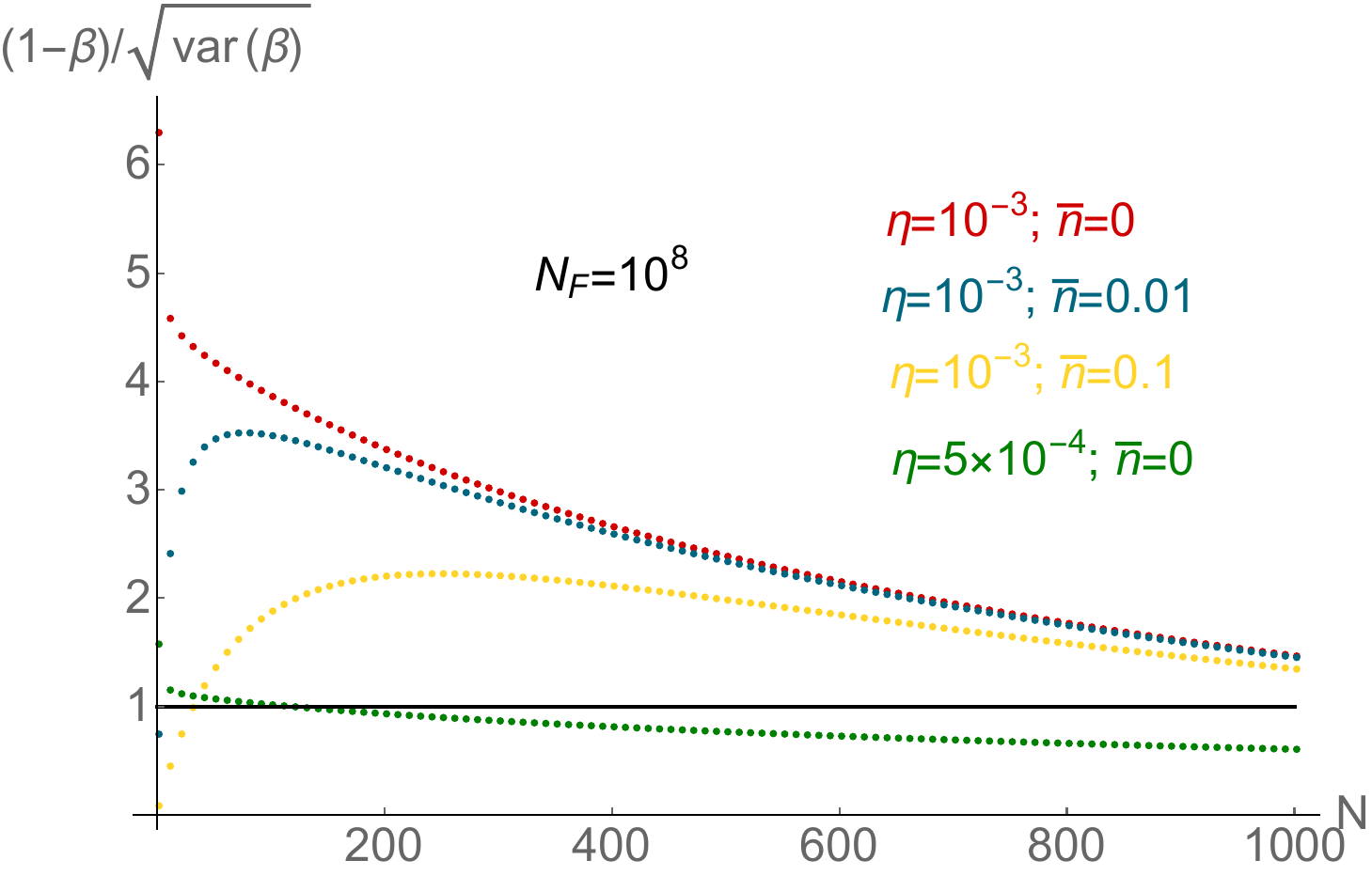}
\caption{A simulation of reliability of detected nonclassicality. The horizontal axis quantifies an amount of single-photon emitters, whereas the vertical one corresponds to parameter $1-\beta$ in units of estimated error bars. The simulations consider $10^8$ experimental runs. The colours distinguish the parameters of the model states. The red and green points are results for simulation without the background noise. In contrast, the blue dots correspond to cases with background noise having $\bar{n}=0.01$ and the yellow dots represent a state deteriorated by background noise with $\bar{n}=0.1$. The efficiencies of radiation were chosen $\eta=10^{-3}$ (red, blue and yellow) and $\eta=5 \times 10^{-4}$ (green). For states above the black horizontal line, a parameter $\beta$ exceeds experimental error bars, so the nonclassicality is observed reliably.}
\label{fig:reliability}
\end{figure}

A remaining aspect that can prevent conclusive detection of nonclassical light is the time needed for sufficient suppression of statistical errors. Although it can appear as a rather technical matter, very weak nonclassicality cannot be observed due to very long measurement.
The click distribution determines error bars stemming from a finite measurement. The standard deviation of the measured parameters quantifies the error bars. To obtain them, let us approximate the no-click distribution in $P_0$-$P_{00}$ space achieved in $M$ measurements by the Gaussian distribution
\begin{equation}
    P(M_0,M_{00})=\frac{1}{\sqrt{2\pi V_c}} e^{-\frac{\left(2 M_0-M_{00}-2 M P_0+M P_{00}\right)^2}{2V_c}} \frac{1}{\sqrt{2\pi V_a}} e^{-\frac{\left( M_0+2M_{00}- M P_0-2 M P_{00}\right)^2}{2V_a}},
    \label{normalDistr}
\end{equation}
where $M_0$ ($M_{00}$) is a number of no-click events in one (both) SPADs. The arguments in the exponentials are set in such a way that the former Gaussian function from the left represents the normal distribution of simultaneous clicks of both SPADs and the later one corresponds to the normal distribution of an auxiliary quantity $M_0+2M_{00}$. The parameters $V_c$ and $V_a$ denote the variance of those events. Employing the distribution (\ref{normalDistr}) leads to evaluation of the variance of the parameters revealing the nonclassicality
\begin{eqnarray}
\mbox{var}(\beta) &=& \frac{V_c P^4_{00}}{P^8_{0}}\left(\frac{\sin \phi}{2 \sqrt{P_{00}}}+\cos \phi\right)^2 \nonumber \\
&+&\frac{V_a P^4_{00}}{P^8_{0}} \left(\frac{\cos \phi}{2 \sqrt{P_{00}}}-\sin \phi\right)^2+\left(\frac{1-\beta}{\beta}\right)^2 \frac{P_{00}^7}{P_0^8}V_{00} \nonumber \\
\mbox{var}(\alpha) &=& \frac{1}{P^4_{s}}\left[V_c\left(\frac{\sin \phi}{2 \sqrt{P_{00}}}+\cos \phi\right)^2 +V_a \left(\frac{\cos \phi}{2 \sqrt{P_{00}}}-\sin \phi\right)^2 \right] \nonumber \\
&+&\left(\frac{1-\beta}{\beta}\right)^2\frac{P_{00}^2 V_s}{P_s^6}
\label{errorbarEst}
\end{eqnarray}
where $\phi=\arctan 1/2$. If the source is weak, i. e. $1-P_{00} \ll 1$, the variance of $\beta$ scales with $\mbox{var}(\beta) \propto V_c$.
Reliability of the nonclassicality in an experimental test can be expressed as a ratio between the parameters in (\ref{probingPar}) and the squared root of its variance. According to (\ref{errorbarEst}), one gets
\begin{equation}
    \mbox{var}(\beta)\approx \mbox{var}(\alpha)
    \label{relaibilityCrit}
\end{equation}
for states close to the boundary with $1-\beta \ll 1$. Fig.~(\ref{fig:reliability}) depicts the ratio (\ref{relaibilityCrit}) quantifying the reliability of the parameter $\beta$ for different sizes of the ensemble of single-photon emitters in a realistic experiment. The ratio $(1-\beta)/\mbox{var}(\beta)$  grows with the efficiency of emission $\eta$. If the background noise does not deteriorate a source, the ratio drops with the number of emitters. It means recognising nonclassical light from larger ensembles requires a longer experimental time. When the background noise contributes to the measured statistics, which often occurs in solid-state sources \cite{Qi2018}, there is an optimal size of the ensemble leading to the greatest ratio when all the remaining parameters are fixed. In that case, the ratio increases for small ensembles because photons coming from the background noise contribute less significantly to the overall statistics of clicks of detectors. Fig.~\ref{fig:reliability} predicts nonclassical light from a large ensemble of single-photon emitters under realistic conditions. They include background noise or low overall efficiency of emission. The limiting factors are only experimental error bars achieved due to finite measurement.

Prospective sources of single-photon states in atoms and solid-state emitters exploit discrete energy levels. Addressing the energy levels by appropriate light beams leads to a spontaneous or controllable single-photon emission. Many physical platforms involving ions, molecules, quantum dots or NV centers manifest such behaviour \cite{Aharonovich2016}. Among them, the quantum dots represent a promising platform due to their easy manipulation and a technological possibility to implement them in nanostructures \cite{Somaschi2016}. However,  a sample contains several different quantum dots in a cluster with background noise from a substrate. The nonclassical light was detected from such a cluster of quantum dots \cite{Qi2018} and NV centers \cite{Moreva2017}. 

Compared with the clusters of the quantum dots or NV centers, ions captured in a Paul trap constitute a platform where the number of ions is controlled accurately. Such a source of nonclassical light exhibits negligible background noise contributing to the light emitted from the ions. However, the detection efficiency is very low, and coupling to a cavity is only in progress. Such an experiment was realized exploiting calcium ions, which were Doppler cooled by two laser beams  \cite{Obsil2018}. A lambda scheme of transitions interacting with the pumping beams allowed the controllable emission of a single photon from each ion. The measurement was performed in a regime where the pumping beams were pulses and, further, when the laser shined continually. After loading the ions and cooling them, the ions formed a crystal with a shell structure \cite{Hasse1991}. The ions could move inside each shell, and thus their behaviour was similar to a two-dimensional liquid. The experiment was repeated for crystals with 12, 55, 125, 204, 275 ions. The number of ions was estimated from a picture from a CCD camera. A lens collected $2\%$ of light radiated from a focus point and directed the light towards a BS and two SPADs, which measured click statistics. The measurement was performed in both regimes of continual and pulsed pumping. The measurement was performed five times, and the error bars were calculated as a standard deviation in these five measurements. The measurement confirmed the nonclassical emission for crystals with up to 55 ions. Then it gets saturated due to the low collection efficiency of light emitted from ions too remote from the focus of the detection. Later, nonclassicality stayed observable even for light emitted from a cluster having up to 275 single-photon emitters.

\section{Quantum non-Gaussian light}

In quantum optics, processes beyond the Hamiltonian quadratic in the annihilation and creation operators turn the Heisenberg equations nonlinear. A paramount example is the emission of light from a driven two-level system emitting light into a cavity \cite{Scully1997}. The detection of nonclassical light is not sufficient to faithfully prove that the process is indeed nonlinear. Therefore, new criteria are required. If nonlinear dynamics produces a pure state, it gets a non-Gaussian Wigner function \cite{Hudson1974}. Such non-Gaussian character has been investigated broadly and even quantified in Ref.~\cite{Genoni2007,Genoni2010, Barbieri2010}. However, mixtures of coherent states also possess the non-Gaussian Wigner function \cite{Genoni2013}. For this substantial reason, this classical non-Gaussianity concept has to be upgraded. 

An unambiguous recognition of {\em quantum} non-Gaussianity has to refuse all stochastic mixtures of Gaussian states, i.e.
\begin{equation}
    \rho \neq \int P(\xi, \alpha)D(\alpha) S(\xi) \vert 0 \rangle \langle 0 \vert S^{\dagger} (\xi)D^{\dagger}(\alpha)\mathrm{d}^2 \alpha \mathrm{d}^2 \xi,
    \label{qngDef}
\end{equation}
where $P(\xi,\alpha)$ is a probability density function. The rejected Gaussian states are coherently displaced squeezed states (\ref{gaussSt}), which can be obtained by linear coherently driven dynamics from a vacuum in the Heisenberg picture. The squeezed states are the simplest examples of states that violate rules of the classical coherence theory. That is a reason why the squeezed states of light were used historically in proof-of-principle experiments \cite{Burnham1970,Heidmann1987,Hong1987,Reid1988}. However, as squeezed states are Gaussian, they are excluded. Therefore, the quantum non-Gaussianity also puts a new benchmark for surpassing these experiments.

According to the definition (\ref{qngDef}), the negativity of the Wigner function reveals the quantum non-Gaussianity. In cases of pure states, the negativity is even a sufficient, and necessary condition \cite{Hudson1974}. Thus, all the Fock states except the vacuum exhibit the negativity. The negativity of the Wigner function has appeared as a crucial feature for quantum computing \cite{Mari2012}. However, the photonic systems often suffer from losses, and the losses above fifty percentages always make the negativity disappear. Thus, the negativity of the Wigner function is too challenging for many atomic and solid-state experiments, especially in their early stage. Therefore, recognising the quantum non-Gaussianity of states with the positive Wigner function is an intermediate step for evaluating the quantum states of light.

A criterion enabling such recognition imposes a constraint on the Wigner function concerning the mean number of photon $\langle n \rangle$ \cite{Genoni2013,Hughes2014}. The criterion inspired such a criterion in Ref. \cite{Filip2011}, which will be discussed later separately. It was shown that all the mixtures of the Gaussian states obey
\begin{equation}
    W(0,0) \geq \frac{1}{2\pi}e^{-2\langle n \rangle(1+\langle n \rangle)}.
    \label{wCritQNG}
\end{equation}
The attenuated Fock state $\eta \vert 1 \rangle \langle 1 \vert+(1-\eta)\vert 0 \rangle \langle 0 \vert$ violates the condition when $\eta>0$, and therefore the condition reveals its quantum non-Gaussianity for that state. Such a method could only detect QNG states if the detectors faithfully resolve all photon numbers without systematic errors. Without such direct detection, this method suffers from the same limitations as a measurement of the $g^{(2)}$ discussed before. Otherwise, quantum state tomography must be used to estimate $W(0,0)$ accurately from homodyne measurements. Criterion (\ref{wCritQNG}) was also extended so that the certification relies on a general $s$-parametrized quasiprobability distribution \cite{Hughes2014} where the parameter $s$ changes representation continuously \cite{Lee1991} from Wigner function ($s=0$) to Husimi $Q$-function \cite{Husimi1940} ($s=-1$). Note that $s<-1$ corresponds formally to representation that is identical to $Q$-function of a state that undergoes losses $T=2/(1-s)$ \cite{Paris1996}. On the contrary, positive $s$ yields quasiprobability distribution with singularities. An example of such quasiprobability distribution is $P$-function, which is gained for $s=1$. For a given value of $s$, all Gaussian states obeying $\langle n \rangle \leq \bar{n}$ are limited in the origin of the related $s$-parametrized quasiprobability distribution. Dependence of the respective thresholds exposing quantum non-Gaussianity on $\bar{n}$ can be achieved numerically. Utilizing such criteria for $s<0$ can bring up advantages against the criterion (\ref{wCritQNG}) when these criteria are applied to higher Fock states deteriorated by losses \cite{Hughes2014}.

Another approach defines a combination of four values of the Wigner function \cite{Park2015}
\begin{equation}
    B=2\pi \sum_{i=0}^1 \sum_{j=0}^1 (-1)^{i j}W(x_i,p_j)
\end{equation}
and proves that parameter $B$ can reveal both the nonclassicality and the quantum non-Gaussianity. In the case of the classical states, the Wigner function can be treated formally as a function providing a correlation between two random variables $a$ and $b$ with values between zero and one, i. e.
\begin{equation}
   2 \pi W(x,p)=\langle a(x) b(p) \rangle,
\end{equation}
and therefore the parameter $B$ is restricted to
\begin{equation}
    B=\langle a(x_0) b(p_0) \rangle+\langle a(x_0) b(p_1) \rangle+\langle a(x_1) b(p_0) \rangle-\langle a(x_1) b(p_1) \rangle \leq 2,
    \label{Bcond}
\end{equation}
which resembles the CHSH inequality \cite{Clauser1969}. Only the nonclassical states violates the condition (\ref{Bcond}). Although the Gaussian states can also break the condition the quantum non-Gaussianity is reached when \cite{Park2015}
\begin{equation}
    B > \frac{8}{3^{9/8}}.
\end{equation}
This condition is useful when it is applied to the state $\eta \vert 2 \rangle \langle 2 \vert+(1-\eta) \vert 0 \rangle \langle 0 \vert$ because it exposes the quantum non-Gaussianity when the condition (\ref{wCritQNG}) fails. Similarly, detection also requires faithfull full-range photon-number resolving detector without systematic errors. The quantum non-Gaussianity can be also detected from an expectation value of an operator
\begin{equation}
    O(\rho)=\langle e^{-c X^2}\rangle+\langle e^{-c P^2} \rangle,
\end{equation}
which can be acquired from the heterodyne measurement. The mixtures of the Gaussian states establish a boundary on the $O(\rho)$ that quantum non-Gaussian states can surpass. Although the criterion does not reveal the quantum non-Gaussianity of the Fock state one, it has appeared as useful for states yielded from some superposition of coherent states and squeezed states \cite{Happ2018}.

A direct detection of the quantum non-Gaussianity for single-photon states can advantageously use a partially resolving measurement of no photon, single photon and more photons, without a need for a photon-number-resolving detector \cite{Filip2011}. The criterion compares a sucess probability of a single-photon with an error probability of multiphoton contribution. The probabilities $P_1=\langle 1 \vert \rho \vert 1 \rangle$ and $P_{2+}=1-\langle 0 \vert \rho \vert 0 \rangle-\langle 1 \vert \rho \vert 1 \rangle$ are inserted into a linear form
\begin{equation}
    F_{a}(\rho)=P_1+a P_{2+}
\end{equation}
and a threshold function $F(a)$ covering all mixtures of the Gaussian states is derived. A criterion states
\begin{equation}
    \exists a: F_a(\rho)>F(a).
    \label{qng:fa}
\end{equation}
This approach uncover the quantum non-Gaussianity of the attenuated Fock state $\eta \vert 1 \rangle \langle 1 \vert +(1-\eta)\vert 0 \rangle \langle 0 \vert$ for $\eta>0$ using photon-number-resolving detector instead of the homodyne detection or full-range photon number resolving detector, which all the other criteria \cite{Genoni2013,Park2015,Happ2018} published later exploited. For comparison, $a=0$ will give an absolute quantum non-Gaussian criterion for single-photon states $P_1>0.4779$. It is the most strict quantum non-Gaussian criterion for direct single-copy measurement of a single photon. Although condition (\ref{qng:fa}) can be evaluated only numerically for a given tuple $(P_{2+},P_1)$ of experimentally gained probabilities, realistic sources radiate typically very weak light with $P_{2+}\ll 1$ on which criterion (\ref{qng:fa}) imposes an approximate requirement
\begin{equation}
    P_1^3>\frac{3}{2}P_{2+}.
    \label{smqngApp}
\end{equation}
A single-photon state with the density matrix $\rho=\left[(1-\eta)|0\rangle \langle 0 |+\eta |1\rangle \langle 0 |\right]\otimes \rho_{\bar{n}}$ with $\rho_{\bar{n}}$ referring to the Poissonian noise fulfills (\ref{smqngApp}) only when $\eta^2>3\bar{n}/2$ with $\bar{n}$ being the mean number of noisy photons. This suggests that 
the quantum non-Gaussianity survives in this limit small but non-zero contributions of the noise which, however, affects its robustness against the losses. Although the single-photon state in the limit $\bar{n}\rightarrow 0$ exhibits absolute robustness, the noise with positive $\bar{n}>0$ always causes the quantum non-Gaussianity to disappear due to the finite losses.
This criterion was also modified for HBT measurement with SPADs \cite{Lachman2013}. The modified criterion imposes qualitatively the same requirement on the realistic single-photon states.

The derived threshold was surpassed experimentally using heralding in SPDC \cite{Jezek2011}, photon subtraction in a squeezed vacuum state \cite{Jezek2012}, emission from quantum dots \cite{Predojevic2014} and from a single atom \cite{Higginbottom2016}. Also, a depth of the robustness of the quantum non-Gaussianity of light against losses was explored \cite{Straka2014}. These experiments confirmed that the directly measurable quantum non-Gaussianity is a measurable advanced feature for many single-photon sources since it is more demanding than the nonclassicality. Still, it is not as strict as the negativity of the Wigner function. Moreover, the quantum non-Gaussianity of the single-photon states has appeared as an indicating aspect for the security of the single-photon quantum key distribution \cite{Lasota2017} and a necessary feature preserved in single photon-phonon-photon transfer \cite{Rakhubovsky2017}.

The following step is disclosing the quantum non-Gaussianity of all the Fock states. For the Fock state $\vert 2 \rangle$, the absolute criterion in \cite{Filip2011} fails when the state is attenuated already above $70 \%$ and recognition for higher Fock states are even more sensitive to the losses. Since significant optical losses are present at an early stage of any multiphoton experiment, reliable detection requires new criteria involving error probability. 

Besides the development of the feasible criteria, current research aims at a resource theory for the quantum non-Gaussianity that is analogous with the resource theories of the entanglement \cite{ContrerasTejada2019} and coherence \cite{Winter2016}. A general resource theory defines a set of free operations $\mathcal{O}$ and a set of free states $\mathcal{S}$ that is closed against any action of the free operators, i.e. any $\rho_f \in \mathcal{S}$ and any $P \in \mathcal{O}$ obey $P(\rho_f)\in \mathcal{S}$ \cite{Brandao2015,Coecke2016}. Thus, the states beyond the set $\mathcal{S}$ represent a resource since they can not be prepared by applying the free operations on the free states. In addition, a resource theory provides a monotone corresponding to a functional that becomes zero for all the free states and does not grow under the actions of the free operations. Specifically, a resource theory for quantum non-Gaussianity considers set of free states $\mathcal{F}$ that are mixtures of multimode Gaussian states, i.e.
\begin{equation}
    \mathcal{F} = \bigg\{ \rho \in \mathcal{H}|\rho=\int P(\boldsymbol{\lambda})|\gamma(\boldsymbol{\lambda})\rangle\langle \gamma(\boldsymbol{\lambda})|\mathrm{d}\boldsymbol{\lambda}\bigg\},
\end{equation}
where $\mathcal{H}$ is the Hilbert space that multimode light occupies, $\boldsymbol{\lambda}$ is a vector of parameters determining any pure Gaussian state $|\gamma(\boldsymbol{\lambda})\rangle$ in $\mathcal{H}$ and $P(\boldsymbol{\lambda})$ is a probability density function. The monotones based on the logarithmic negativity \cite{Takagi2018} or on the relative entropy \cite{Albarelli2018} have been introduced. However, they quantify the resource state from the full tomography of a density matrix.

In contrast, the criteria of quantum non-Gaussianity employ only partial knowledge relying fully on the outcome of a detector. Thus, their investigation can be adapted more easily to the requirements of the current experiments. Particularly, a great experimental effort is made currently to produce states close to the Fock states. The recent experiments \cite{He2013,Ding2016,Somaschi2016} succeeded in the generation of single-photon states even with a high degree of indistinguishability. However, higher Fock states have not been generated yet. Although the negativity of the Wigner function was achieved for up to the Fock state $\vert 3 \rangle$ \cite{Cooper2013,Yukawa2013}, experiments with a higher mean number of photons in a single-mode exhibit only nonclassicality \cite{Laurat2003,Harder2016,Iskhakov2016}. So far, the properties of a source approaching these states can be only simulated by multiplexing single-photon states \cite{Straka2018,Lachman2019}. Although such states occupy many spatial or temporal modes, they share the photon distribution with the Fock states. Therefore they are attractive for proof-of-principle tests of quantum non-Gaussianity on multiphoton light.

\subsection{Hierarchy of criteria for multiphoton quantum non-Gaussianity}

\begin{figure}[t]
\centering
\includegraphics[width=0.65\linewidth]{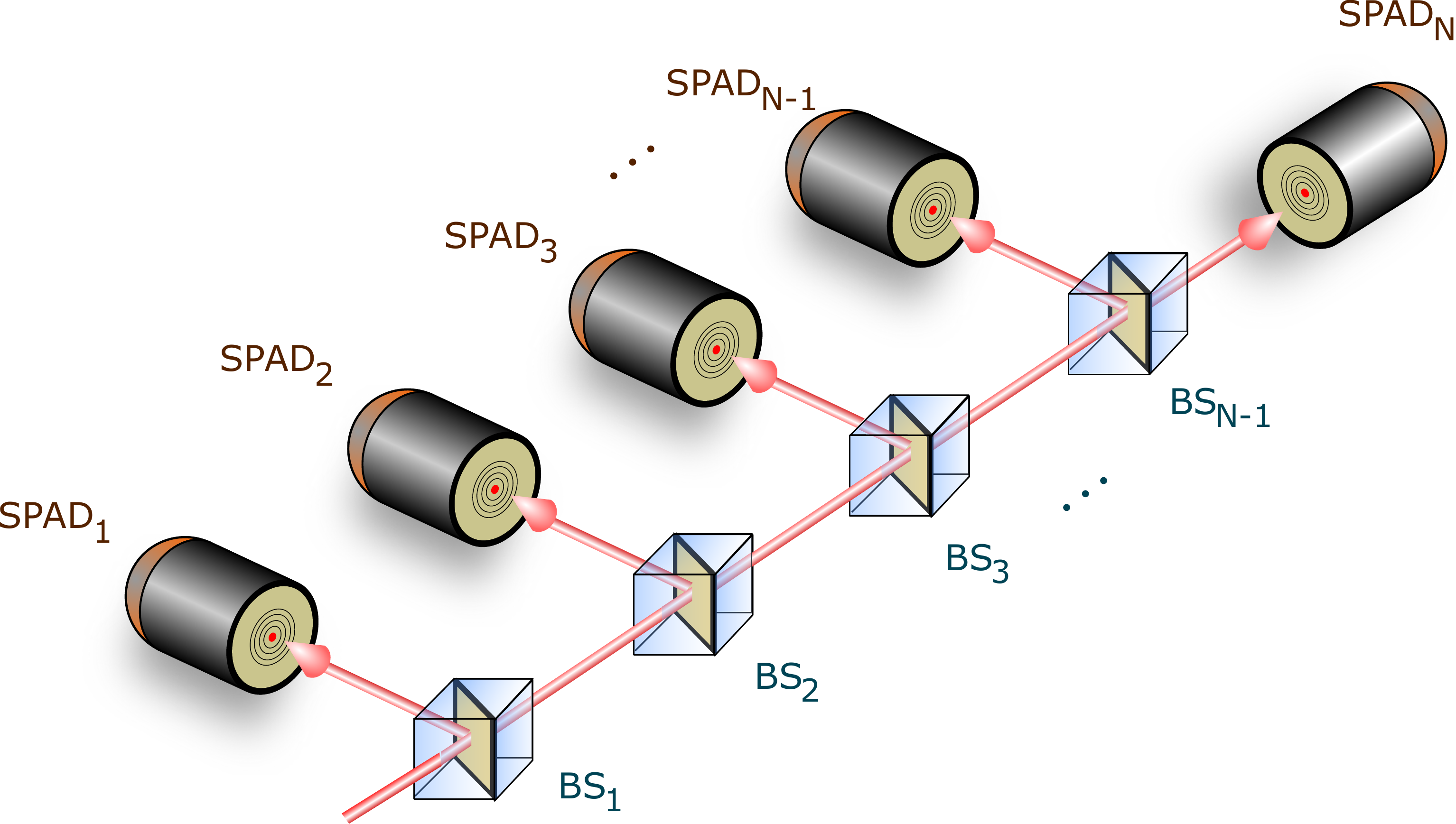}
\caption{Detection of quantum non-Gaussian light by a multi-channel detector. It splits incoming light towards several spatial modes by an array of BSs, and SPAD$_i$, $i=1,\ldots\,N$ detects each mode. For $N$ SPADs, a criterion incorporates a probability of a successful event (a simultaneous click of $N-1$ selected SPADs) and a probability of error event (all $N$ SPADs register a click).}
\label{fig:nonGNetw}
\end{figure}

The method of deriving the criterion of the quantum non-Gaussianity in \cite{Filip2011,Lachman2013} applies to any extended detection schemas. An example is a multi-channel detector using a network of BSs guiding light towards $N$ SPADs as depicted in Fig.~\ref{fig:nonGNetw}. The detector response to light is a sequence of clicks of the SPADs. We define success as simultaneous clicks of $n\leq N-1$ arbitrarily chosen SPADs and an error, as simultaneous clicks of all $n+1$ SPADs to recognize the quantum non-Gaussianity for multiphoton states. The corresponding probabilities quantifying those events are called the success probability $P_n$ ($n$ clicks) and the error probability $P_{n+1}$ ($n+1$ clicks). It comes from an expected detection of the ideal Fock state $\vert n \rangle$.

\begin{figure}[t]
\centering
\includegraphics[width=0.55\linewidth]{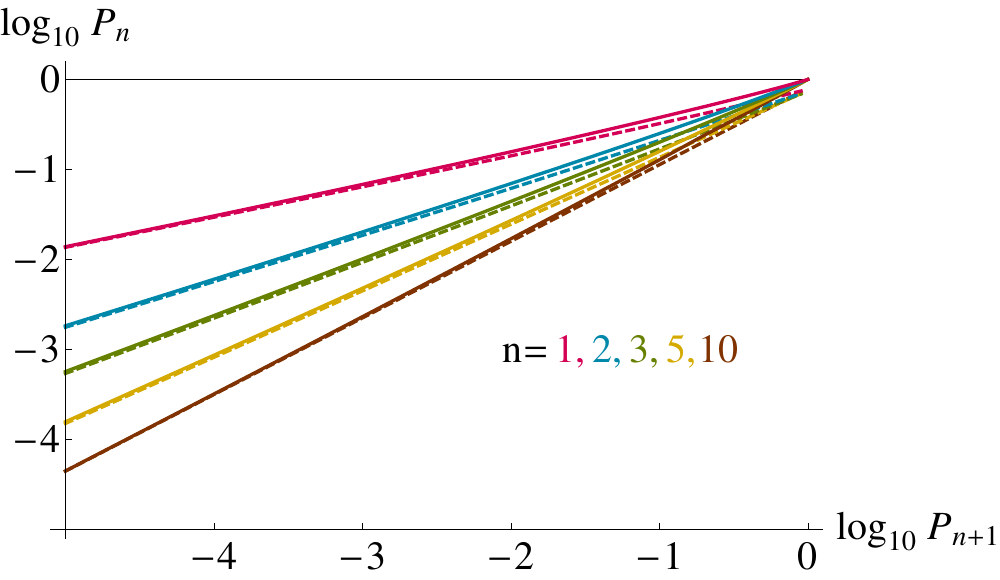}
\caption{Hierarchy of criteria for multiphoton quantum non-Gaussianity. The solid lines represent thresholds for the quantum non-Gaussianity for different $n$ distinguished by the colours. The dashed lines depict the approximate thresholds (\ref{nonGThs}) that are appropriate for highly attenuated states.}
\label{fig:nonGTh}
\end{figure}

We determine click statistics corresponding to a density matrix $\rho$ of a single mode of light from a probability of the vacuum after an attenuation
\begin{equation}
P_0(\tau)=\mbox{Tr} \left[(\vert 0 \rangle \langle 0 \vert \otimes \mathbb{1}) \cdot U(\tau) (\rho \otimes \vert 0 \rangle \langle 0 \vert) \cdot U^{\dagger}(\tau)  \right],
\end{equation}
where $U(\tau)$ is a unitary operation corresponding to a BS with transmission $\tau$. The pure Gaussian states exhibit
\begin{equation}
P_0(\tau)=2 \frac{e^{-\frac{\beta^2 \tau}{2}\left[\frac{\cos^2 \phi}{\mu(1/V)}+\frac{\sin^2 \phi}{\mu(V)}\right]}}{\sqrt{\mu (V)\mu(1/V)}},
\label{p0T}
\end{equation}
with the parameters
\begin{eqnarray}
\beta  e^{i \phi}&=&\frac{1+V}{2\sqrt{V}}\alpha+\frac{1-V}{2\sqrt{V}}\alpha^*, \nonumber \\
\mu(V)&=&2V+\tau (1-V)
\end{eqnarray}
where $\beta$ is real and positive \cite{Yuen1976}. The parameter $V$ is a minimal variance of the light quadrature, i. e. $V=e^{-2 \vert \xi \vert}$. The click statistics is expressed explicitly by a formula
\begin{equation}
P_n=1+ \sum_{k=0}^n \binom{n}{k} (-1)^k P_0(k/N),
\label{rnp0}
\end{equation}
where $N$ is a total number of SPADs in the layout. The criterion for particlar $n\leq N$ is yielded from optimizing a linear functional
\begin{equation}
F_{a,n}(\rho)=P_n+a P_{n+1}.
\label{nonGFucnt}
\end{equation}
Because of the linearity, the maximum over a statistical mixture of Gaussian states is identical with a maximum over all pure states. We maximize over the minimal quadrature variance $V$, the amplitude $\vert \alpha \vert$ and the angle $\phi$ parametrizing the formulas (\ref{p0T}) and (\ref{rnp0}). It apperared that he optimal states are squeezed perpendicularly to the direction of the displacement amplitude, i. e. $\phi=0$. The optimal parameters $\vert \alpha \vert$ and $V$ obey
\begin{equation}
\partial_{\vert \alpha \vert} P_n \partial_{V} P_{n+1}-\partial_{\vert \alpha \vert} P_{n+1} \partial_{V} P_{n}=0.
\label{rV}
\end{equation}
This relation eliminates one of these two parameters. The last parameter can be chosen for binding the pair of probabilities $P_n$ and $P_{n+1}$ in order to parameterize the quantum non-Gaussian thresholds in the probabilities $P_n$ and $P_{n+1}$. The derived thresholds have only numerical solutions, which can be simplified in a limit of states with a low error probability. The numerical solutions of the thresholds are presented in Fig.~\ref{fig:nonGTh}.

Let us focus on the approximation of highly attenuated states. The probabilities of the success and the error can be approximated by
\begin{eqnarray}
    P_n &\approx& r_n t^n \nonumber \\
    P_{n+1} &\approx& r_{n+1}t^{n+1}+r_{n+2}t^{n+2},
    \label{appExp}
\end{eqnarray}
where $t$ is a formal parameter being very small and $r_i$ are some coefficients. The function (\ref{nonGFucnt}) is
\begin{equation}
    F_a(r_n,r_{n+1},r_{n+2},t)=r_n t^n -a(r_{n+1}t^{n+1}+r_{n+2}t^{n+2})
\end{equation}
and a local extreme of this function satisfies $\partial_t F_a=0$. It leads to
\begin{equation}
    t=\frac{-a (1+n)r_{n+1}+\sqrt{a\left[a(1+n)^2r_{n+1}^2+4n(2+n)r_n r_{n+2}\right]}}{2 a(2+n)r_{n+2}}.
    \label{ta}
\end{equation}
The discussed limit of highly attenuated states is relevant to $a$ being very large. The convergence of (\ref{ta}) depends on behaviour of the expression $a(1+n)^2r_{n+1}^2+4n(2+n)r_n r_{n+2}$. If $a r_{n+1}^2 \gg r_n r_{n+2}$, it results in the function
\begin{equation}
    \widetilde{F}(a)\approx \frac{r_n^{n+1} n^n}{(1+n)^{n+1}r_{n+1}^n}\frac{1}{a^n}.
    \label{FaApp}
\end{equation}
However, the Gaussian states allow us to achieve $r_{n+1} \ll r_n$, which entails $P_{n+1}\approx r_{n+2}t^{n+2}$ and, consequently, $a r_{n+1}^2 \ll r_n r_{n+2}$. In this case, the approximate threshold function yields
\begin{equation}
    F(a)\approx \frac{2 r_n}{2+n}\left[\frac{ r_n}{(n+2)r_{n+2}}\right]^{n/2}\frac{1}{a^{n/2}}.
\end{equation}
By comparing $\widetilde{F}(a)$ and $F(a)$ in the limit of large $a$, we determine the function (\ref{FaApp}) as the approximate threshold function covering all the mixtures of the Gaussian states in this limit. Moreover, it gives rise to a constraint on the approximately optimal Gaussian states; they have the expansion (\ref{appExp}) with $r_{n+1} t \ll r_{n+2}$. According to (\ref{gaussPn}), setting $\vert \alpha \vert^2 = x t$ and $V=1-t$ leads to the constraint $H_{n+1}(x)=0$. Finally, excluding the parameter $a$ from the condition $\exists a: P_n+a P_{n+1}>F(a)$ leads to
\begin{equation}
 P_n^{n+2}>H^4_n(x)\left[\frac{P_{n+1}}{2(n+1)^3}\right]^n,
 \label{appNG}
 \end{equation}
where $x$ is the greatest value among those satisfying $H_{n+1}(x)=0$. The accuracy of the approximation can increase when we assume
\begin{eqnarray}
P_n &\approx& r_n t^n+r_{n+1}t^{n+1}\nonumber \\
P_{n+1} &\approx& r_{n+2} t^{n+2}+r_{n+3}t^{n+3}.
\end{eqnarray}
After a technical but straightforward calculation, the approximate thresholds are given by
\begin{eqnarray}
P_{n+1}&\approx& \frac{t^{n+2}}{3\times 2^{7+2n}(1+n)^n}H_n^2(x)\nonumber \\
&\times&\left[x t+24(1+n)^2t+3(1+n)(16-x t+8 t)\right]\nonumber \\
P_n &\approx& \frac{t^n}{2^{1+2n}(1+n)^n}H_n^2(x)(2+ n t)
\label{nonGThs}
\end{eqnarray}
keeping the condition $H_{n+1}(x)=0$. Inequalities (\ref{appNG}) represent a rough but useful approximation. It has to be used carefully, because their right sides are below the exact thresholds and therefore they can lead to a false positive. On the other hand, it very well illustrates the sensitivity of the quantum non-Gaussianity to imperfections in realistic states as will be discussed later. Approximate relations in (\ref{nonGThs}) are more accurate, and therefore they can be applied to a broader set of states. Their convergence to the true thresholds is depicted in Fig.~\ref{fig:nonGTh}.

\begin{figure}
\centering
\includegraphics[width=0.9\linewidth]{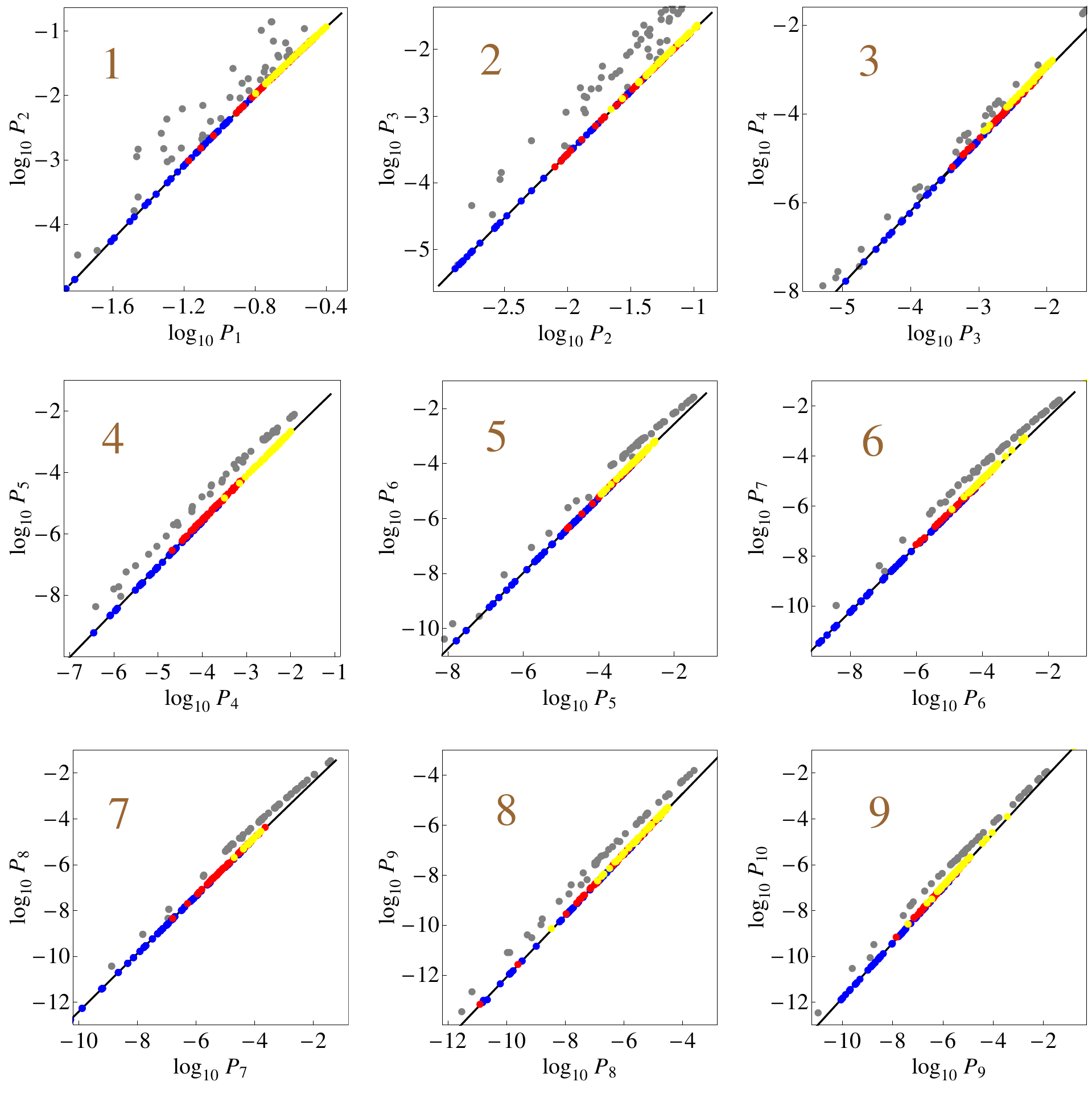}
\caption{Results of a Monte-Carlo simulation for multiphoton quantum non-Gaussianity on multimode states. The simulations demonstrate how thresholds of the quantum non-Gaussianity are covered by single-mode Gaussian states, two-mode Gaussian states, and three-mode Gaussian states. The fifty states closest to the thresholds are depicted by blue (single-mode), red (two modes) and yellow (three modes) points. The total number of cycles in the simulation was $10^5$ (a single-mode), $10^6$ (two modes) and $10^7$ (three modes).  The grey points are yielded from fifty randomly generated single-mode Gaussian states. The black lines depict the thresholds.}
\label{nonGTable}
\end{figure}

The thresholds of the quantum non-Gaussianity preserve their form also for two independent modes $a$ and $b$ filled by a mixture of Gaussian states
\begin{equation}
    S_a(\xi_1) S_b(\xi_2) D_a(\alpha_1)D_b(\alpha_2)\vert 0 \rangle_a \vert 0 \rangle_b.
    \label{twoModeG}
\end{equation}
A Monte-Carlo simulation can demonstrate this conjecture. Since the functions $F_{a,n}$ are linear in a state, the optimal state is necessarily a pure state even in the case of two modes. Therefore, we simulate the results over six parameters identifying the state  (\ref{twoModeG}). Also, the simulation confirms that the thresholds cover the separable Gaussian states occupying even three modes. Fig.~\ref{nonGTable} summarizes the results of the simulations for criteria using up to ten SPADs to measure the probability of the error.

\begin{figure}[t]
\centering
\includegraphics[width=0.9\linewidth]{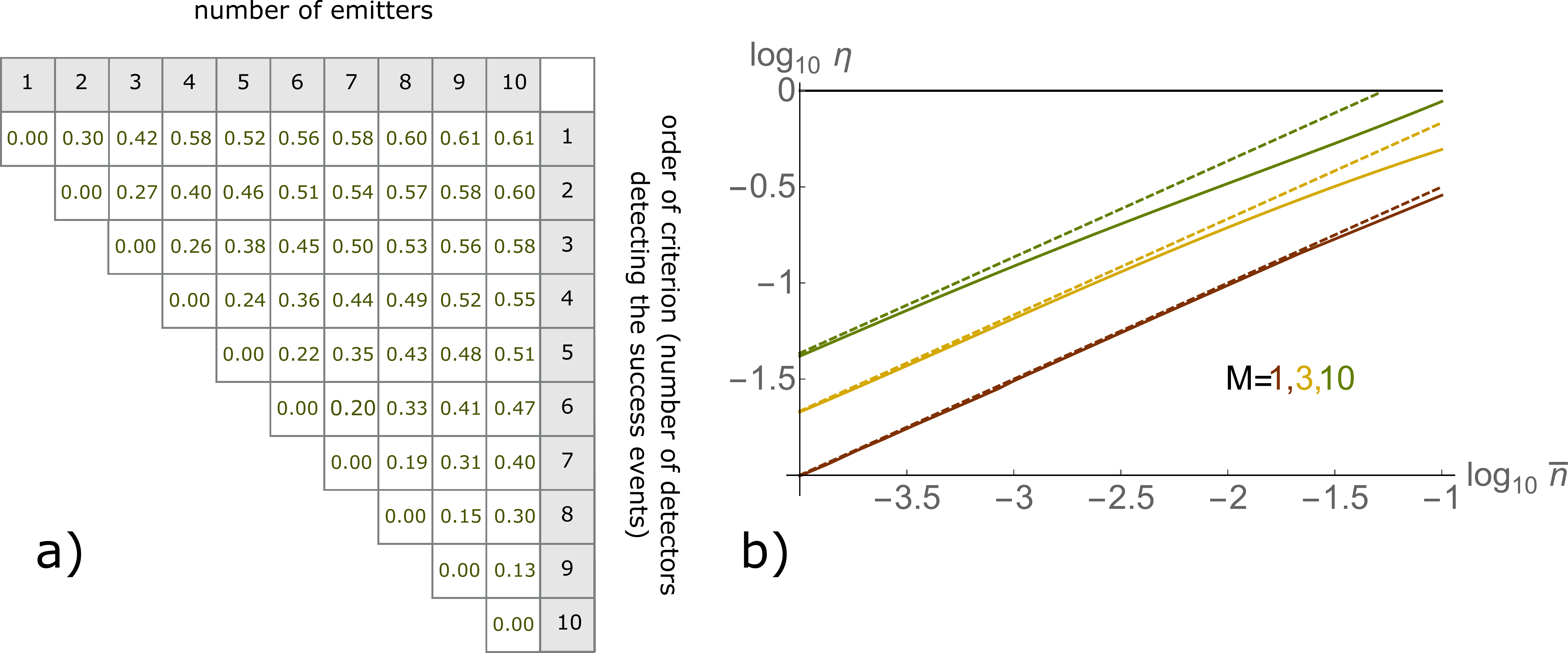}
\caption{\emph{a}) The table states minimal efficiencies $\eta$ in an ideal state $\rho_{\eta,M}$  required for the detection of quantum non-Gaussianity by the criterion with the success probability quantifying clicks of $n$ SPADs. \emph{b}) Quantum non-Gaussianity of multiphoton states from independent emitters with background noise. The state $\rho_{\eta,M}\otimes \rho_{\bar{n}}$ with parameters above the solid lines exhibit quantum non-Gaussianity according to the criterion for $M+1$ SPADs. The colours differentiate the number of emitters. The dashed lines show convergence of conditions in (\ref{modelApCond}) to the exact thresholds.}
\label{fig:MCQNG}
\end{figure}

We apply the criteria to reveal the quantum non-Gaussianity of multiphoton light from $M$ independent emitters with a density matrix
\begin{equation}
\rho_{\eta,M}=\left[\eta \vert 1 \rangle \langle 1 \vert +(1-\eta) \vert 0 \rangle \langle 0 \vert \right]^{\otimes M},
\label{modelNonG}
\end{equation}
where $\eta$ is a product of emission and detection efficiency of a single emitter. The state $\rho_{\eta,M}$ shares the same photon distribution with an attenuated Fock state. Since the states $\rho_{\eta,M}$ are without multiphoton contribution over $M$ photons, their quantum non-Gaussianity is always observable by a criterion with $N=M+1$ SPADs.  When the number of SPADs is lower, the criterion imposes a condition on the parameter $\eta$ as shown in Fig.~\ref{fig:MCQNG} \emph{a}). Therefore, observing the quantum non-Gaussianity requires a complex detector that reveals the truncation of photon statistics.

The involvement of the independent background noise leads to a more realistic model for a multiphoton emission.  We consider $\rho_{\eta,M} \otimes \rho_{\bar{n}}$ with $\rho_{\bar{n}}$ representing background noise with a Poissonian distribution of photons parametrised by a mean number of photons $\bar{n}$. When the background noise affects the state (\ref{modelNonG}), the quantum non-Gaussianity can be lost even if the detector contains many SPADs. The critical amount of the noise for the quantum non-Gaussianity is depicted in Fig.~\ref{fig:MCQNG} \emph{b}). Apparently, the condition gets stricter when $M$ grows. For the states with strongly suppressed noise with $\bar{n} \ll 1$, the quantum non-Gaussianity can be observed when
\begin{equation}
\eta > \frac{H_M^{2/M}(x)}{\sqrt[M]{M!}}\sqrt{\frac{M \bar{n}}{2(M+1)}},
\label{modelApCond}
\end{equation}
where $H_M(x)$ is the same as in relation (\ref{nonGThs}). The approximate condition (\ref{appNG}) is used for the inequality in (\ref{modelApCond}).

The background noise also affects the quantum non-Gaussianity under an optical loss. Although the ideal states without any noise tolerate arbitrary losses, the presence of a small noise results in a sensitivity of quantum non-Gaussianity to attenuation. For small noise, the quantum non-Gaussianity approximately tolerates losses with
\begin{equation}
T>\frac{M \bar{n} H_N^{4/M}(x)}{2\eta^2 (M+1)(M!)^{2/M}}.
\end{equation}
It shows that the robustness of the quantum non-Gaussianity is inversely proportional to the mean number of noise photons. This methodology substantially improves the robustness of the quantum non-Gaussianity to losses compared to the negativity the Wigner function as quantum non-Gaussianity witness.

\section{Experimental test of the multiphoton quantum non-Gaussianity}

The quantum non-Gaussianity criteria were tested experimentally for $M$ independent emission events using spontaneous parametric down-conversion in a periodically poled KTP crystal. The multiphoton light was emitted in $n$ successive time windows, where the heralding detector registered a signal in an idler down-conversion mode. The quantum non-Gaussianity was measured only when the source operated in a regime with low gain, which suppressed the heralding of more than one photon in a signal mode. The challenge was to find a trade-off between sufficiently good statistics of the heralded light and the time necessary to acquire experimental data.

\begin{figure}[t]
\centering
\includegraphics[width=0.5 \linewidth]{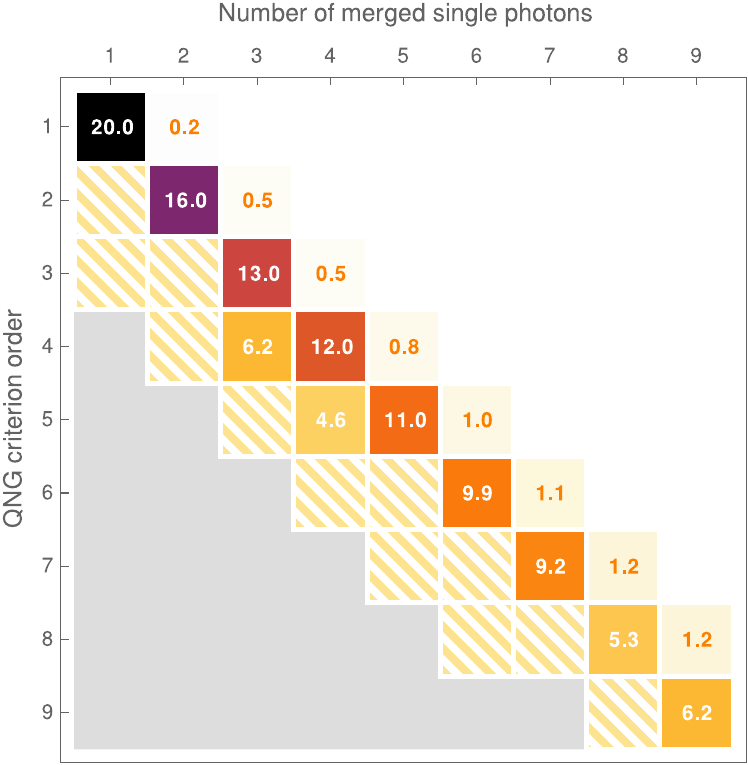}
\caption{The table presents the robustness of the quantum non-Gaussianity against optical losses for the criteria \cite{Straka2018}. The horizontal axis quantifies how many heralded states were merged, and the vertical one shows the employed criterion. The solid boxes corresponded to cases when the quantum non-Gaussianity was recognized. The numbers in these boxes stand for maximal attenuation in decibels that preserve the quantum non-Gaussianity. The orange stripes below the diagonal identify inconclusive cases when error bars cross the thresholds. The grey region stand for situations when no data was acquired. The white area above the diagonal represents combinations when the criteria fail in the recognition.}
\label{fig:nonGExp}
\end{figure}

The detector consisted of a network of polarizing BSs and SPADs. Because the SPADs had different quantum efficiencies, the network was equipped with half-wave plates that were adjusted such that the light was split among each SPAD equally. Such a detector is characterized only by an overall quantum efficiency. Importantly, the efficiency brings only additional optical losses, which cannot produce false quantum non-Gaussianity. The total number of SPADs in the realized detector was $M=10$. It rendered the criteria from functional (\ref{nonGFucnt}) to be tested up to $n=9$. The tested multi-photon light was yielded from temporal mixing of the heralded single-photon states. The quantum non-Gaussianity manifested itself on such a state having a mean number of photons up to five despite detection losses.

The criteria applied on the states yielded from temporal mixing of the heralded single-photon states revealed the quantum non-Gaussianity of a state with a mean number of photons up to five despite detection losses.

Relevant information associated with the quantum non-Gaussianity of light is its robustness against optical losses. The table in Fig.~\ref{fig:nonGExp} summarizes which criteria revealed the quantum non-Gaussianity of merged heralded states together with predicted robustness against failures in decibels. It demonstrates that the property is most resistant when the number of SPADs measuring success events agrees with the number of merged single-photon states. If it is higher, the noise contributes to the measured events dominantly. These cases are inconclusive because the experimental uncertainty did not determine if the measured states surpassed the thresholds. If the number of SPADs is lower than the number of merged states, the tests are not passed mainly due to optical losses.

The experiment confirmed that the criteria of quantum non-Gaussianity are feasible and that they represent stimulating tests for sources of the quantum light involving molecules \cite{Chu2016} or solid-state sources \cite{Somaschi2016,Lund-Hansen2008}. Quantum non-Gaussian criteria can be directly applied to a diagnosis of the Fock states prepared in superconducting circuits \cite{Hofheinz2008} or in the motional degree of freedom of ions captured in the Paul trap \cite{Meekhof1996}.

\section{Quantum non-Gaussian coincidences} 

So far, the quantum non-Gaussian features have been measured over many modes without addressing individual ones. However, the quantum technologies need photonic states approaching a pair of distinguishable single photons \cite{Clauser1974,Kuzmich2003,Kurpiers2018}. They become crucial for many applications since such states can carry the entanglement \cite{Kwiat1999,Brendel1999}. Simultaneously, a new hybrid quantum physics deals with correlation between optical or microwave photons and excitations in atoms \cite{Chou2005,Ritter2012}, solid-state emitters \cite{Yilmaz2010,Usmani2012,Bernien2013}, superconducting circuits \cite{Narla2016} and mechanical systems \cite{Riedinger2016,Riedinger2018}. Before developing these physical platforms for future applications, which can be challenging on many aspects of the engineered states, the generation of photon pairs can be examined according to a capability to manifest quantum non-Gaussian coincidences in a simple detection layout.

\begin{figure}
\centering
\includegraphics[width=0.6 \linewidth]{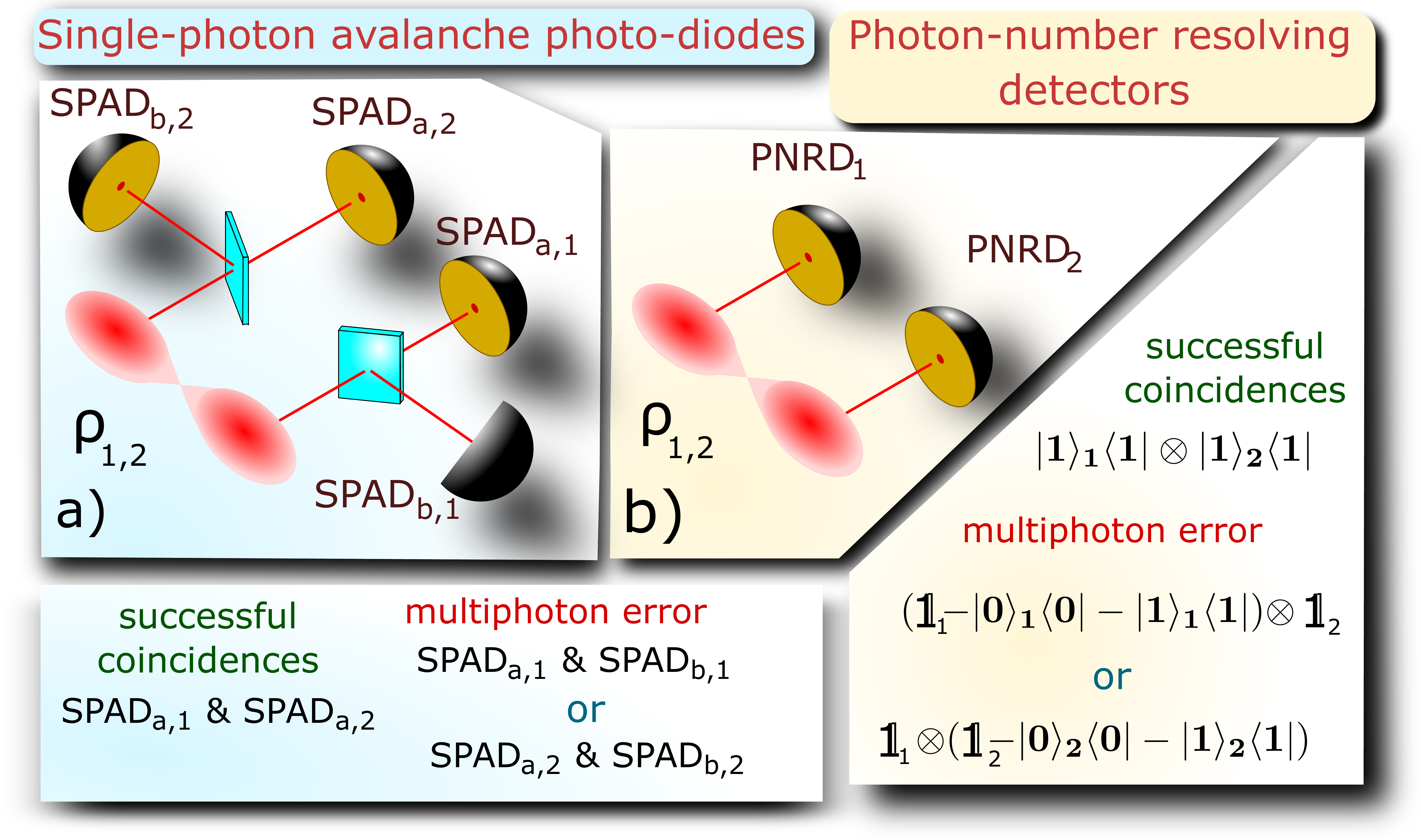}
\caption{Detection of the quantum non-Gaussian coincidences. \emph{a}) The incident state occupying two modes labelled by $1$ and $2$ propagates through a beam-splitter in each mode. Both reflected and transmitted modes are measured by four single-photon avalanche photo-diodes denoted by SPAD$_{a,i}$ and SPAD$_{b,i}$ with $i=1,2$. Whereas the probability $P_s$ of success quantifies coincidences clicks between SPAD$_{a,1}$, the error probabilities $P_{e,i}$ with $i=1,2$ quantify coincidence clicks between SPAD$_{a,i}$ and SPAD$_{b,i}$. \emph{b}) The incident state is measured by two PNRDs. In this case, the success probability $P_s$ refers to the probability given by the projection $|1\rangle_1 \langle 1 |\otimes |1\rangle_2 \langle 1|$ and the error probabilities $P_{e,1}$ and $P_{e,2}$ correspond to the probabilities achieved from the projection $($\(\mathds{1}\)$_1-|0\rangle_1 \langle 0 |-|1\rangle_1 \langle 1 |)\otimes $\(\mathds{1}\)$_2$ and $($\(\mathds{1}\)$_2-|0\rangle_2 \langle 0 |-|1\rangle_2 \langle 1 |)\otimes $\(\mathds{1}\)$_1$, respectively. }
\label{fig:schemeQNGC}
\end{figure}

The quantum non-Gaussian coincidences represent a more challenging feature than nonclassical coincidences of a pair of single-photon states. The criterion for nonclassical coincidences uses the correlation functions 
\begin{equation}
    g_{i,j}=\frac{\langle a_i^{\dagger}a_j^{\dagger}a_i a_j \rangle}{\langle a^{\dagger}_i a_i\rangle \langle a^{\dagger}_j a_j \rangle}
    \label{gij}
\end{equation}
with $i=1,2$ differentiating the modes of light. The Cauchy-Schwarz inequality gives a criterion of nonclassical coincidences \cite{Kuzmich2003}
\begin{equation}
    g_{1,2}^2>g_{1,1}g_{2,2}
    \label{CSgs}
\end{equation}
However, a scheme with SPAD gives the correlation functions (\ref{gij}) only approximately in a limit of a low mean number of photons. The scheme with SPADs is depicted in Fig.~\ref{fig:schemeQNGC} \emph{a}). It comprises two BSs which divide the incident light between two SPADs in each modes. Thus, the scheme contains four SPADs with responses allowing us to approximate any correlation function (\ref{gij}) by coincidence clicks between a respective pair of SPADs. Let us introduce the probability $P_s$ of coincidence clicks between the detectors SPAD$_{a,1}$ and SPAD$_{a,2}$ and probabilities $P_{e,i}$ of coincidence clicks between the detectors SPAD$_{a,i}$ and SPAD$_{b,i}$ with $i=1,2$. We recognize $P_s$ as the success probability and $P_{e,i}$ as the error probability according to the expected response of the detection of the target state $|1\rangle_1 | 1\rangle_2$. Then, the Cauchy-Schwarz inequality gives a direct criterion
\begin{equation}
    P_s^2>P_{e,1}P_{e,2},
    \label{SCP1}
\end{equation}
for the  detection scheme in Fig.~\ref{fig:schemeQNGC}. This condition (\ref{SCP1}) converges to (\ref{CSgs}) for low mean photon numbers in both the modes. A criterion utilizing the probabilities instead of the correlation functions can be also derived from the linear functional $P_s+a (P_{e,1}+P_{e,2})$, which leads to a new criterion 
\begin{equation}
    \frac{2 P_s}{P_{e,1}+P_{e,2}}>1.
    \label{SCP2}
\end{equation}
Criterion (\ref{SCP2}) imposes an identical condition as criterion (\ref{SCP1}) only when a state exhibits $P_{e,1}=P_{e,2}$. Beyond that case, the criterion (\ref{SCP2}) is always more demanding to be fulfilled than (\ref{SCP1}) but the target $|1\rangle_1 | 1\rangle_2$ always satisfies both the criteria. It illustrates that even nonclassical aspects of such coincidences can be better certified.

\begin{figure}
\centering
\includegraphics[width=0.9 \linewidth]{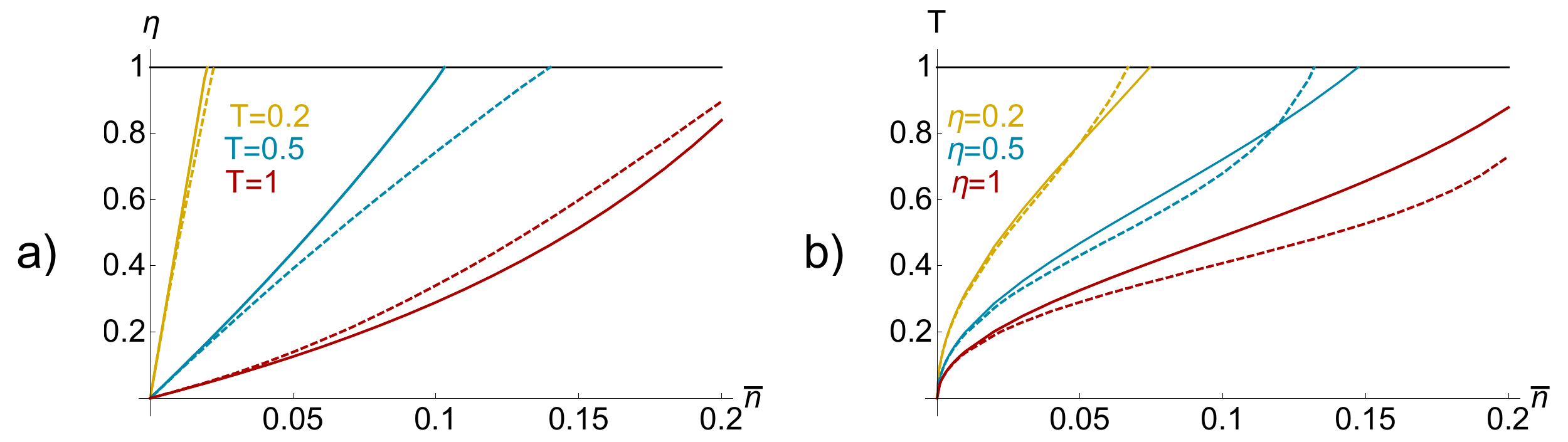}
\caption{Manifestation of the quantum non-Gaussian coincidences in terms of the parameters $\eta$, $T$ and $\bar{n}$ identifying the states with density matrix (\ref{QNGC:model}). \emph{a}) Thresholds in parameters $\eta$ and $\bar{n}$ enabling certification of the quantum non-Gaussian coincidences are shown for various fixed values of the remaining parameter $T$. The colours distinguish the individual values of $T$. The solid lines present the thresholds derived for the scheme with SPADs. The dashed lines correspond to the thresholds yielded from the criterion based on detection employing PNRDs. \emph{b}) The same thresholds are depicted in parameters $T$ and $\bar{n}$ with fixed values of $\eta$. The meaning of the solid and dashed lines remains the same as in \emph{a}).}
\label{fig:QNGCmodel}
\end{figure}

The quantum non-Gaussianity rejects any statistical mixture of Gaussian states. Formally, the quantum non-Gaussianity of states occupying two modes is defined as
\begin{equation}
\rho \neq \int P\left( G \right) \vert G \rangle_{1,2} \langle G \vert \mathrm{d}^2G,
\label{2mQNG}
\end{equation}
where $|G\rangle_{1,2}$ stands for a general pure two-mode Gaussian state and $P(G)$ refers to a probability density function of parameters determining the state $\vert G\rangle_{1,2}$. Employing the Bloch-Messiah reduction of two-mode state, $|G\rangle_{1,2}$ can be always expressed as \cite{Braunstein2005}
\begin{equation}
    |G\rangle_{1,2}=D(\alpha_1)D(\alpha_2)U_{BS}(\tau)S(\xi_1)S(\xi_2)|0\rangle |0\rangle,
\end{equation}
where $S(\xi_i)$ is a squeezing operator acting on the $i$th mode, $U_{BS}(\tau)$ is a unitary evolution operator describing the interaction on a beam-splitter and $D(\alpha_i)$ is a displacement operator acting on the $i$th mode. Since the global phase of any state is irrelevant for a direct detection, $|G\rangle_{1,2}$ has eight parameters unambiguously identifying all features. To witness  quantum non-Gaussian coincidences we use the same layouts in Fig.~\ref{fig:schemeQNGC}. A criterion revealing this feature of coincidences is therefore also based on a response from either SPADs or PNRDs as depicted in Fig.~\ref{fig:schemeQNGC} \emph{a}) and Fig.~\ref{fig:schemeQNGC} \emph{b}), respectively. Let us focus firstly on the former detection layout with SPADs. This layout is identical to the layout described for more strict criterion of the nonclassical coincidences and a criterion incorporates the same probabilities $P_s$, $P_{e,1}$ and $P_{e,2}$ defined for that detection scheme. Maximizing their linear combination
\begin{equation}
    F_{a}(\rho)=P_s+a (P_{e,1}+P_{e,2})
    \label{QNGC:F2G}
\end{equation}
over the states $|G_{1,2}\rangle$ gives a function covering all the mixture of states $|G\rangle_{1,2}$. 
Thus, a crucial step is to determine the involved probabilities for a general state $|G\rangle_{1,2}$. Let us introduce the covariance matrix $\boldsymbol{\sigma}$ according to
\begin{eqnarray}
\sigma_{2i-1,2j-1}&=&\frac{1}{2}(\langle X_i X_j \rangle+\langle X_j X_i \rangle) -\langle X_i \rangle \langle X_j \rangle \nonumber \\
\sigma_{2i,2j}&=&\frac{1}{2}(\langle P_i P_j \rangle+\langle P_j P_i \rangle) -\langle P_i \rangle \langle P_j \rangle \nonumber \\
\sigma_{2i-1,2j}&=&\frac{1}{2}(\langle X_i P_j \rangle+\langle P_j X_i \rangle) -\langle X_i \rangle \langle P_j \rangle \nonumber \\
\sigma_{2i,2j-1}&=&\frac{1}{2}(\langle P_i X_j \rangle+\langle X_j P_i \rangle) -\langle P_i \rangle \langle X_j \rangle,
\label{covMat}
\end{eqnarray}
where $i=1,2$ and $X_i$ and $P_i$ denote the coordinate and momentum operator acting on the $i$th mode. Further, let $\boldsymbol{v}$ denotes a vector with elements
\begin{eqnarray}
v_{2i-1}&=&\langle X_i \rangle \nonumber \\
v_{2i}&=& \langle P_i \rangle.
\end{eqnarray}
The covariance matrix $\boldsymbol{\sigma}$ together with $\boldsymbol{v}$ specify any Gaussian state. The squeezing $S(\xi)$, displacement $D(\alpha)$ and beam-splitter interaction $U_{BS}(\tau)$ effect only linear transformation of $\boldsymbol{\sigma}$ and $\boldsymbol{v}$. This allows us to express
\begin{equation}
    \begin{aligned}
    \boldsymbol{\sigma}=\boldsymbol{O}\boldsymbol{\sigma}_0 \boldsymbol{O}^{T},
    \end{aligned}
\end{equation}
where $\boldsymbol{\sigma}_0$ corresponds to the covariance matrix of the vacuum in four modes, $\boldsymbol{\sigma}$ stands for the covariance matrix of the state that results from propagation of the state $|G \rangle_{1,2}$ through two BSs in Fig.~\ref{fig:schemeQNGC} \emph{a}) and $\boldsymbol{O}$ is some symplectic matrix. The covariance matrix allows us to express the probabilities yielded from projection on the vacuum. Let $\boldsymbol{M}$ denote a matrix with off-diagonal elements being zero and the diagonal elements fulfilling the following. When a projector on the vacuum acts on the $i$th mode, $M_{2i-1,2i-1}=1$ and, simultaneously, $M_{2i,2i}=1$. If the $i$th mode is left without any detection, $M_{2i-1,2i-1}=0$ and, simultaneously, $M_{2i,2i}=0$. The probability $P_{\boldsymbol{M}}$ resulting from such a measurement reads
\begin{equation}
    P_{\boldsymbol{M}}=\frac{\exp \left[ \frac{\boldsymbol{v}(\boldsymbol{\sigma}+\boldsymbol{M})^{-1}\boldsymbol{v}^T-\boldsymbol{v}\boldsymbol{\sigma} \boldsymbol{v}^T}{2}\right]}{\sqrt{\det (\boldsymbol{\sigma}+\boldsymbol{M})}}.
    \label{p0M}
\end{equation}
All the probabilities $P_s$, $P_{e,1}$ and $P_{e,2}$ are expressed by linear combinations of $P_{\boldsymbol{M}}$ of different $\boldsymbol{M}$. This allows us to gain the relations for the function $F_a(|G\rangle_{1,2})$ analytically and perform its maximizing over the states $|G\rangle_{1,2}$. The maximizing state works out to be 
\begin{equation}
    \vert G_{r} \rangle_{1,2} = \sqrt{1- r^2}\sum_{n=0}^{\infty} r^n \vert n\rangle_1 \vert n \rangle_2.
    \label{QNGC:twoModeSQ}
\end{equation}
This induces the criterion
\begin{equation}
    P_s>\frac{1}{2}\sqrt{\frac{P_e}{8+P_e}}\left[2 +P_e+\sqrt{P_e(8+P_e)}\right],
    \label{QNGC:thresSym}
\end{equation}
where $P_e=(P_{e,1}+P_{e,2})/2$. This criterion is necessary condition for quantum non-Gaussian coincidences derived for this detection layout.

Similarly, the quantum non-Gaussian coincidences can be detected in the layout depicted in Fig.~\ref{fig:schemeQNGC} \emph{b}), which comprises two PNRDs responding on both the modes. PNRD distinguishes an exact number of arriving photons. Therefore this layout enables measurement of the success probability $P_s=\langle 1|\langle 1|\rho|1\rangle |1\rangle$ and error probabilities $P_{e,i}$ of having two or more photons in the individual modes. The formalism based on the covariance matrix allows us to gain the probabilities of no photon in the individual modes or no photons in both modes by a simple application of relation (\ref{p0M}) for Gaussian states in two modes. The whole photon number distribution can be expressed from the derivation of (\ref{p0M}) according to the elements of the covariance matrix. Thus, the procedure yielding criterion (\ref{QNGC:thresSym}) can be applied analogously to this detection scheme with PNRDs. The state maximizing the function $F_a(|G\rangle_{1,2})=P_s+a (P_{e,1}+P_{e,2})$ is the state in (\ref{QNGC:twoModeSQ}) again, which results in the criterion
\begin{equation}
    P_s>\sqrt{P_e}-P_e
    \label{QNGC:thresSym2}
\end{equation}
with $P_e=(P_{e,1}+P_{e,2})/2$. Criteria (\ref{QNGC:thresSym}) or (\ref{QNGC:thresSym2}) can be employed for reliable tests of quantum non-Gaussian coincidences in many platforms involving the optical homodyne tomography \cite{Makino2016}, microwave experiments \cite{Gao2018} or trap ions experiments \cite{Ding2017}.

To verify robustness quantum non-Gaussian coincidences, we allow for a model of a realistic state suffering from losses and noise. The model is related to the modern sources in quantum technologies exploiting the emission from a cascade in solid state emitters or atoms inside a two-mode cavity. Energy levels  of such sources allows radiation of states with a density matrix approaching
\begin{equation}
    \rho_{1,2}=\eta |1\rangle_1 \langle 1|\otimes |1\rangle_2 \langle 1|+(1-\eta)|0\rangle_1 \langle 0|\otimes |0\rangle_2 \langle 0|,
\end{equation}
where $\eta$ represents the efficiency of the emission. The processes deteriorating the state $\rho_{1,2}$ are modeled by the map
\begin{equation}
\begin{aligned}
    \rho &=\mbox{Tr}_{3,4}\{ L_{2,4}(T)L_{1,3}(T) \cdot \left[\mathcal{N}_{\bar{n},\bar{n}}(\rho_{1,2}) \otimes \vert 0 \rangle_3 \langle 0 \vert \otimes \vert 0 \rangle_4 \langle 0 \vert \right] \cdot L_{2,4}^{\dagger}(T)L_{1,3}^{\dagger}(T)\},
\end{aligned}
\label{QNGC:model}
\end{equation}
where $L_{i,j}(T)$ stands for the unitary operator of the beam-splitter mixing the modes $i$ and $j$ with transmission $T$. Tracing the modes $3$ and $4$ induces the losses affecting the transmitted state. The map $\mathcal{N}_{\bar{n},\bar{n}}(\rho_{1,2})$ describes effects of noise with $\bar{n}$ being the mean number of noisy photons in the individual modes. We consider the noise to be a result of acting displacement operators on both modes such that the amplitudes of the displacement have random and uncorrelated phases. Thus, model state (\ref{QNGC:model}) involve the relevant realistic imperfections to which the quantum non-Gaussian coincidences are sensitive. Fig.~\ref{fig:QNGCmodel} illustrates the thresholds imposed by criteria (\ref{QNGC:thresSym}) and (\ref{QNGC:thresSym2}) in terms of the parameters $T$, $\eta$ and $\bar{n}$.
Considering the weak emission rates in the layout in Fig.~\ref{fig:schemeQNGC}, the probabilities of the success and error follow the approximation
$P_s \approx T^2 \eta\left[1+2\bar{n}(1-T)\right]/4 +T^2 \bar{n}^2/4$ and $P_{e,1}=P_{e,2} \approx \eta T^2 \bar{n}+T^2\bar{n}^2/4$. Inserting these relations in (\ref{QNGC:twoModeSQ}) induces approximate expressions of the thresholds from Fig.~\ref{fig:QNGCmodel} according to
\begin{equation}
T \approx \sqrt{\frac{\bar{n}}{\eta}},
\label{QNGC:condNonG1}
\end{equation}
which holds for $\bar{n}\ll 1$. Eq.~(\ref{QNGC:condNonG1}) expresses robustness of the quantum non-Gaussian coincidences against losses. Similarly, we can employ the approximate formulas for the probabilities $P_s$, $P_{e,1}$ and $P_{e,2}$ and derive the approximate condition exposing the nonclassical coincidences. It results in the requirement $T \eta>0$, which is independent of the noise. Therefore, the certification of the quantum non-Gaussian coincidences imposes stricter demands on the quality of the cascade source than the recognition of the nonclassical coincidences discussed before.

\begin{figure}
\centering
\includegraphics[width=0.5 \linewidth]{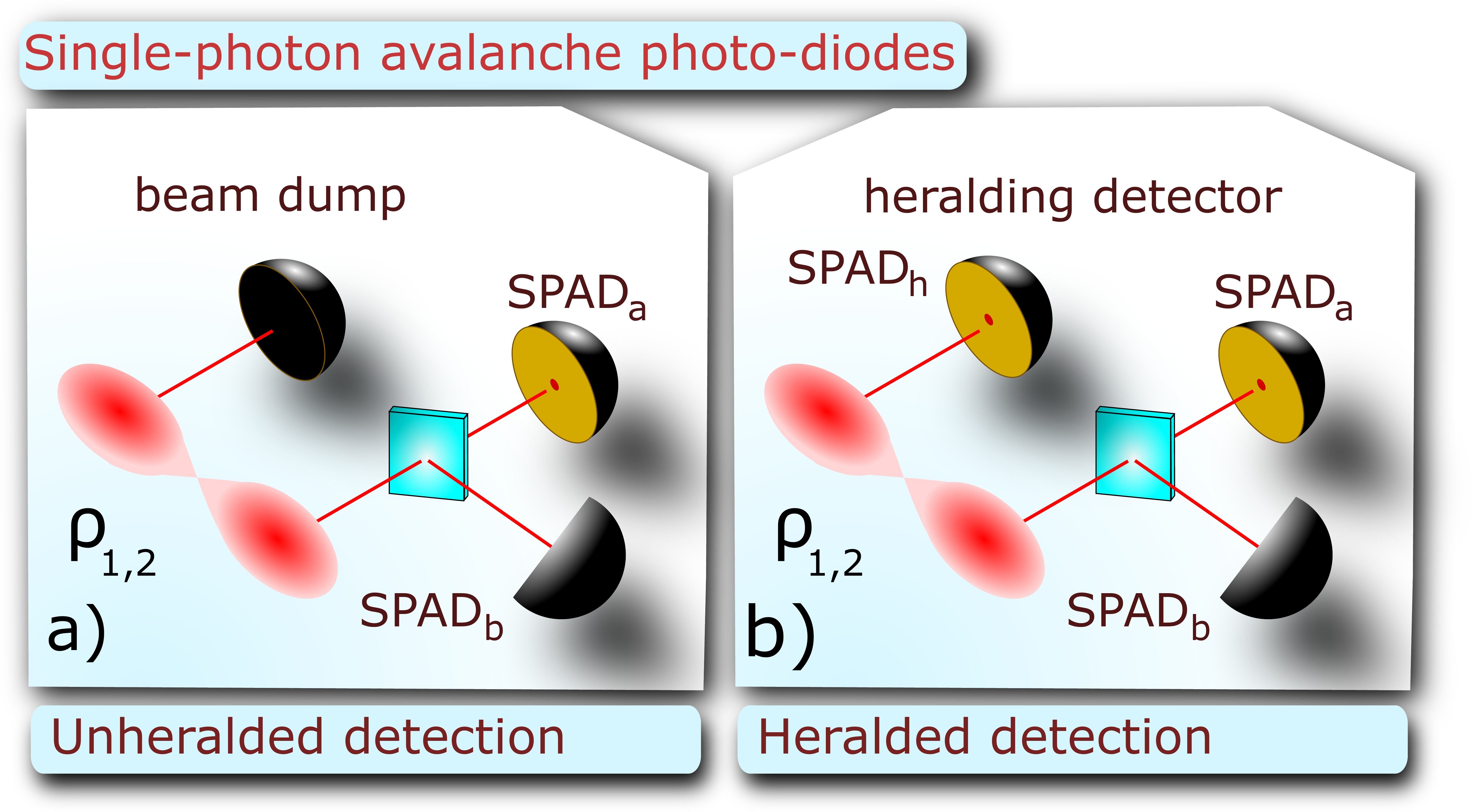}
\caption{The unheralded and heralded certification of the single-mode quantum non-Gaussianity from the coincidences. The incident state $\rho_{1,2}$ exhibits a correlation between the idler and signal mode (labelled by $1$ and $2$). The idler mode is either ignored (\emph{a}) or used for heralding on a click of SPAD$_h$ (\emph{b}). A beam-splitter split the signal mode to the detectors SPAD$_a$ and SPAD$_b$.}
\label{fig:QNGCheralded}
\end{figure}

The model state (\ref{QNGC:model}) assumes a photon number correlation in the two modes (called further idler and signal mode). Thus, heralding on at least one photon in the idler mode can improve or even produce the quantum non-Gaussian aspect of the heralded state in the signal mode. The certification of the single-mode quantum non-Gaussianity uses two SPADs, depicted in Fig.~\ref{fig:QNGCheralded}. Recall, the criterion imposes a threshold for the success probability, being a probability of a click provided by one SPAD, defined by the error probability of simultaneous clicks of both SPADs as described in \cite{Lachman2013}.

\begin{figure}
\centering
\includegraphics[width=0.8 \linewidth]{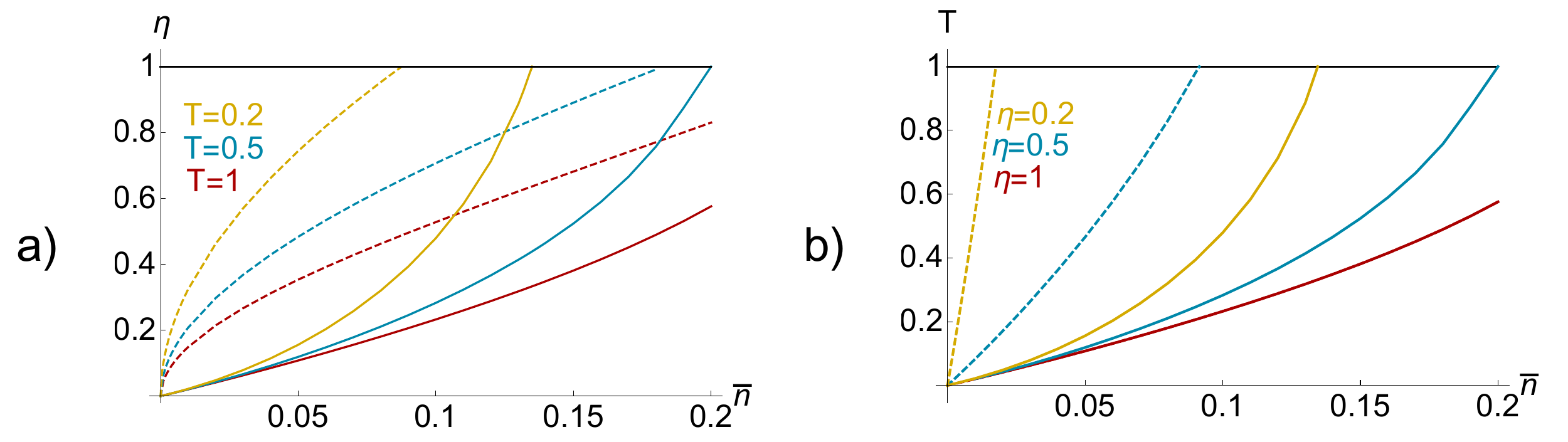}
\caption{\emph{a}) Threshold parameters for certification of the single-mode quantum non-Gaussianity manifested in the signal mode of state (\ref{QNGC:model}). Whereas the solid lines correspond to the threshold of the heralded quantum non-Gaussianity, the dashed lines represent thresholds of the unheralded quantum non-Gaussianity. The colours differentiate a value of losses $T$ that is fixed for a particular threshold. \emph{b}) Robustness of single-mode quantum non-Gaussianity exhibited by state (\ref{QNGC:model}) in the signal mode against losses. }
\label{fig:QNGCHeraldedmodel}
\end{figure}

Let us compare the capabilities of the model state (\ref{QNGC:model}) to exhibit the heralded or unheralded single-mode quantum non-Gaussianity with the manifestation of the quantum non-Gaussian coincidences by this state. In both the scenarios with and without heralding, the density matrix in the signal mode achieves formally the same form
\begin{equation}
    \rho=\mbox{Tr}_2\left \{ L_{1,2}(T)\cdot \left[\mathcal{N}_{\bar{n}}(\rho_{\eta_s}) \otimes \vert 0 \rangle_2 \langle 0 \vert \right]\cdot L_{1,2}^{\dagger}(T)\right \},
    \label{QNGC:reduced}    
\end{equation}
where $\rho_{\eta_s}=\eta_s \vert 1 \rangle_1 \langle 1 \vert+(1-\eta_s)\vert 0 \rangle_1 \langle 0 \vert$ and $L_{1,2}(T)$ and $\mathcal{N}_{\bar{n}}$ are defined as in (\ref{QNGC:model}). The parameter $\eta_s$ is identical with $\eta$ in (\ref{QNGC:model}) when the idler mode is dumped without heralding. In contrast, the heralding effects the parameter $\eta_s$ becomes
\begin{equation}
    \eta_s=\eta T \frac{1-e^{-\bar{n} T}(1-T+\bar{n} T^2)}{1-e^{-\bar{n} T}\left(1-\eta T+\eta\bar{n} T^2\right)}.
\end{equation}
It induces an increase of the single-photon component in the heralded state. Fig.~\ref{fig:QNGCHeraldedmodel} shows thresholds in the model (\ref{QNGC:reduced}) that the single-mode criterion yields. In the regime of low noise contributions, the thresholds of the unheralded quantum non-Gaussianity obey
\begin{equation}
    \eta T=\sqrt{2\bar{n}}.
    \label{QNGC:singleApp}
\end{equation}
Comparing (\ref{QNGC:singleApp}) with (\ref{QNGC:condNonG1}), one can conclude that the quantum non-Gaussian coincidences are more sensitive to the losses than the single-mode quantum non-Gaussianity of the unheralded states. Still, they are more robust against dropping of $\eta$. In contrast, the single mode-quantum non-Gaussianity of the heralded state manifests itself for $T>2\bar{n}$ regardless of the parameter $\eta$. Therefore the quantum non-Gaussianity, in this case, is revealed more efficiently than both the quantum non-Gaussian coincidences and unheralded single-mode quantum non-Gaussianity. Note, criteria (\ref{QNGC:thresSym}) and (\ref{QNGC:thresSym2}) can be satisfied misleadingly by the multi-mode Gaussian states but the single-mode quantum non-Gaussianit threshold covers all the multi-mode Gaussian states. This makes certification of the quantum non-Gaussian coincidences applicable only for states occupying exactly two modes. Currently, quantum non-Gaussian coincidences of light are investigated experimentally in a cascade of quantum dots \cite{Schimpf2021}.

\section{Genuine n-photon quantum non-Gaussianity of light}

Since the Fock states represent building stones of quantum optics, their generation and faithful direct recognition have been attractive targets for many decades. Each Fock state exhibits different negative regions in the Wigner function in phase space. While the negative values of the Fock state $\vert 1 \rangle$ form a simple circle, the Fock state $\vert 2 \rangle $ manifests a negative ring. The higher Fock states show several concentric annuli \cite{Zapletal2017}. The squeezing and displacement operations can deform or shift these negativities, but they cannot change the topology. However, mixing squeezed and displaced superposition of Fock states might make different negative regions.  It gives rise to a question if by such mixing of these states one can mimic higher Fock states \cite{D.F.Walls2008}. Although the answer is subject to further investigation, it stimulates a formulation of a hierarchy of higher quantum attributes that are possessed by individual Fock states. The quantum attributes can be labelled by an order $n$, meaning that the feature is not achieved from definition by any displaced and squeezed superposition of the Fock states lower than $\vert n \rangle $. The squeezing or displacement do not increase the ordered quantum attribute of any state. Besides the classification of states showing negativity of the Wigner function, the hierarchy can be extended to realistic states affected by attenuation that still manifest the quantum non-Gaussianity. A sequence of quantum features that meets these requirements is called genuine $n$-photon quantum non-Gaussianity to recognize it from previous basic quantum non-Gaussianity.

One can believe naively that the criteria of the quantum non-Gaussianity from the previous Chapter constitute the hierarchy. Indeed, these criteria impose conditions that sort somehow the ideal states $\rho_{\eta,M}$ without the noise. However, the criteria always recognize that a state is only not a mixture of Gaussians states, without any further refinement. Criteria recognizing the genuine $n$-photon quantum non-Gaussianity are derived from optimizing over a broader set of states. Therefore, they impose stricter conditions on the truncation of photon statistics of the multiphoton light. It can become an important tool for an analysis of the future sources of quantum multiphoton light.

\subsection{Genuine n-photon quantum non-Gaussianity criteria}

\begin{figure}
\centering
\includegraphics[width=0.45 \linewidth]{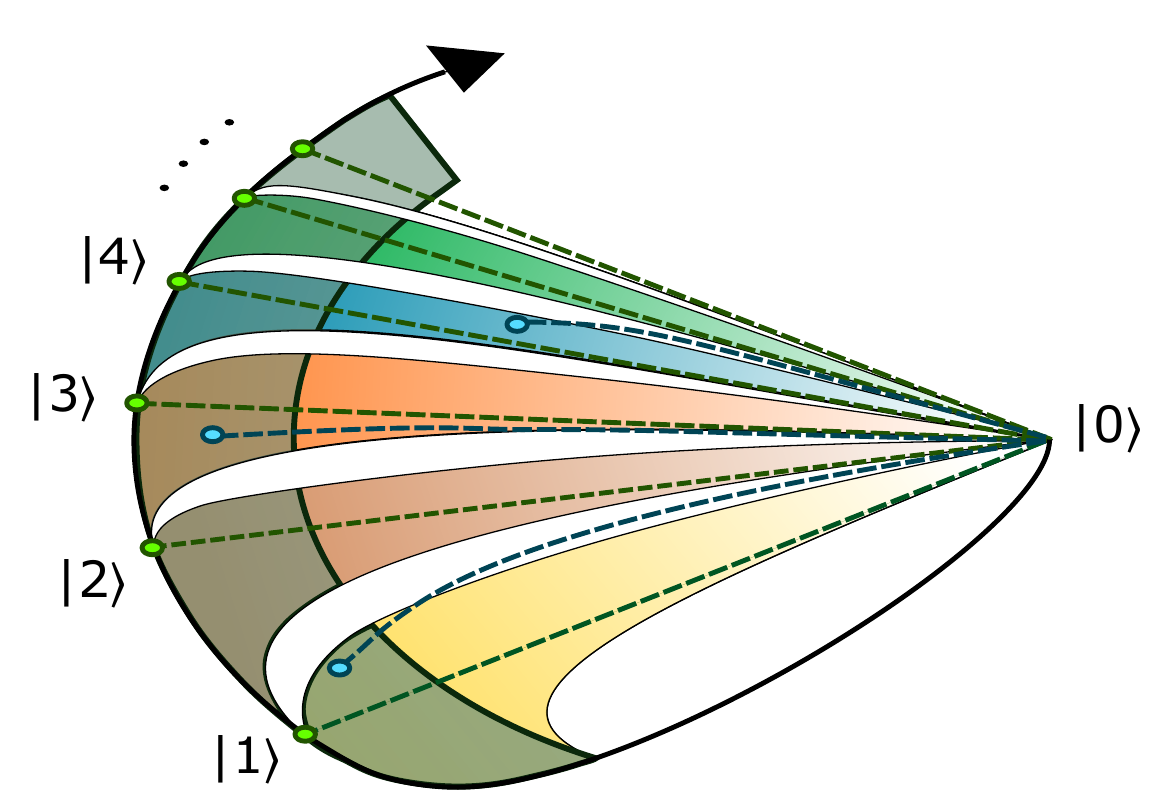}
\caption{A scheme demonstrating the genuine $n$-photon quantum non-Gaussianity. The white area stands for the mixtures of Gaussian states. All color regions correspond to states beyond those mixtures. Different colors represent a hierarchy of new quantum properties, which classify the quantum non-Gaussian states with negative (opaque region) and even positive Wigner function. Each property is inherent to a respective Fock state and cannot be achieved by the lower Fock states and their superposition. The genuine quantum non-Gaussianity of ideal Fock states exhibit absolute robustness against losses, but realistic states can lose the attribute when they are attenuated. The dashed lines depict an impact of attenuation on the ideal Fock states (green points) and realistic states (blue points).}
\label{fig:schemeGQNG}
\end{figure}

The genuine $n$-photon quantum non-Gaussianity of a pure state $\vert \psi \rangle$ is identified with inequality
\begin{equation}
\vert \psi \rangle \neq S(\xi)D(\alpha) \vert \widetilde{\psi}_{n-1} \rangle,
\label{genNG}
\end{equation}
where the core state $\vert \widetilde{\psi}_{n-1} \rangle$ is any superpositions of the Fock states $\vert 0 \rangle$,..., $\vert n-1 \rangle$. The Gaussian transformation $S(\xi)D(\alpha)$ changes a shape of the Wigner function, breaks a sharp truncation in distribution of photons but cannot produce the core state $\vert \widetilde{\psi}_{n} \rangle$ associated with the following order. The definition can be extended to mixtures of states. A state with a density matrix $\rho$ possesses the genuine $n$-photon quantum non-Gaussianity if $\rho$ is not identical with any statistical mixture of the right side of inequality (\ref{genNG}). The lowest order attribute is identical to a basic quantum non-Gaussianity discussed until now since it refuses all squeezed coherent states. The second-order quantum non-Gaussianity means that a state is beyond any mixtures of state $S(\xi)D(\alpha)(c_1 \vert 1 \rangle + c_0 \vert 0 \rangle)$ with complex $c_0$ and $c_1$ satisfying $\vert c_0 \vert^2+\vert c_1 \vert^2=1$. A scheme illustrating this new hierarchy is depicted in Fig.~\ref{fig:schemeGQNG}.

\begin{figure}
\centering
\includegraphics[width=0.9 \linewidth]{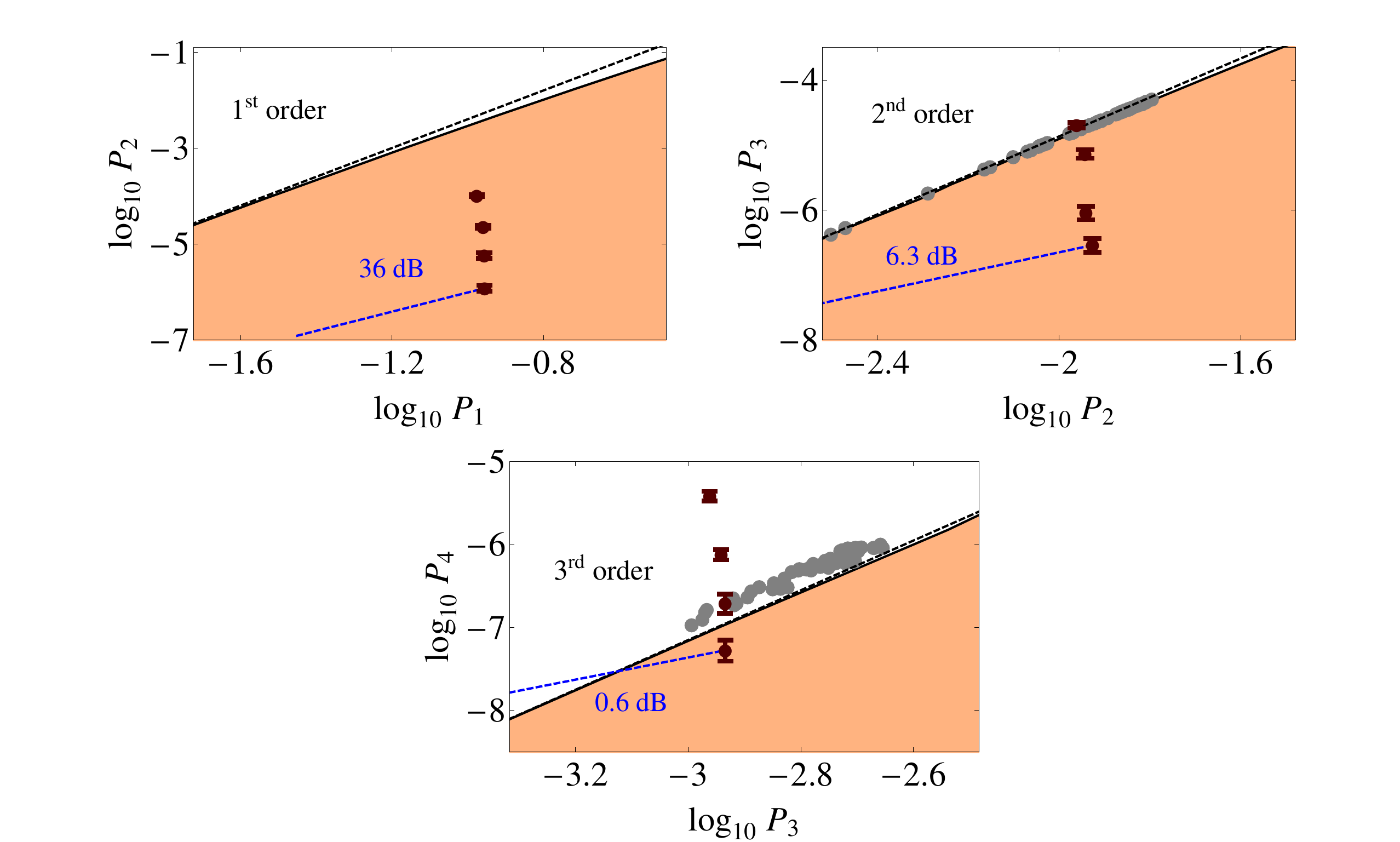}
\caption{Genuine $n$-photon quantum non-Gaussianity. The threshold of the genuine $n$-photon quantum non-Gaussianity up to order three can be compared with experimentally achieved data and results of a Monte-Carlo simulation. The states surpassing the criteria are in the orange regions. The solid black lines in the boundary of the orange regions are thresholds derived exactly, and the dashed black lines are the approximate solutions (\ref{gqngApp}). The grey points correspond to fifty points generated in the Monte-Carlo simulation closest to the threshold. Brown points represent the measured states. The sequences of the brown points in the vertical direction demonstrate an impact of background noise on the measured states. The noise exhibits Poissonian statistics with the mean number of photons $\bar{n}=0, 4 \times 10^{-5}, 2 \times 10^{-4}, 10^{-3}$ in the detection window. The dashed blue lines predict a movement of the states without deterioration by the background noise when they are affected by attenuation. The theoretical robustness is shown above the blue dashed lines.}
\label{fig:gqng}
\end{figure}

\begin{figure}
\centering
\includegraphics[width=0.9 \linewidth]{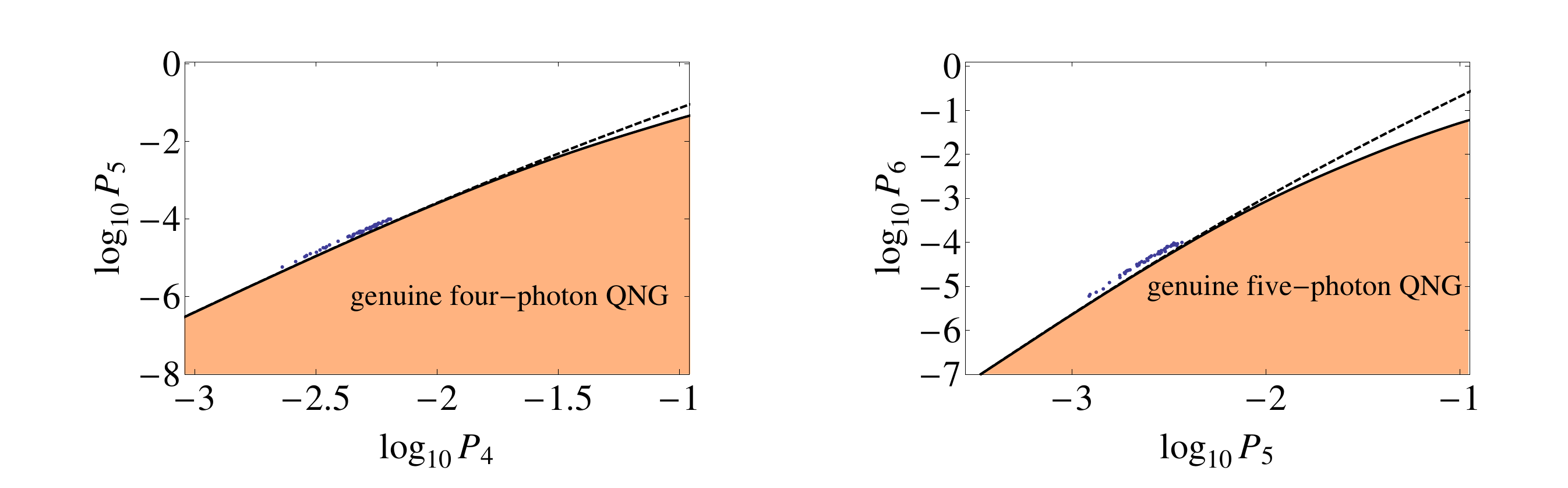}
\caption{Thresholds for genuine four and five-photon quantum non-Gaussianity. States possessing those quantum attributes are in the orange regions. The blue points represent fifty results of a Monte-Carlo simulation that were generated closest to the thresholds.}
\label{fig:gqng45}
\end{figure}

Detection of the genuine $n$-photon quantum non-Gaussianity utilizes the same layout that is exploited for the recognition of thebasic  quantum non-Gaussianity of the multiphoton light before. Also, the criteria consider success and error probabilities that are uniform with the probabilities used in that hierarchy of conditions, i. e. $P_n$ refers to the probability of simultaneous clicks of $n$ SPADs and $P_{n+1}$ denotes the probability of clicks of $n+1$ SPADs.  However, the thresholds differ since they are yielded from optimizing of the linear functional 
\begin{equation}
    F_{a,n}(\rho)=P_n+a P_{n+1}
    \label{gqngF}
\end{equation}
over mixtures of the states $S(\xi)D(\alpha) \vert \widetilde{\psi}_{n-1} \rangle$. Solving the optimizing is technically more difficult because it has to be done over the core states $\vert \widetilde{\psi}_{n-1} \rangle$ and squeezing and displacement operations. The core state $\vert \widetilde{\psi}_{n-1} \rangle =\sum_{k=0}^{n-1}c_k \vert k \rangle$ is described formally by $n$ complex coefficients, which hold normalization. Since a global phase does not differentiate the quantum states, the state $\vert \widetilde{\psi}_{n-1} \rangle =\sum_{k=0}^{n-1}c_k \vert k \rangle$ is determined by $2(n-2)$ parameters. Together with four more parameters characterizing the squeezing and the displacement operations, the right side in inequality (\ref{genNG}) is determined by $2(n+1)$ parameters over which the optimizing of function (\ref{gqngF}) was carried out.
It was assumed that the optimal squeezing $\xi$ and displacement $\alpha$ are real and also the optimal core state has a form $\vert \widetilde{\psi}_{n-1} \rangle = c \vert n-1 \rangle + \sqrt{1-c^2} \vert n -2 \rangle$ with $c$ real. These premises were verified by a Monte-Carlo simulation. The used algorithm eliminates successively all the parameters characterizing the optimal state besides the minimal variance of the quadrature in time $V$, which was left as a parameter determining a curve $\left[P_n(V), P_{n+1}(V) \right]$ corresponding to the threshold in the employed probabilities of success and error. The algorithm exploits a relation that is holded by the optimal squeezing and displacement parameters
\begin{equation}
\partial_{V}P_n \partial_{\alpha} P_{n+1}-\partial_{V}P_{n+1} \partial_{\alpha} P_{n}=0.
\label{eqalf}
\end{equation}
Similar identities can be retrieved for the coefficient $c$
\begin{equation}
\partial_{c}P_n \partial_{\alpha} P_{n+1}-\partial_{c} P_{n+1} \partial_{\alpha} P_{n}=0.
\label{eqck}
\end{equation}
The algorithm converges to the optimal state as follows, it initially sets $\vert \widetilde{\psi}_{n-1} \rangle = \vert n-1 \rangle $ and generates the optimal $\alpha$ for a fixed $V$ through relation (\ref{eqalf}). For those $\alpha$ and $V$ it applies the identity (\ref{eqck}) to acquire a corrected state $\vert \widetilde{\psi}_{n-1} \rangle$. With that state, it solves equation (\ref{eqalf}) again. This can be repeated several times. It reduces all the parameters to the remaining $V$, which parametrizes the threshold in the detected probabilities. When the optimal squeezing is very small, i. e. $1-V \ll 1$, the optimal parameters gain $c \approx 1$ and $\alpha^2 \approx 2 (1-V) + (3+2n+n^2) (1-V)^2/3$ and the approximate thresholds read as
\begin{eqnarray}
P_{n+1} &\approx & \frac{n n!(2+n)^2}{55296 (n+1)^{n-1}}t^3 \left[384+t(896+307 n + 99 n^2) \right]\nonumber \\
P_n & \approx & \frac{n n!}{12 (n+1)^n} t \left[ 6+t (6+12 n +n^2)\right],
\label{gqngApp}
\end{eqnarray}
where $t$ parametrizes the thresholds. Fig.~\ref{fig:gqng} depicts the exactly resolved thresholds for the second and third order and compare them with an approximate solution (\ref{gqngApp}). The figure also shows results of the Monte-Carlo simulation that verifies the thresholds. It was performed by generating randomly squeezing, displacement and the core state $\vert \widetilde{\psi}_{n-1} \rangle$ in $10^6$ (2nd order) and $10^8$ (3rd order) cycles. The intervals where the parameters were randomly generated were set such that the respective simulated probabilities fill the region where the experimental data was acquired. Fig.~\ref{fig:gqng} presents thresholds for genuine four and five-photon quantum non-Gaussianity together with approximate thresholds (\ref{gqngApp}) as well. Again, both the thresholds were verified by a Monte-Carlo simulation with $10^8$ cycles. The Figs.~\ref{fig:gqng} and \ref{fig:gqng45} show the accuracy of the approximation is dropping for higher $n$.

\begin{figure}
\centering
\includegraphics[width=0.5\linewidth]{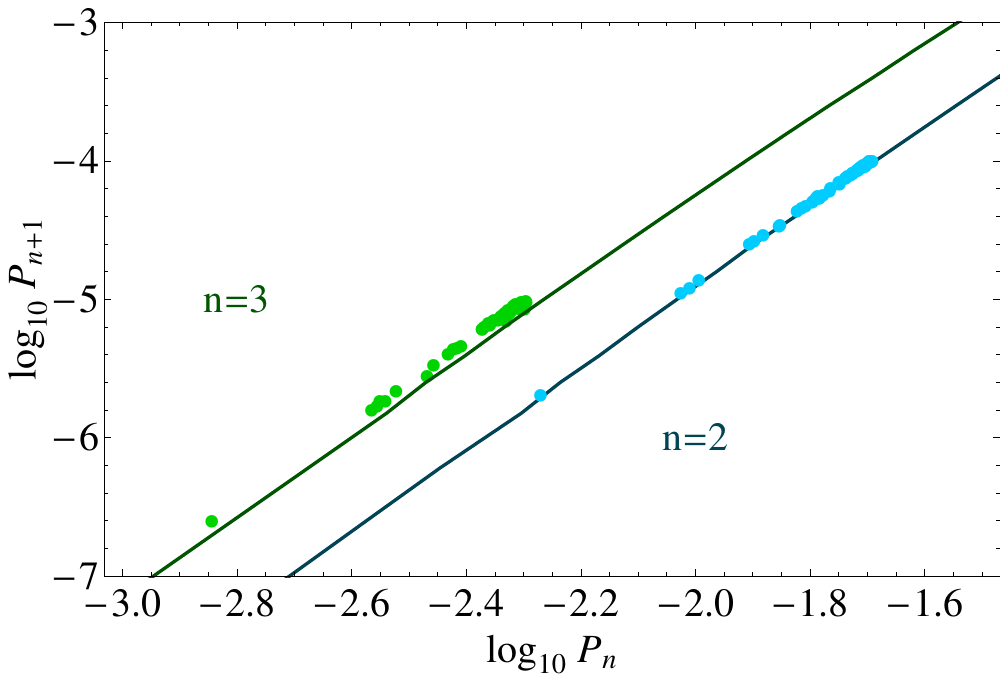}
\caption{Results of a Monte-Carlo simulation confirming that two-mode states in the right side of inequality (\ref{twoModeG}) do not surpass the thresholds of genuine two and three-photon quantum non-Gaussianity. The blue points correspond to fifty points closest to the threshold of genuine two-photon quantum non-Gaussianity. The green points represent the best fifty attempts to the threshold of genuine three-photon quantum non-Gaussianity. The number of generated states was $10^8$ in both cases.}
\label{fig:twoMC}
\end{figure}

The genuine $n$-photon quantum non-Gaussianity simulated by multi-mode states of light requires a theoretical confirmation that the thresholds remain the same when they are derived from optimizing over multi-mode states. A multi-mode core state $\vert \widetilde{\psi}_{n-1} \rangle$ obeys a condition
\begin{equation}
    \langle m_1 \vert \otimes ... \otimes \langle m_M \vert \widetilde{\psi}_{n-1} \rangle \neq 0
\end{equation}
only if $\sum_{i=1}^M m_i<n$, where $\langle m_i \vert$ is the Fock state occupying the $i$th mode and $M$ denotes a number of considered modes. Whether a core state in a single-mode case has a form $\vert \widetilde{\psi}_{n-1}\rangle=\sum_{k=0}^{n-1} c_k \vert k \rangle$, the core states occupying two modes are expressed as  $\vert \widetilde{\psi}_{n-1}\rangle=\sum_{k=0}^{n-1}\sum_{l=0}^{n-k-1} C_{k,l} \vert k \rangle \vert l \rangle$. The higher photon contribution can be produced as a consequence of the squeezing or displacement acting on the core state. Let $S_{i}(\xi_i)$ and $D_i(\alpha_i)$ denote squeezing and displacement operators acting on the $i$th mode with $\xi_i$ and $\alpha_i$ being the parameters determning the operators. A pure state $\vert \psi \rangle$ exhibits the genuine $n$-photon quantum non-Gaussianity when
\begin{equation}
    \vert \psi \rangle \neq S_M(\boldsymbol{\xi})D_M(\boldsymbol{\alpha})\vert \widetilde{\psi}_{n-1}\rangle,
    \label{nmodeGNQNG}
\end{equation}
where $\boldsymbol{\xi}$, $\boldsymbol{\alpha}$ are vectors $\boldsymbol{\xi}=(\xi_1,...,\xi_M)$, $\boldsymbol{\alpha}=(\alpha_1,...,\alpha_M)$ and $S_M(\boldsymbol{\xi})$, $D_M(\boldsymbol{\alpha})$ read
\begin{eqnarray}
S_M(\boldsymbol{\xi})&=&\Pi_{i=1}^M \otimes S_i(\xi_i)\nonumber \\
D_M(\boldsymbol{\alpha})&=&\Pi_{i=1}^M \otimes D_i(\alpha_i).
\end{eqnarray}
The genuine $n$-photon quantum non-Gaussianity of a general state also reject all statistical mixtures of the right side of inequality (\ref{nmodeGNQNG}). Again, a Monte-Carlo simulation certified that these states do not exceed the thresholds. Fig.~\ref{fig:twoMC} demonstrates the thresholds of the genuine two and three-photon quantum non-Gaussianity cover all the states that were generated in the simulation.

An application of the criteria can be analysed on the considered model of multiphoton light $\rho_{\eta,M}\otimes \rho_{\bar{n}}$, where $\rho_{\eta,M}$ is expressed in (\ref{modelNonG}) and $\rho_{\bar{n}}$ has the Poissonian photon distribution with the mean number of photons $\bar{n}$. Assuming, the state is deteriorated by very low noise with $\bar{n}\ll \eta^{2M}$, where $M$ is a number of emitters in the ensemble, the genuine $M$-photon quantum non-Gausssianity requires
\begin{equation}
    \eta^M>\frac{12 M}{\sqrt{M^3+5M^2+8M+4}}\sqrt{\bar{n}}.
\end{equation}
Since the inequality compares $M$ power of $\eta$ with the square root of $\bar{n}$, achieving the genuine $M$-photon quantum non-Gaussianity becomes very sensitive to noise with a growing number of emitters $M$.

The experimental feasibility of achieving the genuine $n$-photon quantum non-Gaussianity was investigated. The quantum light source exploited the spontaneous parametric down-conversion process in a crystal. The generation of multiphoton light was identical to the previous experiment. The genuine $n$-photon quantum non-Gaussianity was observed on the light where three heralded states were merged. Fig.~\ref{fig:gqng} demonstrates the experimental results. The data was deteriorated by background noise artificially to explore the impact of the noise. The robustness against losses was also estimated theoretically to analyse the feasibility of the genuine $n$-photon quantum non-Gaussianity. Whereas the genuine one-photon quantum non-Gaussianity survives attenuation $36$ dB, the genuine three-photon quantum non-Gaussianity is lost already for $0.6$ dB.

Because the strictness of the criteria on the unwanted heralding of two or more photons increases with the order in the hierarchy, fulfilling these criteria requires decreasing the gain of the parametric process and a time window for the coincidence events so that the quality of the heralded single-photon state becomes high. The low gain means that the heralding events rarely occur, prolonging the experimental time. When the gain is deficient, deterioration of the click statistics by the dark counts of the detector becomes relevant. The limiting factors for detecting the following genuine four-photon quantum non-Gaussianity appeared both the dark counts of the detector and the measurement time, which would take several months.

This experimental test revealed the genuine $n$-photon quantum non-Gaussianity only up to the order three corresponding to the Fock state $|3\rangle$. Promising platforms where the quantum $n$-photon quantum non-Gaussianity can be manifested over this limit involve the quantum dots in nanophotonics structures \cite{Somaschi2016,Lund-Hansen2008}, molecules in an antenna \cite{Chu2016}, superconducting circuits \cite{Hofheinz2008}, or motional excitation of ions in the Paul trap \cite{Meekhof1996}, where the Fock states can be generated without the heralding.

\section{Quantum non-Gaussian criteria for phononic Fock states of ions}

Mechanical oscillator states of atoms can exhibit genuine quantum non-Gaussian properties in a single motional mode. They find direct application in quantum sensing \cite{Gessner2019,Wolf2019}, quantum computing and error correction \cite{Mari2012,Niset2009,Michael2016,CampagneIbarcq2020}. Moreover, motional states of ions captured in the Paul trap do not suffer from losses compared to optical platforms. Advantageously, they experience little deterioration due the noise from interaction with the environment. This makes this platform almost ideal for studying the tests of genuine $n$-phonon quantum non-Gaussianity and comparing it to phonon statistics sufficient for quantum sensing.

The genuine $n$-phonon quantum non-Gaussianity of the mechanical oscillator follows the definition of this quantum property for the photonic states, i.e. it rejects all the mixtures of the state
\begin{equation}
    |\psi_{n-1} \rangle = D(\alpha)S(\xi)\sum_{k=0}^{n-1}c_k |k\rangle,
\end{equation}
where the displacement operator $D(\alpha)$ and the squeezing operator $S(\xi)$ perform a Gaussian modulation of the core state $\sum_{k=0}^{n-1}c_k |k\rangle$. This property can also be defined equivalently in the stellar representation as a number of roots of the Husimi $Q$ function \cite{Chabaud2020}. We allow for the most strict condition for a single copy measurement of the phononic states near the Fock states of the trapped ions. It aims to define a demanding threshold that is exceeded only by a narrow set of realistic states that closely approach the target Fock states. A criterion based only on the success probability $P_n=\langle n|\rho|n \rangle$ yields such a tight requirement. Such an absolute criterion is, therefore, a simplified version of the previous single-parameter criterion (\ref{gqngF}). The new criterion based only on $P_n$ will be more challenging to fulfil; however, it will conclusively witness that $P_n$ itself is not achievable by any lower state in the hierarchy. From this perspective, criterion (\ref{gqngF}) is a helpful navigator when damping affects the Fock states.

\begin{figure}
\centering
\includegraphics[width=0.6\linewidth]{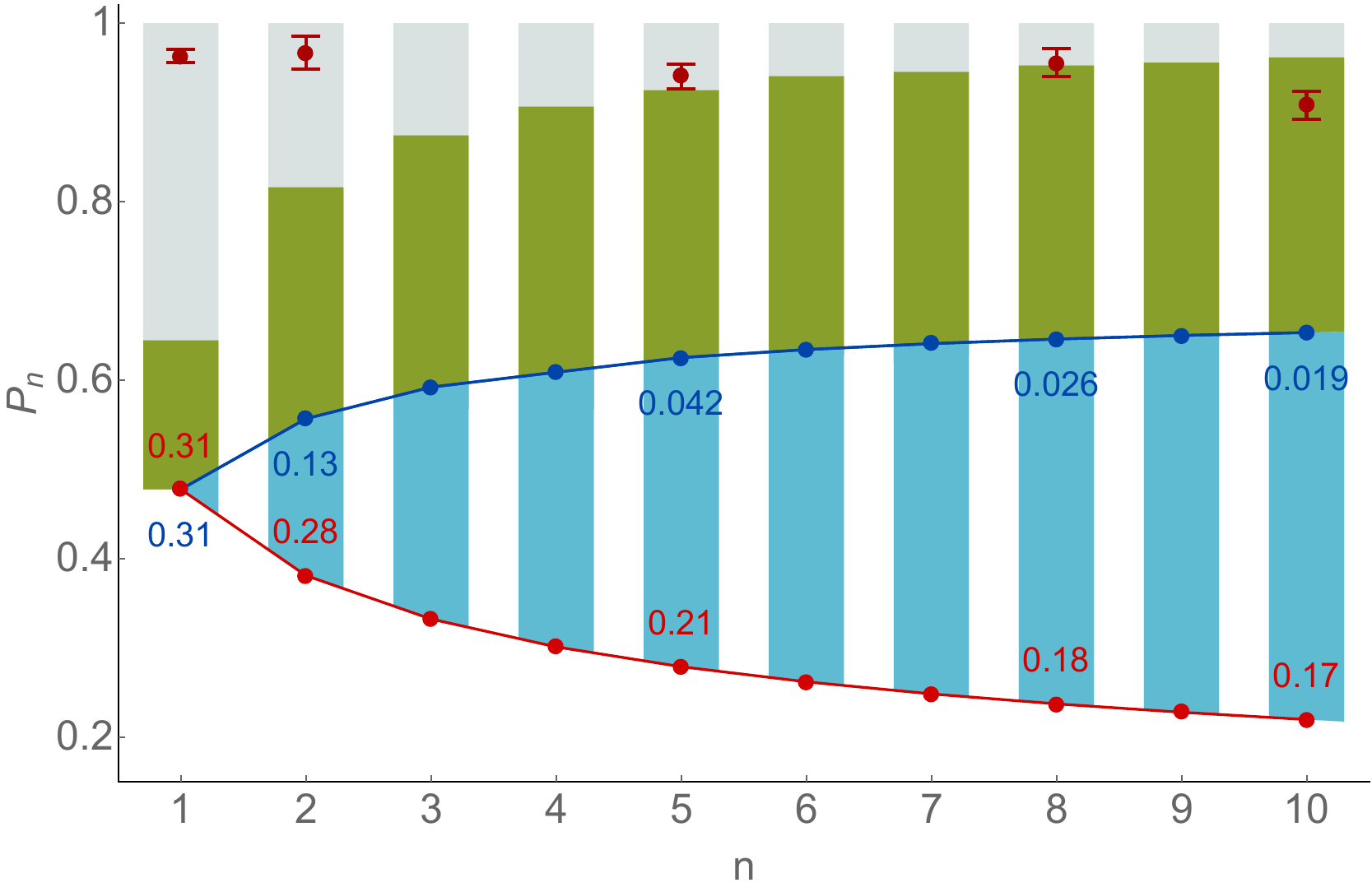}
\caption{Evaluation of realistic states of motion approaching the Fock states $|n\rangle$ up to $n=10$. The brown points with error bars depict the experimental results. The blue points show the thresholds $\bar{P}_n$ for the genuine $n$-phonon quantum non-Gaussianity. Similarly, the red points correspond to the lowest probability $P_n$ for which  the model of the noisy Fock states exhibit the proper number of negative regions in Winger function. Thus, the light blue stripes signify regions where overcoming the thresholds is more demanding than the conditions relying on the negativity of the Wigner function. In contrast, a probability $P_n$ above the green stripes guarantees a metrological advantage of a state against the lower ideal Fock state.
The green, blue and red numbers associated with those thresholds quantify the thermal depth of the measured states - the maximal mean number of the noisy phonons preserving the related quantum aspects.}
\label{fig:Fock1}
\end{figure}

The absolute criterion relies on surpassing the thresholds
\begin{equation}
    \bar{P}_n=\max_{|\psi_{n-1}\rangle}|\langle n |\psi_{n-1}\rangle|^2,
\end{equation}i.e. genuine $n$-phonon quantum non-Gaussianity manifests itself when $\langle n|\rho|n\rangle>\bar{P}_n$.
Since the state $|\psi_{n-1}\rangle$ is determined by $2n+2$ parameters, the complexity of the maximized states grows up with $n$, which makes the maximizing task harder to solve for higher $n$. All the parameters identifying the maximizing state are real and, further, they obey
\begin{equation}
\begin{aligned}
     \partial_{\alpha}\vert \langle n|D(\alpha)S(\xi)\sum_{k=0}^{n-1}c_k |k\rangle \vert ^2=0\\
     \partial_{\alpha}\vert \langle n|D(\alpha)S(\xi)\sum_{k=0}^{n-1}c_k |k\rangle \vert ^2=0\\
     \partial_{c_{n-1}}\vert \langle n|D(\alpha)S(\xi)\sum_{k=0}^{n-1}c_k |k\rangle \vert ^2=0\\
     \vdots \\
     \partial_{c_{1}}\vert \langle n|D(\alpha)S(\xi)\sum_{k=0}^{n-1}c_k |k\rangle \vert ^2=0,
\end{aligned}
\label{MF:eqsArray}
\end{equation}
which holds together with the constraint $c_0=\sqrt{1-\sum_{k=1}^{n-1} c_k^2}$. Sequentially solving (\ref{MF:eqsArray}) converges quickly to the maximizing parameters because they fulfill $c_k \gg c_{k-1}$. The maximal $\bar{P}_n$ can be further checked by a sufficiently large number of attempts in a Monte-Carlo simulation generating random states $|\psi_{n-1}\rangle$ in order to confirm that the maximizing state $|\psi_{n-1}\rangle$ is given by the real parameters. Although this approach can be applied for deriving even more complex criteria, the thresholds in this particular maximizing task read
\begin{equation}
    \bar{P}_n=\max_{\alpha,\xi} \mbox{Tr}\left[D(\alpha)S(\xi)|n-1\rangle \langle n-1|S(\xi)^{\dagger}D(\alpha)^{\dagger} \right],
    \label{theorem2}
\end{equation}
which was analytically proved in \cite{Chabaud2021}. Identity (\ref{theorem2}) always simplifies the maximizing task to searching for the maximum over just two complex parameters. The values of $\bar{P}_n$ are depicted in Fig.~\ref{fig:Fock1}.

An appropriate model of realistic motional states of ions approaching the Fock states allows for contribution of noise. In the simplest description, the noise is a result of the map
\begin{equation}
    \mathcal{M}_{\bar{n}}(\rho)=\frac{1}{2\pi}\int e^{-\frac{|\alpha|^2}{\bar{n}}}D(\alpha)\rho D^{\dagger}(\alpha)\mathrm{d}^2\alpha,
    \label{Focks:noiseMap}
\end{equation}
where $D(\alpha)$ is a displacement operator and $\bar{n}$ corresponds to the mean number of phonons that $\mathcal{M}_{\bar{n}}$ produces on the ground state. In the limit of small $\bar{n}$, this map can be simplified to
\begin{equation}
    \mathcal{M}_{\bar{n}}(\rho)\approx \rho+ \bar{n}^2 \left[a \rho a^{\dagger}+a^{\dagger} \rho a-(a^{\dagger}a+1/2)\rho-\rho(a^{\dagger}a+1/2)\right],
    \label{FockMapN}
\end{equation}
where the terms proportional to $\bar{n}^4$ are neglected. This approximation describes a deterioration process caused by scattering of a single-photon on an ion with $\bar{n}$ being the Lamb-Dicke parameter \cite{Leibfried2003}. Thus, the effects of $\mathcal{M}_{\bar{n}}$ on the Fock states serve as a simulation of the realistic states. Simultaneously, the parameter $\bar{n}$ can be exploited to define the depth of genuine $n$-phonon quantum non-Gaussianity as the maximal $\bar{n}$ that preserves recognition of this quantum property on the state $\mathcal{M}_{\bar{n}}(\rho)$. 
Model (\ref{Focks:noiseMap}) also allows us to compare the manifestation of the genuine $n$-phonon quantum non-Gaussianity with an exhibition of the corresponding number of negative annuli in the Wigner representation of the model state. Fig.~\ref{fig:Fock1} demonstrates clearly that exceeding the threshold probabilities $\bar{P}_n$ imposes a stricter requirement on the statistics than the negativity of the Wigner function.

A more complex model of realistic motional states assumes evolution dictated by the damping of the states $|n\rangle$ to $|n-1\rangle$ and incoherent excitation of $|n-1\rangle$ to $|n\rangle$ according to the rate equations \cite{Leibfried2003}
\begin{equation}
    \frac{d}{dt}P_n=A (n+1)P_{n+1}+B n P_{n-1}-\left[A n+B(n+1)\right]P_n,
    \label{Focks:ModelEq}
\end{equation}
where $P_m$ stands for the phonon number distribution and $A$ and $B$ corresponds to the rates of the damping and excitation, respectively. Equations (\ref{Focks:ModelEq}) provides a steady state solution $P_n=\bar{n}^n/(1+\bar{n})^{n+1}$ with $\bar{n}=B/(A-B)$ only for $A>B$. The opposite case $A \leq B$ leads to divergence of the mean number of phonons due to the heating process. Moreover, equations (\ref{Focks:ModelEq}) preserve the Bose-Einstein statistics and simplify themselves to a single equation
\begin{equation}
    \frac{d}{d t}\bar{n}=A \bar{n}-B(\bar{n}+1)
\end{equation}
for the mean number of phonons $\bar{n}$. The dynamics induced by (\ref{Focks:noiseMap}) is adequate even for high Fock states, where the model based on  map (\ref{FockMapN}) does not remain valid anymore for the realistic states of captured ions. Map $\mathcal{M}_{\bar{n}}$, which depends only on a single parameter $\bar{n}$, exploits the related deteriorating process for a pessimistic prediction of the genuine $n$-phonon quantum non-Gaussian depth exhibited by states approaching the high Fock states. 

\begin{figure}
\centering
\includegraphics[width=0.8\linewidth]{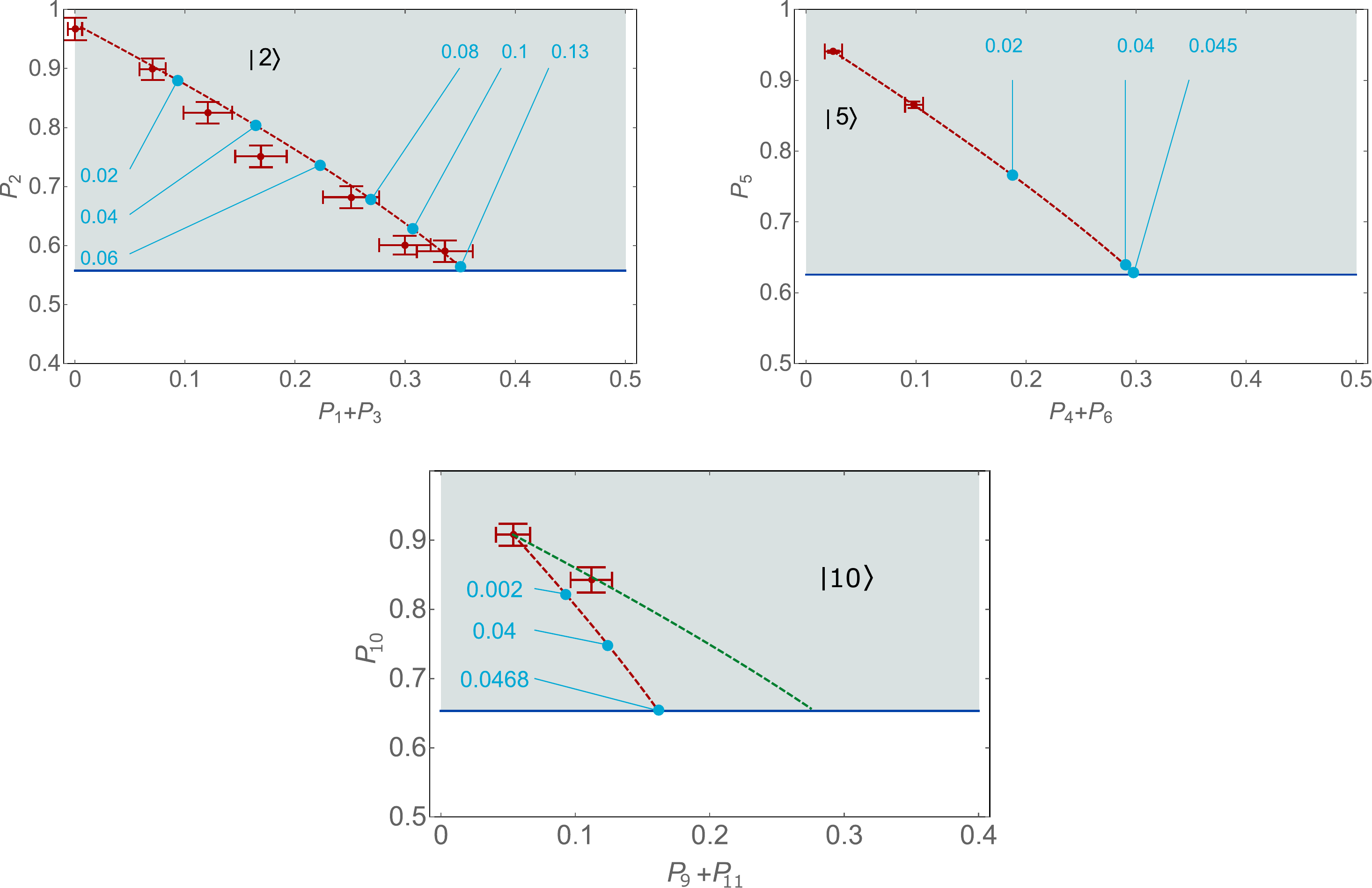}
\caption{Evaluation of the experimental states close to the states $|2\rangle$, $|5\rangle$ and $|10\rangle$ with respect to the thresholds exposing the relevant genuine $n$-phonon quantum non-Gaussianity. Each figure utilizes the tuple of probabilities $(P_{n-1}+P_{n+1},P_n)$ to present experimental results. The horizontal blue lines depict the thresholds probabilities $\bar{P}_n$. The brown points correspond to these probabilities of experimental states that are deteriorated by various effects of noise. The dashed brown lines depict simulation of noise relying on map (\ref{Focks:noiseMap})  that acts on the initial state with greatest probability $P_n$. The blue numbers represent the mean number of noisy phonons induced by this map. The thermal depth of the genuine $n$-phonon quantum non-Gaussianity is given by the greatest blue numbers in each figure. The dashed green line in the figure presenting results of the Fock state $|10\rangle$ shows the impact of the dynamics driven by rate equation (\ref{Focks:ModelEq}), which fits the data better.}
\label{fig:Fock:expRes}
\end{figure}

Experimental implementation exploited the axial motional mode of a single $^{40}$Ca$^+$ ion captured in a linear Paul trap. The states were prepared by applying a sequence of $n$ blue and red sideband $\pi$-pulses causing the transition between the electronic ground state $|g\rangle  \sim 4 S_{1/2}(m=-1/2)$ and excited state $|e \rangle \sim 3 S_{5/2}(m=-1/2)$ that was accompanied by adding a phonon to the motional state after each action of the blue and red sideband $\pi$-pulse on the ion \cite{Leibfried1996,Roos1999,McCormick2019}. The reading of the motional states was performed by precise measurement of the Rabi oscillation of the first blue motional sideband \cite{Leibfried2003}. The experimental results of states close to Fock states $|2\rangle$, $|5\rangle$ and $|10\rangle$ are depicted in Fig.~\ref{fig:Fock:expRes}. These states exhibit their respective genuine $n$-phonon quantum non-Gaussianity. For analysis of the thermal QNG depth, these states were deteriorated by a heating mechanism, which increased the noise contributions. The probabilities of these noisy realistic states were compared with the introduced models. This confirmed that the map (\ref{Focks:noiseMap}) is appropriate only for low Fock states.

\subsection{Force estimation capability}

\begin{figure}
\centering
\includegraphics[width=0.5\linewidth]{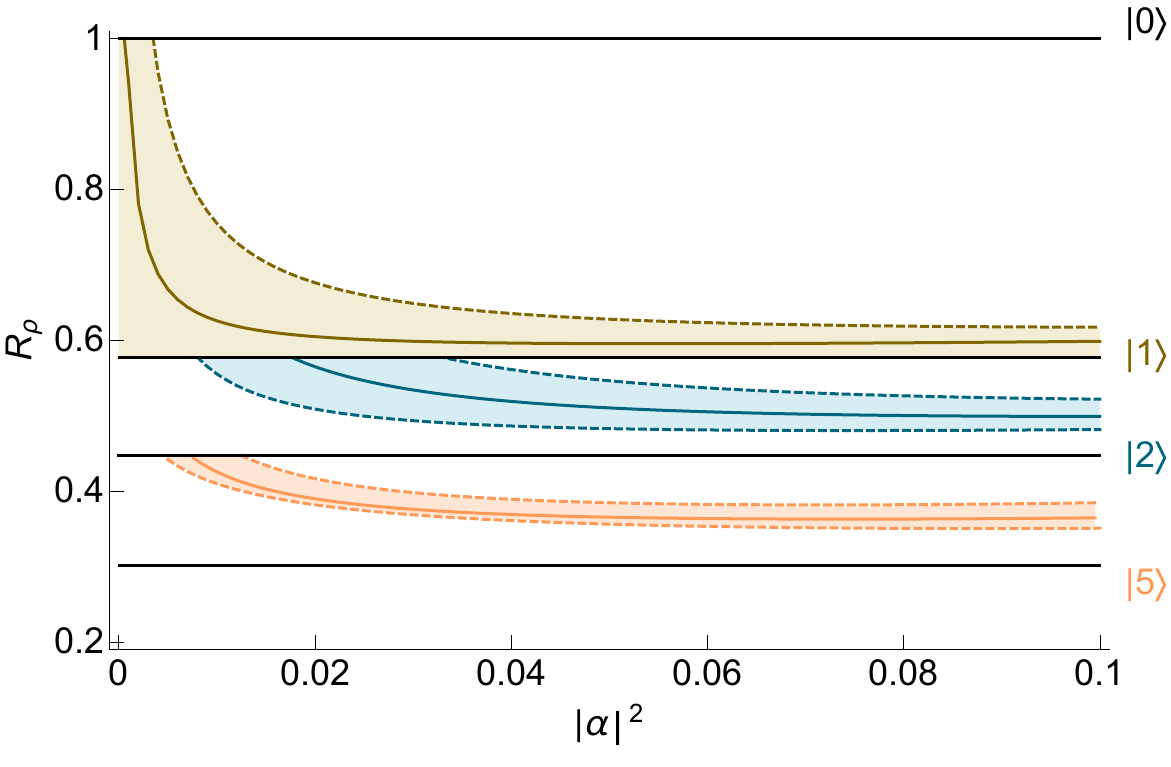}
\caption{Figure presents the metrological advantage $R_{\rho}$ of experimentally achieved states. Whereas the black horizontal lines correspond to the metrological advantage exhibited by the Fock states, the yellow, blue and orange solid lines depict $R_{\rho}$ of the realized states $\rho$ approaching the Fock states $|1\rangle$, $|2\rangle$ and $|5\rangle$, respectively. The colored regions bounded by the dashed lines represent $R_{\rho}$ within the experimental error bars of the measured phonon number distribution.}
\label{fig:Fock:sensing}
\end{figure}

Real states close to the Fock states can be directly employed for enhanced phase independent sensing of a weak force that exerts a tiny displacement on the trapped ion motion. This applies directly to precise measurements of a small, radio-frequency noise or quantum logic spectroscopy \cite{Wan2014}. Let $D(\alpha)$ denote the displacement operator corresponding to the action of the force on an initial state $\rho$. The Fisher information achieved in the estimation of the intensity $|\alpha|^2$ of the displacement is defined as
\begin{equation}
    F=\sum_{n=0}^{\infty}\frac{1}{P_n\left(|\alpha|^2\right)}\left[\frac{d P_n\left(|\alpha|^2\right)}{d|\alpha|^2}\right]^2,
\end{equation}
where $P_n\left(|\alpha|^2\right)=\langle n|D(|\alpha|)\rho D^{\dagger}(|\alpha|)|n \rangle$ is the underlying phonon distribution of a motional state affected by the displacement. The Cram\' er-Rao bound imposes a lower limit on the standard deviation $\sigma$ of the estimation according to $\sigma^2 \geq 1/(N F)$ with $N$ being a number of samples in the sensing. Let $R_{\rho}$ denotes the ratio $R_{\rho}=\sigma/\sigma_0$, where $\sigma_0$ stands for the standard deviation in the estimation when the displacement affects the  ground state. Thus, $R_{\rho}<1$ quantifies the metrological advantage of the state $\rho$ against the ground state. The ideal Fock states achieve
\begin{equation}
    R_{|n\rangle}=\frac{1}{\sqrt{2n+1}},
\end{equation}
which is independent of $|\alpha|^2$ and drops with $n$. It suggests a higher Fock state enhances the advantage. However, this advantage can be lost due to noise. Thus, we can evaluate the quality of a state $\rho$ approaching $|n\rangle$ according to its capability to provide an advantage against a lower but ideal Fock state in the sensing, i.e. $R_{\rho}<R_{|n\rangle}$. The capability of the experimental state at Fig.~\ref{fig:Fock1} can be compared with the genuine $n$-phonon quantum non-Gaussianity on the model of thermalized Fock states. It suggests that the condition for supremacy of higher Fock states in quantum sensing imposes a more challenging threshold on the statistics than the condition of the genuine $n$-phonon quantum non-Gaussianity. Fig.~\ref{fig:Fock:sensing} depicts the metrological advantage $R_{\rho}$ of the experimentally yielded states for the trapped ions experiment, and it demonstrates enhanced capability in sensing of the realized state approaching the Fock state $|5\rangle$ compared to the Fock state $|2\rangle$.

In summary, the platforms based on a cold ion trapped in Paul trap allow us to prepare motional high Fock states and explore their quantum non-Gaussian properties. The certification of genuine $n$-phonon quantum non-Gaussianity can rely on the strictest criterion. An even more demanding hierarchy of conditions, which are still experimentally feasible, is related to the metrological advantage against a given Fock state. Thus, these QNG conditions are necessary for the quality of realistic phononic states before the sensing applications. 

\section{Conclusion}

To conclude, nonclassicality and quantum non-Gaussianity were reviewed, focusing on the states of the light and motion of atoms. Loss-tolerant criteria revealing these two properties were derived for a direct detection technique, using the methodology introduced in \cite{Filip2011}. Firstly, a criterion revealing the nonclassicality in the HBT layout was introduced. The criterion was applied to a relevant model allowing for a cluster of single-photon emitters radiating light under background noise with the Poissonian statistics. The criterion recognizes the nonclassicality of the state regardless of the number of single-photon emitters or the amount of the Poissonian background noise deteriorating the source. The only limiting factor is the experimental time needed for sufficient suppression of error bars, which grows with decreasing the collection and detection efficiency. Thus, the nonclassicality represents a feasible test for a broad group of experimental platforms.

Formally, the nonclassical criterion represents a condition imposed on measurable quantities. Two such quantities faithfully obtainable from the HBT measurement setup were analysed to characterise two aspects of a source radiating the nonclassical light. Specifically, we used the $\alpha$ parameter derived ab initio, which converges to the second-order correlation function for weak light emission. In this approximation, the $\alpha$ parameter is independent of the optical losses. On the contrary, a growing number of single-photon emitters radiating the nonclassical light and increasing contributions of the Poissonian background noise increase the $\alpha$ parameter. Therefore, both changes influence the $\alpha$ parameter similarly. Consequently, we introduced another parameter $\beta$ independent of the Poissonian background noise. Although it increases with the optical losses, it drops with the number of single-photon emitters, deteriorated with the background Poissonian noise. Due to the different behaving of those two parameters, their combination can be exploited for advanced evaluation of the nonclassical light. The nonclassicality of light radiated from a source of 275 ions, which were captured in the Paul trap, was demonstrated experimentally.

Further, the quantum non-Gaussianity of the multiphoton light was explored on a network splitting light among many spatial modes. A sequence of conditions recognizing this quantum property was derived. They were exploited for revealing the quantum non-Gaussianity of the light emitted from an ensemble of single-photon emitters and affected by the Poissonian background noise. It was proved that any number of ideal single-photon emitters that the noise does not deteriorate exhibit the quantum non-Gaussianity if they are split among more single-photon detectors than is the number of the emitters. When a number of the detectors is lower, the criteria impose a condition on the efficiency of the emission from individual emitters. An imperfect source deteriorated with the background noise exhibits the quantum non-Gaussianity only when the noise is suppressed below a threshold determined from the criteria. Another imperfection affecting the realistic sources are losses. The quantum non-Gaussianity appeared as a property that is reasonably sensitive to the losses without a fundamental limit, which the negativity of Wigner function has. The robustness of the quantum non-Gaussianity against the losses is getting lower when the contributions of the background noise are growing. All these conclusions indicate, the quantum non-Gaussianity is a more stimulating quantum property for state preparation and detection than the nonclassicality measured in the HBT layout. It represents an appropriate test when observation of the nonclassicality is too easy but the negativity of the Wigner function cannot be achieved due to the losses. The feasibility of the quantum non-Gaussianity for realistic states was verified by measuring the property on a state produced by multiplexing up to nine heralded single-photon states.

The quantum non-Gaussianity was utilized for discrimination of quantum features that only some Fock states exhibit. These features establish an ordered hierarchy called genuine $n$-photon quantum non-Gaussianity, where $n$ denotes the lowest Fock state that possesses the quantum property. Criteria recognizing the genuine $n$-photon quantum non-Gaussianity were derived and achieved experimentally up to order three corresponding to three photons. The photon statistics of the Fock states was simulated by multiplexing heralded states radiated from the spontaneous parametric down-conversion process. The limit was therefore not probability of success but truly statistical features of generated states. We predicted theoretically and demonstrated experimentally that this challenging quantum features of multiphoton light tolerate only small but non-zero noise contribution and realistic optical losses. 

In summary, the Review introduces the criteria of the nonclassicality and the quantum non-Gaussianity. These criteria are analysed with respect to currently developing sources of quantum light. These two properties appeared as useful for a diagnosis and comparison of the quantum states of the light and, simultaneously, they stimulate the current progress in the quantum technologies. Experiments supported many of the theoretical proposals and proved the feasibility of these analyses. Currently, the hot subject is an investigation of interference effects occurring on a detector and their impacts on the nonclassical aspects. Such effects were explored concerning the effect on the correlation functions \cite{Bhatti2015} without a deeper discussion about the nonclassical behaviour. Simultaneously, criteria of nonclassicality involving triplets of measured probabilities are surveyed. They extend criteria derived in Ref. \cite{Klyshko1996}. Further, the quantum non-Gaussianity of states occupying two distinguishable modes is being explored. The aim is to recognize quantum non-Gaussian correlation between these two modes, which are responsible for photon coincidences. Finally, the genuine $n$-photon quantum non-Gaussianity has been experimentally recognized in the motional states of trapped ions. We are currently investigating different criteria that are appropriate for this platform.

Future next targets follow the path of exploring both the nonclassicality and the quantum non-Gaussianity. Firstly, extended nonclassical criteria that include more error events are going to be investigated. Since these criteria incorporate more parameters, they will be able to recognize the nonclassicality for a broader set of states. 
Another ongoing research of the quantum non-Gaussianity aims at recognition of this quantum feature on light emitted from the cavity where the cavity mediates an interaction between a quantum dot or an atom. Depending on the parameters of the interaction, one can analyse bad cavity regime commonly exploited for efficient generation of the single-photon states \cite{Kuhn2015}, strong coupling regimes leading to non-trivial distribution of photons inside and outside the cavity or other regimes beyond these two scenarios.

This review used a bottom-to-top approach to present quantum non-Gaussian states of photons and phonons. It focused on the essential but instructive and applicable set of less or more available Fock states and the conclusive detection of quantum non-Gaussian aspects despite the optical loss and mechanical noise. The extensions of this approach are diverse. In theory, this approach can be applied to derive, although mostly numerically, conclusive criteria based on prior knowledge of the detection scheme for a diverse set of different quantum non-Gaussian states, including their imperfections. The hierarchy for the Fock states can inspire theorists to find another hierarchical set of quantum non-Gaussian states. The thresholds defined by criteria can be versatile and tailored to be less or more strict using different probabilities than success. The experimental effort can continue in quantum optics and mechanics of atoms and further in quantum optomechanics \cite{Wallucks2020,Enzian2021} and electromechanics \cite{Chu2018,Gely2019} and substantially in quantum superconducting circuits \cite{Axline2018, Hu2019,Magnard2020}.

\section{Corrigendum to “Quantum non-Gaussianity of light and atoms”
[Progress Quant. Electron. 83 (2022) 100395]}
The authors regret missing a reference in this review about the material presented in section 8 related to the preprint L. Podhora, L. Lachman, T. Pham, A. Lešundák, O. Číp, L. Slodička, R. Filip, Quantum non-Gaussianity of multi-phonon states of a single atom, arXiv:2111.10129v1 including the Figs. 19,20 and 21. The entire material has been published in L. Podhora, L. Lachman, T. Pham, A. Lešundák, O. Číp, L. Slodička, and R. Filip, Phys. Rev. Lett. 129, 013602 (2022), https://doi.org/10.1103/PhysRevLett.129.013602. Reprinted figures with permission from this paper. Copyright (2022) by the American Physical Society. The authors would like to apologise for any inconvenience caused.

The authors regret the typo in Eq.~(\ref{splittingFormula}). In its correct form, this equation is given by:
\begin{equation}
P_n=1+\sum_{k=1}^{n}\binom{n}{k}(-1)^k P_0(n/N)
\end{equation}
and the probability $P_n$ refers to the probability that a group of $n$ SPADs registers $n$ clicks irrespective of an output of others $N-n$ SPADs.

The authors also regret incorrect Eq.~(\ref{theorem2}) in section 8 and the misleading description that follows. The correct form of this equation is
\begin{equation}
P_n=\max_{\xi,\alpha,c_1,...,c_{n-1}}|\langle n|D(\alpha)S(\xi)\sum_{k=0}^{n-1}c_k|k\rangle|^2.
\end{equation}
The appropriate text below Eq.~(\ref{theorem2}) follows: "Assuming that the maximum occurs for real amplitudes $c_{n-1}\gg c_k$ and $c_{n-2}\gg c_k$ with $k<n-2$, we allow for a core state $\sum_{k=n-3}^{n-1}c_k|k\rangle$, where $c_{n-3}=\sqrt{1-c_{n-1}^2-c_{n-2}^2}$. For fixed $\alpha$ and $\xi$, we sequentially solve equations given by Eq.~(\ref{MF:eqsArray}). We employ the equation in Eq.~(\ref{MF:eqsArray}), in which $P_n$ is differentiated with respect to $c_{n-1}$, and find a solution for $c_{n-1}$ considering the initial amplitudes $c_{n-1}=1$ and $c_{n-2}=0$. We put this solution into following equation in Eq.~(\ref{MF:eqsArray}) and determine the amplitude $c_{n-2}$. We repeat the procedure taking the derived amplitudes $c_{n-1}$ and $c_{n-2}$ as new initial amplitudes. This numerical approach converges quickly to a solution. The maximal probability $P_n$ was indepdendently derived in \cite{Chabaud2021}.

\section*{Acknowledgement}
We thank D. Moore for a discussion and careful reading of the manuscript. We acknowledge the project 21-13265X of Czech Science Foundation and EU H2020-WIDESPREAD-2020-5 project NONGAUSS (951737) under the CSA - Coordination and support action.








\bibliographystyle{elsarticle-num}

\bibliography{nonclass}

\begin{thebibliography}{100}
\expandafter\ifx\csname url\endcsname\relax
  \def\url#1{\texttt{#1}}\fi
\expandafter\ifx\csname urlprefix\endcsname\relax\def\urlprefix{URL }\fi
\expandafter\ifx\csname href\endcsname\relax
  \def\href#1#2{#2} \def\path#1{#1}\fi

\bibitem{Loudon2000}
R.~Loudon, The Quantum Theory of Light, Oxford University Press, 2000.

\bibitem{D.F.Walls2008}
G.~F.~M. D.~F.~Walls, Quantum Optics, Berlin: Springer, 1994.

\bibitem{Harvey1984}
J.~E. Harvey, J.~L. Forgham, The spot of arago: New relevance for an old
  phenomenon, American Journal of Physics 52~(3) (1984) 243--247.
\newblock \href {https://doi.org/10.1119/1.13681} {\path{doi:10.1119/1.13681}}.

\bibitem{Born1999}
M.~Born, E.~Wolf, A.~B. Bhatia, P.~C. Clemmow, D.~Gabor, A.~R. Stokes, A.~M.
  Taylor, P.~A. Wayman, W.~L. Wilcock, Principles of Optics, Cambridge
  University Press, 1999.
\newblock \href {https://doi.org/10.1017/cbo9781139644181}
  {\path{doi:10.1017/cbo9781139644181}}.

\bibitem{Glauber1963b}
R.~J. Glauber, Photon correlations, Physical Review Letters 10~(3) (1963)
  84--86.
\newblock \href {https://doi.org/10.1103/physrevlett.10.84}
  {\path{doi:10.1103/physrevlett.10.84}}.

\bibitem{Sudarshan1963}
E.~C.~G. Sudarshan, Equivalence of semiclassical and quantum mechanical
  descriptions of statistical light beams, Physical Review Letters 10~(7)
  (1963) 277--279.
\newblock \href {https://doi.org/10.1103/physrevlett.10.277}
  {\path{doi:10.1103/physrevlett.10.277}}.

\bibitem{Glauber1963}
R.~J. Glauber, Coherent and incoherent states of the radiation field, Physical
  Review 131~(6) (1963) 2766--2788.
\newblock \href {https://doi.org/10.1103/physrev.131.2766}
  {\path{doi:10.1103/physrev.131.2766}}.

\bibitem{Kimble1977}
H.~J. Kimble, M.~Dagenais, L.~Mandel, Photon antibunching in resonance
  fluorescence, Physical Review Letters 39~(11) (1977) 691--695.
\newblock \href {https://doi.org/10.1103/physrevlett.39.691}
  {\path{doi:10.1103/physrevlett.39.691}}.

\bibitem{Davidovich1996}
L.~Davidovich, Sub-poissonian processes in quantum optics, Reviews of Modern
  Physics 68~(1) (1996) 127--173.
\newblock \href {https://doi.org/10.1103/revmodphys.68.127}
  {\path{doi:10.1103/revmodphys.68.127}}.

\bibitem{Slusher1985}
R.~E. Slusher, L.~W. Hollberg, B.~Yurke, J.~C. Mertz, J.~F. Valley, Observation
  of squeezed states generated by four-wave mixing in an optical cavity,
  Physical Review Letters 55~(22) (1985) 2409--2412.
\newblock \href {https://doi.org/10.1103/physrevlett.55.2409}
  {\path{doi:10.1103/physrevlett.55.2409}}.

\bibitem{Andersen2016}
U.~L. Andersen, T.~Gehring, C.~Marquardt, G.~Leuchs, 30 years of squeezed light
  generation, Physica Scripta 91~(5) (2016) 053001.
\newblock \href {https://doi.org/10.1088/0031-8949/91/5/053001}
  {\path{doi:10.1088/0031-8949/91/5/053001}}.

\bibitem{Bennett2014}
C.~H. Bennett, G.~Brassard, Quantum cryptography: Public key distribution and
  coin tossing, Theoretical Computer Science 560 (2014) 7--11.
\newblock \href {https://doi.org/10.1016/j.tcs.2014.05.025}
  {\path{doi:10.1016/j.tcs.2014.05.025}}.

\bibitem{Boto2000}
A.~N. Boto, P.~Kok, D.~S. Abrams, S.~L. Braunstein, C.~P. Williams, J.~P.
  Dowling, Quantum interferometric optical lithography: Exploiting entanglement
  to beat the diffraction limit, Physical Review Letters 85~(13) (2000)
  2733--2736.
\newblock \href {https://doi.org/10.1103/physrevlett.85.2733}
  {\path{doi:10.1103/physrevlett.85.2733}}.

\bibitem{Knill2001}
E.~Knill, R.~Laflamme, G.~J. Milburn, A scheme for efficient quantum
  computation with linear optics, Nature 409~(6816) (2001) 46--52.
\newblock \href {https://doi.org/10.1038/35051009}
  {\path{doi:10.1038/35051009}}.

\bibitem{Mandel1986}
L.~Mandel, Non-classical states of the electromagnetic field, Physica Scripta
  T12 (1986) 34--42.
\newblock \href {https://doi.org/10.1088/0031-8949/1986/t12/005}
  {\path{doi:10.1088/0031-8949/1986/t12/005}}.

\bibitem{Filip2011}
R.~Filip, L.~Mi{\v{s}}ta, Detecting quantum states with a positive {W}igner
  function beyond mixtures of {G}aussian states, Physical Review Letters
  106~(20) (may 2011).
\newblock \href {https://doi.org/10.1103/physrevlett.106.200401}
  {\path{doi:10.1103/physrevlett.106.200401}}.

\bibitem{Lasota2017a}
M.~Lasota, R.~Filip, V.~C. Usenko, Sufficiency of quantum non-{G}aussianity for
  discrete-variable quantum key distribution over noisy channels, Physical
  Review A 96~(1) (jul 2017).
\newblock \href {https://doi.org/10.1103/physreva.96.012301}
  {\path{doi:10.1103/physreva.96.012301}}.

\bibitem{Rakhubovsky2017}
A.~A. Rakhubovsky, R.~Filip, Photon-phonon-photon transfer in optomechanics,
  Scientific Reports 7~(1) (apr 2017).
\newblock \href {https://doi.org/10.1038/srep46764}
  {\path{doi:10.1038/srep46764}}.

\bibitem{Yuen1976}
H.~P. Yuen, Two-photon coherent states of the radiation field, Physical Review
  A 13~(6) (1976) 2226--2243.
\newblock \href {https://doi.org/10.1103/physreva.13.2226}
  {\path{doi:10.1103/physreva.13.2226}}.

\bibitem{Boyd2008}
R.~W. Boyd, Nonlinear Optics, Third Edition, 3rd Edition, Academic Press, Inc.,
  USA, 2008.

\bibitem{Friberg1985}
S.~Friberg, C.~K. Hong, L.~Mandel, Measurement of time delays in the parametric
  production of photon pairs, Physical Review Letters 54~(18) (1985)
  2011--2013.
\newblock \href {https://doi.org/10.1103/physrevlett.54.2011}
  {\path{doi:10.1103/physrevlett.54.2011}}.

\bibitem{JohnDepartmentofPhysics2008}
R.~Chiao, J.~Garrison, Quantum Optics, Oxford University Press, 2008.

\bibitem{Parigi2007}
V.~Parigi, A.~Zavatta, M.~Kim, M.~Bellini, Probing quantum commutation rules by
  addition and subtraction of single photons to/from a light field, Science
  317~(5846) (2007) 1890--1893.
\newblock \href {https://doi.org/10.1126/science.1146204}
  {\path{doi:10.1126/science.1146204}}.

\bibitem{Wigner1932}
E.~Wigner, On the quantum correction for thermodynamic equilibrium, Physical
  Review 40~(5) (1932) 749--759.
\newblock \href {https://doi.org/10.1103/physrev.40.749}
  {\path{doi:10.1103/physrev.40.749}}.

\bibitem{Kenfack2004}
A.~Kenfack, K.~yczkowski, Negativity of the {W}igner function as an indicator
  of non-classicality, Journal of Optics B: Quantum and Semiclassical Optics
  6~(10) (2004) 396--404.
\newblock \href {https://doi.org/10.1088/1464-4266/6/10/003}
  {\path{doi:10.1088/1464-4266/6/10/003}}.

\bibitem{Lvovsky2009}
A.~I. Lvovsky, M.~G. Raymer, Continuous-variable optical quantum-state
  tomography, Reviews of Modern Physics 81~(1) (2009) 299--332.
\newblock \href {https://doi.org/10.1103/revmodphys.81.299}
  {\path{doi:10.1103/revmodphys.81.299}}.

\bibitem{Royer1977}
A.~Royer, Wigner function as the expectation value of a parity operator,
  Physical Review A 15~(2) (1977) 449--450.
\newblock \href {https://doi.org/10.1103/physreva.15.449}
  {\path{doi:10.1103/physreva.15.449}}.

\bibitem{BROWN1956}
R.~H. Brown, R.~Q. Twiss, A test of a new type of stellar interferometer on
  sirius, Nature 178~(4541) (1956) 1046--1048.
\newblock \href {https://doi.org/10.1038/1781046a0}
  {\path{doi:10.1038/1781046a0}}.

\bibitem{Glauber1963a}
R.~J. Glauber, The quantum theory of optical coherence, Physical Review 130~(6)
  (1963) 2529--2539.
\newblock \href {https://doi.org/10.1103/physrev.130.2529}
  {\path{doi:10.1103/physrev.130.2529}}.

\bibitem{Kiesel2008}
T.~Kiesel, W.~Vogel, V.~Parigi, A.~Zavatta, M.~Bellini, Experimental
  determination of a nonclassical {G}lauber-{S}udarshan {P} function, Physical
  Review A 78~(2) (aug 2008).
\newblock \href {https://doi.org/10.1103/physreva.78.021804}
  {\path{doi:10.1103/physreva.78.021804}}.

\bibitem{Volovich2016}
I.~V. Volovich, Cauchy{\textendash}{S}chwarz inequality-based criteria for the
  non-classicality of sub-{P}oisson and antibunched light, Physics Letters A
  380~(1-2) (2016) 56--58.
\newblock \href {https://doi.org/10.1016/j.physleta.2015.09.011}
  {\path{doi:10.1016/j.physleta.2015.09.011}}.

\bibitem{Mandel1979}
L.~Mandel, Sub-{P}oissonian photon statistics in resonance fluorescence, Optics
  Letters 4~(7) (1979) 205.
\newblock \href {https://doi.org/10.1364/ol.4.000205}
  {\path{doi:10.1364/ol.4.000205}}.

\bibitem{Skornia2001}
C.~Skornia, J.~von Zanthier, G.~S. Agarwal, E.~Werner, H.~Walther, Nonclassical
  interference effects in the radiation from coherently driven uncorrelated
  atoms, Physical Review A 64~(6) (nov 2001).
\newblock \href {https://doi.org/10.1103/physreva.64.063801}
  {\path{doi:10.1103/physreva.64.063801}}.

\bibitem{Shchukin2005}
E.~V. Shchukin, W.~Vogel, Nonclassical moments and their measurement, Physical
  Review A 72~(4) (oct 2005).
\newblock \href {https://doi.org/10.1103/physreva.72.043808}
  {\path{doi:10.1103/physreva.72.043808}}.

\bibitem{Eisaman2011}
M.~D. Eisaman, J.~Fan, A.~Migdall, S.~V. Polyakov, Invited review article:
  Single-photon sources and detectors, Review of Scientific Instruments 82~(7)
  (2011) 071101.
\newblock \href {https://doi.org/10.1063/1.3610677}
  {\path{doi:10.1063/1.3610677}}.

\bibitem{Achilles2003}
D.~Achilles, C.~Silberhorn, C.~{\'{S}}liwa, K.~Banaszek, I.~A. Walmsley,
  Fiber-assisted detection with photon number resolution, Optics Letters
  28~(23) (2003) 2387.
\newblock \href {https://doi.org/10.1364/ol.28.002387}
  {\path{doi:10.1364/ol.28.002387}}.

\bibitem{Hoepker2019}
J.~P. Höpker, T.~Gerrits, A.~Lita, S.~Krapick, H.~Herrmann, R.~Ricken,
  V.~Quiring, R.~Mirin, S.~W. Nam, C.~Silberhorn, T.~J. Bartley, Integrated
  transition edge sensors on titanium in-diffused lithium niobate waveguides,
  {APL} Photonics 4~(5) (2019) 056103.
\newblock \href {https://doi.org/10.1063/1.5086276}
  {\path{doi:10.1063/1.5086276}}.

\bibitem{Sperling2012}
J.~Sperling, W.~Vogel, G.~S. Agarwal, True photocounting statistics of multiple
  on-off detectors, Physical Review A 85~(2) (feb 2012).
\newblock \href {https://doi.org/10.1103/physreva.85.023820}
  {\path{doi:10.1103/physreva.85.023820}}.

\bibitem{Short1983}
R.~Short, L.~Mandel, Observation of sub-{P}oissonian photon statistics,
  Physical Review Letters 51~(5) (1983) 384--387.
\newblock \href {https://doi.org/10.1103/physrevlett.51.384}
  {\path{doi:10.1103/physrevlett.51.384}}.

\bibitem{Grangier1986}
P.~Grangier, G.~Roger, A.~Aspect, Experimental evidence for a photon
  anticorrelation effect on a beam splitter: A new light on single-photon
  interferences, Europhysics Letters ({EPL}) 1~(4) (1986) 173--179.
\newblock \href {https://doi.org/10.1209/0295-5075/1/4/004}
  {\path{doi:10.1209/0295-5075/1/4/004}}.

\bibitem{Stoler1970}
D.~Stoler, Equivalence classes of minimum uncertainty packets, Physical Review
  D 1~(12) (1970) 3217--3219.
\newblock \href {https://doi.org/10.1103/physrevd.1.3217}
  {\path{doi:10.1103/physrevd.1.3217}}.

\bibitem{Lemonde2014}
M.-A. Lemonde, N.~Didier, A.~A. Clerk, Antibunching and unconventional photon
  blockade with {G}aussian squeezed states, Physical Review A 90~(6) (dec
  2014).
\newblock \href {https://doi.org/10.1103/physreva.90.063824}
  {\path{doi:10.1103/physreva.90.063824}}.

\bibitem{Kral1990}
P.~Kr{\'{a}}l, Displaced and squeezed {F}ock states, Journal of Modern Optics
  37~(5) (1990) 889--917.
\newblock \href {https://doi.org/10.1080/09500349014550941}
  {\path{doi:10.1080/09500349014550941}}.

\bibitem{Scully1997}
M.~O. Scully, M.~S. Zubairy, Quantum optics, Cambridge University Press, 1997.

\bibitem{Hlousek2019}
J.~Hlou{\v{s}}ek, M.~Dudka, I.~Straka, M.~Je{\v{z}}ek, Accurate detection of
  arbitrary photon statistics, Physical Review Letters 123~(15) (oct 2019).
\newblock \href {https://doi.org/10.1103/physrevlett.123.153604}
  {\path{doi:10.1103/physrevlett.123.153604}}.

\bibitem{Nehra2019}
R.~Nehra, K.~V. Jacob, Characterizing quantum detectors by {W}igner functions,
  arXiv:1909.10628\href
  {http://arxiv.org/abs/http://arxiv.org/abs/1909.10628v1}
  {\path{arXiv:http://arxiv.org/abs/1909.10628v1}}.

\bibitem{Mosley2008}
P.~J. Mosley, J.~S. Lundeen, B.~J. Smith, P.~Wasylczyk, A.~B. U'Ren,
  C.~Silberhorn, I.~A. Walmsley, Heralded generation of ultrafast single
  photons in pure quantum states, Physical Review Letters 100~(13) (apr 2008).
\newblock \href {https://doi.org/10.1103/physrevlett.100.133601}
  {\path{doi:10.1103/physrevlett.100.133601}}.

\bibitem{Fulconis2007}
J.~Fulconis, O.~Alibart, J.~L. O'Brien, W.~J. Wadsworth, J.~G. Rarity,
  Nonclassical interference and entanglement generation using a photonic
  crystal fiber pair photon source, Physical Review Letters 99~(12) (sep 2007).
\newblock \href {https://doi.org/10.1103/physrevlett.99.120501}
  {\path{doi:10.1103/physrevlett.99.120501}}.

\bibitem{Mika2018}
J.~Mika, L.~Podhora, L.~Lachman, P.~Ob{\v{s}}il, J.~Hlou{\v{s}}ek,
  M.~Je{\v{z}}ek, R.~Filip, L.~Slodi{\v{c}}ka, Generation of ideal thermal
  light in warm atomic vapor, New Journal of Physics 20~(9) (2018) 093002.
\newblock \href {https://doi.org/10.1088/1367-2630/aadc9d}
  {\path{doi:10.1088/1367-2630/aadc9d}}.

\bibitem{Aharonovich2016}
I.~Aharonovich, D.~Englund, M.~Toth, Solid-state single-photon emitters, Nature
  Photonics 10~(10) (2016) 631--641.
\newblock \href {https://doi.org/10.1038/nphoton.2016.186}
  {\path{doi:10.1038/nphoton.2016.186}}.

\bibitem{Qi2018}
L.~Qi, M.~Manceau, A.~Cavanna, F.~Gumpert, L.~Carbone, M.~de~Vittorio,
  A.~Bramati, E.~Giacobino, L.~Lachman, R.~Filip, M.~Chekhova, Multiphoton
  nonclassical light from clusters of single-photon emitters, New Journal of
  Physics 20~(7) (2018) 073013.
\newblock \href {https://doi.org/10.1088/1367-2630/aacf21}
  {\path{doi:10.1088/1367-2630/aacf21}}.

\bibitem{Sperling2012a}
J.~Sperling, W.~Vogel, G.~S. Agarwal, Sub-binomial light, Physical Review
  Letters 109~(9) (aug 2012).
\newblock \href {https://doi.org/10.1103/physrevlett.109.093601}
  {\path{doi:10.1103/physrevlett.109.093601}}.

\bibitem{Filip2013}
R.~Filip, L.~Lachman, Hierarchy of feasible nonclassicality criteria for
  sources of photons, Physical Review A 88~(4) (oct 2013).
\newblock \href {https://doi.org/10.1103/physreva.88.043827}
  {\path{doi:10.1103/physreva.88.043827}}.

\bibitem{Klyshko1996}
D.~Klyshko, The nonclassical light, Uspekhi Fizicheskih Nauk 166~(6) (1996)
  613--638.
\newblock \href {https://doi.org/10.3367/ufnr.0166.199606b.0613}
  {\path{doi:10.3367/ufnr.0166.199606b.0613}}.

\bibitem{Marek2016}
P.~Marek, L.~Lachman, L.~Slodi{\v{c}}ka, R.~Filip, Deterministic
  nonclassicality for quantum-mechanical oscillators in thermal states,
  Physical Review A 94~(1) (jul 2016).
\newblock \href {https://doi.org/10.1103/physreva.94.013850}
  {\path{doi:10.1103/physreva.94.013850}}.

\bibitem{Vogel2000}
W.~Vogel, Nonclassical states: An observable criterion, Physical Review Letters
  84~(9) (2000) 1849--1852.
\newblock \href {https://doi.org/10.1103/physrevlett.84.1849}
  {\path{doi:10.1103/physrevlett.84.1849}}.

\bibitem{Richter2002}
T.~Richter, W.~Vogel, Nonclassicality of quantum states: A hierarchy of
  observable conditions, Physical Review Letters 89~(28) (dec 2002).
\newblock \href {https://doi.org/10.1103/physrevlett.89.283601}
  {\path{doi:10.1103/physrevlett.89.283601}}.

\bibitem{Somaschi2016}
N.~Somaschi, V.~Giesz, L.~D. Santis, J.~C. Loredo, M.~P. Almeida, G.~Hornecker,
  S.~L. Portalupi, T.~Grange, C.~Ant{\'{o}}n, J.~Demory, C.~G{\'{o}}mez,
  I.~Sagnes, N.~D. Lanzillotti-Kimura, A.~Lema{\'{i}}tre, A.~Auffeves, A.~G.
  White, L.~Lanco, P.~Senellart, Near-optimal single-photon sources in the
  solid state, Nature Photonics 10~(5) (2016) 340--345.
\newblock \href {https://doi.org/10.1038/nphoton.2016.23}
  {\path{doi:10.1038/nphoton.2016.23}}.

\bibitem{Moreva2017}
E.~Moreva, P.~Traina, J.~Forneris, I.~P. Degiovanni, S.~D. Tchernij,
  F.~Picollo, G.~Brida, P.~Olivero, M.~Genovese, Direct experimental
  observation of nonclassicality in ensembles of single-photon emitters,
  Physical Review B 96~(19) (nov 2017).
\newblock \href {https://doi.org/10.1103/physrevb.96.195209}
  {\path{doi:10.1103/physrevb.96.195209}}.

\bibitem{Obsil2018}
P.~Ob{\v{s}}il, L.~Lachman, T.~Pham, A.~Le{\v{s}}und{\'{a}}k, V.~Hucl,
  M.~{\v{C}}{\'{i}}{\v{z}}ek, J.~Hrabina, O.~{\v{C}}{\'{i}}p,
  L.~Slodi{\v{c}}ka, R.~Filip, Nonclassical light from large ensembles of
  trapped ions, Physical Review Letters 120~(25) (jun 2018).
\newblock \href {https://doi.org/10.1103/physrevlett.120.253602}
  {\path{doi:10.1103/physrevlett.120.253602}}.

\bibitem{Hasse1991}
R.~W. Hasse, V.~V. Avilov, Structure and {M}adelung energy of spherical coulomb
  crystals, Physical Review A 44~(7) (1991) 4506--4515.
\newblock \href {https://doi.org/10.1103/physreva.44.4506}
  {\path{doi:10.1103/physreva.44.4506}}.

\bibitem{Hudson1974}
R.~Hudson, When is the {W}igner quasi-probability density non-negative?,
  Reports on Mathematical Physics 6~(2) (1974) 249--252.
\newblock \href {https://doi.org/10.1016/0034-4877(74)90007-x}
  {\path{doi:10.1016/0034-4877(74)90007-x}}.

\bibitem{Genoni2007}
M.~G. Genoni, M.~G.~A. Paris, K.~Banaszek, Measure of the non-{G}aussian
  character of a quantum state, Physical Review A 76~(4) (oct 2007).
\newblock \href {https://doi.org/10.1103/physreva.76.042327}
  {\path{doi:10.1103/physreva.76.042327}}.

\bibitem{Genoni2010}
M.~G. Genoni, M.~G.~A. Paris, Quantifying non-{G}aussianity for quantum
  information, Physical Review A 82~(5) (nov 2010).
\newblock \href {https://doi.org/10.1103/physreva.82.052341}
  {\path{doi:10.1103/physreva.82.052341}}.

\bibitem{Barbieri2010}
M.~Barbieri, N.~Spagnolo, M.~G. Genoni, F.~Ferreyrol, R.~Blandino, M.~G.~A.
  Paris, P.~Grangier, R.~Tualle-Brouri, Non-{G}aussianity of quantum states: An
  experimental test on single-photon-added coherent states, Physical Review A
  82~(6) (dec 2010).
\newblock \href {https://doi.org/10.1103/physreva.82.063833}
  {\path{doi:10.1103/physreva.82.063833}}.

\bibitem{Genoni2013}
M.~G. Genoni, M.~L. Palma, T.~Tufarelli, S.~Olivares, M.~S. Kim, M.~G.~A.
  Paris, Detecting quantum non-{G}aussianity via the {W}igner function,
  Physical Review A 87~(6) (jun 2013).
\newblock \href {https://doi.org/10.1103/physreva.87.062104}
  {\path{doi:10.1103/physreva.87.062104}}.

\bibitem{Burnham1970}
D.~C. Burnham, D.~L. Weinberg, Observation of simultaneity in parametric
  production of optical photon pairs, Physical Review Letters 25~(2) (1970)
  84--87.
\newblock \href {https://doi.org/10.1103/physrevlett.25.84}
  {\path{doi:10.1103/physrevlett.25.84}}.

\bibitem{Heidmann1987}
A.~Heidmann, R.~J. Horowicz, S.~Reynaud, E.~Giacobino, C.~Fabre, G.~Camy,
  Observation of quantum noise reduction on twin laser beams, Physical Review
  Letters 59~(22) (1987) 2555--2557.
\newblock \href {https://doi.org/10.1103/physrevlett.59.2555}
  {\path{doi:10.1103/physrevlett.59.2555}}.

\bibitem{Hong1987}
C.~K. Hong, Z.~Y. Ou, L.~Mandel, Measurement of subpicosecond time intervals
  between two photons by interference, Physical Review Letters 59~(18) (1987)
  2044--2046.
\newblock \href {https://doi.org/10.1103/physrevlett.59.2044}
  {\path{doi:10.1103/physrevlett.59.2044}}.

\bibitem{Reid1988}
M.~D. Reid, P.~D. Drummond, Quantum correlations of phase in nondegenerate
  parametric oscillation, Physical Review Letters 60~(26) (1988) 2731--2733.
\newblock \href {https://doi.org/10.1103/physrevlett.60.2731}
  {\path{doi:10.1103/physrevlett.60.2731}}.

\bibitem{Mari2012}
A.~Mari, J.~Eisert, Positive {W}igner functions render classical simulation of
  quantum computation efficient, Physical Review Letters 109~(23) (dec 2012).
\newblock \href {https://doi.org/10.1103/physrevlett.109.230503}
  {\path{doi:10.1103/physrevlett.109.230503}}.

\bibitem{Hughes2014}
C.~Hughes, M.~G. Genoni, T.~Tufarelli, M.~G.~A. Paris, M.~S. Kim, Quantum
  non-{G}aussianity witnesses in phase space, Physical Review A 90~(1) (jul
  2014).
\newblock \href {https://doi.org/10.1103/physreva.90.013810}
  {\path{doi:10.1103/physreva.90.013810}}.

\bibitem{Lee1991}
C.~T. Lee, Measure of the nonclassicality of nonclassical states, Physical
  Review A 44~(5) (1991) R2775--R2778.
\newblock \href {https://doi.org/10.1103/physreva.44.r2775}
  {\path{doi:10.1103/physreva.44.r2775}}.

\bibitem{Husimi1940}
K.~Husimi, Some formal properties of the density matrix, Proceedings of the
  Physico-Mathematical Society of Japan. 3rd Series 22~(4) (1940) 264--314.
\newblock \href {https://doi.org/10.11429/ppmsj1919.22.4_264}
  {\path{doi:10.11429/ppmsj1919.22.4_264}}.

\bibitem{Paris1996}
M.~G.~A. Paris, Quantum state measurement by realistic heterodyne detection,
  Physical Review A 53~(4) (1996) 2658--2663.
\newblock \href {https://doi.org/10.1103/physreva.53.2658}
  {\path{doi:10.1103/physreva.53.2658}}.

\bibitem{Park2015}
J.~Park, J.~Zhang, J.~Lee, S.-W. Ji, M.~Um, D.~Lv, K.~Kim, H.~Nha, Testing
  nonclassicality and non-{G}aussianity in phase space, Physical Review Letters
  114~(19) (may 2015).
\newblock \href {https://doi.org/10.1103/physrevlett.114.190402}
  {\path{doi:10.1103/physrevlett.114.190402}}.

\bibitem{Clauser1969}
J.~F. Clauser, M.~A. Horne, A.~Shimony, R.~A. Holt, Proposed experiment to test
  local hidden-variable theories, Physical Review Letters 23~(15) (1969)
  880--884.
\newblock \href {https://doi.org/10.1103/physrevlett.23.880}
  {\path{doi:10.1103/physrevlett.23.880}}.

\bibitem{Happ2018}
L.~Happ, M.~A. Efremov, H.~Nha, W.~P. Schleich, Sufficient condition for a
  quantum state to be genuinely quantum non-{G}aussian, New Journal of Physics
  20~(2) (2018) 023046.
\newblock \href {https://doi.org/10.1088/1367-2630/aaac25}
  {\path{doi:10.1088/1367-2630/aaac25}}.

\bibitem{Lachman2013}
L.~Lachman, R.~Filip, Robustness of quantum nonclassicality and
  non-{G}aussianity of single-photon states in attenuating channels, Physical
  Review A 88~(6) (dec 2013).
\newblock \href {https://doi.org/10.1103/physreva.88.063841}
  {\path{doi:10.1103/physreva.88.063841}}.

\bibitem{Jezek2011}
M.~Je{\v{z}}ek, I.~Straka, M.~Mi{\v{c}}uda, M.~Du{\v{s}}ek,
  J.~Fiur{\'{a}}{\v{s}}ek, R.~Filip, Experimental test of the quantum
  non-{G}aussian character of a heralded single-photon state, Physical Review
  Letters 107~(21) (nov 2011).
\newblock \href {https://doi.org/10.1103/physrevlett.107.213602}
  {\path{doi:10.1103/physrevlett.107.213602}}.

\bibitem{Jezek2012}
M.~Je{\v{z}}ek, A.~Tipsmark, R.~Dong, J.~Fiur{\'{a}}{\v{s}}ek, L.~Mi{\v{s}}ta,
  R.~Filip, U.~L. Andersen, Experimental test of the strongly nonclassical
  character of a noisy squeezed single-photon state, Physical Review A 86~(4)
  (oct 2012).
\newblock \href {https://doi.org/10.1103/physreva.86.043813}
  {\path{doi:10.1103/physreva.86.043813}}.

\bibitem{Predojevic2014}
A.~Predojevi{\'{c}}, M.~Je{\v{z}}ek, T.~Huber, H.~Jayakumar, T.~Kauten, G.~S.
  Solomon, R.~Filip, G.~Weihs, Efficiency vs multi-photon contribution test for
  quantum dots, Optics Express 22~(4) (2014) 4789.
\newblock \href {https://doi.org/10.1364/oe.22.004789}
  {\path{doi:10.1364/oe.22.004789}}.

\bibitem{Higginbottom2016}
D.~B. Higginbottom, L.~Slodi{\v{c}}ka, G.~Araneda, L.~Lachman, R.~Filip,
  M.~Hennrich, R.~Blatt, Pure single photons from a trapped atom source, New
  Journal of Physics 18~(9) (2016) 093038.
\newblock \href {https://doi.org/10.1088/1367-2630/18/9/093038}
  {\path{doi:10.1088/1367-2630/18/9/093038}}.

\bibitem{Straka2014}
I.~Straka, A.~Predojevi{\'{c}}, T.~Huber, L.~Lachman, L.~Butschek,
  M.~Mikov{\'{a}}, M.~Mi{\v{c}}uda, G.~S. Solomon, G.~Weihs, M.~Je{\v{z}}ek,
  R.~Filip, Quantum non-{G}aussian depth of single-photon states, Physical
  Review Letters 113~(22) (nov 2014).
\newblock \href {https://doi.org/10.1103/physrevlett.113.223603}
  {\path{doi:10.1103/physrevlett.113.223603}}.

\bibitem{Lasota2017}
M.~Lasota, R.~Filip, V.~C. Usenko, Robustness of quantum key distribution with
  discrete and continuous variables to channel noise, Physical Review A 95~(6)
  (jun 2017).
\newblock \href {https://doi.org/10.1103/physreva.95.062312}
  {\path{doi:10.1103/physreva.95.062312}}.

\bibitem{ContrerasTejada2019}
P.~Contreras-Tejada, C.~Palazuelos, J.~I. de~Vicente, Resource theory of
  entanglement with a unique multipartite maximally entangled state, Physical
  Review Letters 122~(12) (mar 2019).
\newblock \href {https://doi.org/10.1103/physrevlett.122.120503}
  {\path{doi:10.1103/physrevlett.122.120503}}.

\bibitem{Winter2016}
A.~Winter, D.~Yang, Operational resource theory of coherence, Physical Review
  Letters 116~(12) (mar 2016).
\newblock \href {https://doi.org/10.1103/physrevlett.116.120404}
  {\path{doi:10.1103/physrevlett.116.120404}}.

\bibitem{Brandao2015}
F.~G. Brand{\~{a}}o, G.~Gour, Reversible framework for quantum resource
  theories, Physical Review Letters 115~(7) (aug 2015).
\newblock \href {https://doi.org/10.1103/physrevlett.115.070503}
  {\path{doi:10.1103/physrevlett.115.070503}}.

\bibitem{Coecke2016}
B.~Coecke, T.~Fritz, R.~W. Spekkens, A mathematical theory of resources,
  Information and Computation 250 (2016) 59--86.
\newblock \href {https://doi.org/10.1016/j.ic.2016.02.008}
  {\path{doi:10.1016/j.ic.2016.02.008}}.

\bibitem{Takagi2018}
R.~Takagi, Q.~Zhuang, Convex resource theory of non-gaussianity, Physical
  Review A 97~(6) (jun 2018).
\newblock \href {https://doi.org/10.1103/physreva.97.062337}
  {\path{doi:10.1103/physreva.97.062337}}.

\bibitem{Albarelli2018}
F.~Albarelli, M.~G. Genoni, M.~G.~A. Paris, A.~Ferraro, Resource theory of
  quantum non-gaussianity and wigner negativity, Physical Review A 98~(5) (nov
  2018).
\newblock \href {https://doi.org/10.1103/physreva.98.052350}
  {\path{doi:10.1103/physreva.98.052350}}.

\bibitem{He2013}
Y.-M. He, Y.~He, Y.-J. Wei, D.~Wu, M.~Atatüre, C.~Schneider, S.~Höfling,
  M.~Kamp, C.-Y. Lu, J.-W. Pan, On-demand semiconductor single-photon source
  with near-unity indistinguishability, Nature Nanotechnology 8~(3) (2013)
  213--217.
\newblock \href {https://doi.org/10.1038/nnano.2012.262}
  {\path{doi:10.1038/nnano.2012.262}}.

\bibitem{Ding2016}
X.~Ding, Y.~He, Z.-C. Duan, N.~Gregersen, M.-C. Chen, S.~Unsleber, S.~Maier,
  C.~Schneider, M.~Kamp, S.~Höfling, C.-Y. Lu, J.-W. Pan, On-demand single
  photons with high extraction efficiency and near-unity indistinguishability
  from a resonantly driven quantum dot in a micropillar, Physical Review
  Letters 116~(2) (jan 2016).
\newblock \href {https://doi.org/10.1103/physrevlett.116.020401}
  {\path{doi:10.1103/physrevlett.116.020401}}.

\bibitem{Cooper2013}
M.~Cooper, L.~J. Wright, C.~Söller, B.~J. Smith, Experimental generation of
  multi-photon fock states, Optics Express 21~(5) (2013) 5309.
\newblock \href {https://doi.org/10.1364/oe.21.005309}
  {\path{doi:10.1364/oe.21.005309}}.

\bibitem{Yukawa2013}
M.~Yukawa, K.~Miyata, T.~Mizuta, H.~Yonezawa, P.~Marek, R.~Filip, A.~Furusawa,
  Generating superposition of up-to three photons for continuous variable
  quantum information processing, Optics Express 21~(5) (2013) 5529.
\newblock \href {https://doi.org/10.1364/oe.21.005529}
  {\path{doi:10.1364/oe.21.005529}}.

\bibitem{Laurat2003}
J.~Laurat, T.~Coudreau, N.~Treps, A.~Ma{\^{i}}tre, C.~Fabre, Conditional
  preparation of a quantum state in the continuous variable regime: Generation
  of a sub-{P}oissonian state from twin beams, Physical Review Letters 91~(21)
  (nov 2003).
\newblock \href {https://doi.org/10.1103/physrevlett.91.213601}
  {\path{doi:10.1103/physrevlett.91.213601}}.

\bibitem{Harder2016}
G.~Harder, T.~J. Bartley, A.~E. Lita, S.~W. Nam, T.~Gerrits, C.~Silberhorn,
  Single-mode parametric-down-conversion states with 50 photons as a source for
  mesoscopic quantum optics, Physical Review Letters 116~(14) (apr 2016).
\newblock \href {https://doi.org/10.1103/physrevlett.116.143601}
  {\path{doi:10.1103/physrevlett.116.143601}}.

\bibitem{Iskhakov2016}
T.~S. Iskhakov, V.~C. Usenko, U.~L. Andersen, R.~Filip, M.~V. Chekhova,
  G.~Leuchs, Heralded source of bright multi-mode mesoscopic sub-{P}oissonian
  light, Optics Letters 41~(10) (2016) 2149.
\newblock \href {https://doi.org/10.1364/ol.41.002149}
  {\path{doi:10.1364/ol.41.002149}}.

\bibitem{Straka2018}
I.~Straka, L.~Lachman, J.~Hlou{\v{s}}ek, M.~Mikov{\'{a}}, M.~Mi{\v{c}}uda,
  M.~Je{\v{z}}ek, R.~Filip, Quantum non-{G}aussian multiphoton light, npj
  Quantum Information 4~(1) (jan 2018).
\newblock \href {https://doi.org/10.1038/s41534-017-0054-y}
  {\path{doi:10.1038/s41534-017-0054-y}}.

\bibitem{Lachman2019}
L.~Lachman, I.~Straka, J.~Hlou{\v{s}}ek, M.~Je{\v{z}}ek, R.~Filip, Faithful
  hierarchy of genuine n -photon quantum non-{G}aussian light, Physical Review
  Letters 123~(4) (jul 2019).
\newblock \href {https://doi.org/10.1103/physrevlett.123.043601}
  {\path{doi:10.1103/physrevlett.123.043601}}.

\bibitem{Chu2016}
X.-L. Chu, S.~Götzinger, V.~Sandoghdar, A single molecule as a high-fidelity
  photon gun for producing intensity-squeezed light, Nature Photonics 11~(1)
  (2016) 58--62.
\newblock \href {https://doi.org/10.1038/nphoton.2016.236}
  {\path{doi:10.1038/nphoton.2016.236}}.

\bibitem{Lund-Hansen2008}
T.~Lund-Hansen, S.~Stobbe, B.~Julsgaard, H.~Thyrrestrup, T.~Sünner, M.~Kamp,
  A.~Forchel, P.~Lodahl, Experimental realization of highly efficient broadband
  coupling of single quantum dots to a photonic crystal waveguide, Physical
  Review Letters 101~(11) (sep 2008).
\newblock \href {https://doi.org/10.1103/physrevlett.101.113903}
  {\path{doi:10.1103/physrevlett.101.113903}}.

\bibitem{Hofheinz2008}
M.~Hofheinz, E.~M. Weig, M.~Ansmann, R.~C. Bialczak, E.~Lucero, M.~Neeley,
  A.~D. O'Connell, H.~Wang, J.~M. Martinis, A.~N. Cleland, Generation of fock
  states in a superconducting quantum circuit, Nature 454~(7202) (2008)
  310--314.
\newblock \href {https://doi.org/10.1038/nature07136}
  {\path{doi:10.1038/nature07136}}.

\bibitem{Meekhof1996}
D.~M. Meekhof, C.~Monroe, B.~E. King, W.~M. Itano, D.~J. Wineland, Generation
  of nonclassical motional states of a trapped atom, Physical Review Letters
  76~(11) (1996) 1796--1799.
\newblock \href {https://doi.org/10.1103/physrevlett.76.1796}
  {\path{doi:10.1103/physrevlett.76.1796}}.

\bibitem{Clauser1974}
J.~F. Clauser, Experimental distinction between the quantum and classical
  field-theoretic predictions for the photoelectric effect, Physical Review D
  9~(4) (1974) 853--860.
\newblock \href {https://doi.org/10.1103/physrevd.9.853}
  {\path{doi:10.1103/physrevd.9.853}}.

\bibitem{Kuzmich2003}
A.~Kuzmich, W.~P. Bowen, A.~D. Boozer, A.~Boca, C.~W. Chou, L.-M. Duan, H.~J.
  Kimble, Generation of nonclassical photon pairs for scalable quantum
  communication with atomic ensembles, Nature 423~(6941) (2003) 731--734.
\newblock \href {https://doi.org/10.1038/nature01714}
  {\path{doi:10.1038/nature01714}}.

\bibitem{Kurpiers2018}
P.~Kurpiers, P.~Magnard, T.~Walter, B.~Royer, M.~Pechal, J.~Heinsoo,
  Y.~Salath{\'{e}}, A.~Akin, S.~Storz, J.-C. Besse, S.~Gasparinetti, A.~Blais,
  A.~Wallraff, Deterministic quantum state transfer and remote entanglement
  using microwave photons, Nature 558~(7709) (2018) 264--267.
\newblock \href {https://doi.org/10.1038/s41586-018-0195-y}
  {\path{doi:10.1038/s41586-018-0195-y}}.

\bibitem{Kwiat1999}
P.~G. Kwiat, E.~Waks, A.~G. White, I.~Appelbaum, P.~H. Eberhard, Ultrabright
  source of polarization-entangled photons, Physical Review A 60~(2) (1999)
  R773--R776.
\newblock \href {https://doi.org/10.1103/physreva.60.r773}
  {\path{doi:10.1103/physreva.60.r773}}.

\bibitem{Brendel1999}
J.~Brendel, N.~Gisin, W.~Tittel, H.~Zbinden, Pulsed energy-time entangled
  twin-photon source for quantum communication, Physical Review Letters 82~(12)
  (1999) 2594--2597.
\newblock \href {https://doi.org/10.1103/physrevlett.82.2594}
  {\path{doi:10.1103/physrevlett.82.2594}}.

\bibitem{Chou2005}
C.~W. Chou, H.~de~Riedmatten, D.~Felinto, S.~V. Polyakov, S.~J. van Enk, H.~J.
  Kimble, Measurement-induced entanglement for excitation stored in remote
  atomic ensembles, Nature 438~(7069) (2005) 828--832.
\newblock \href {https://doi.org/10.1038/nature04353}
  {\path{doi:10.1038/nature04353}}.

\bibitem{Ritter2012}
S.~Ritter, C.~Nölleke, C.~Hahn, A.~Reiserer, A.~Neuzner, M.~Uphoff, M.~Mücke,
  E.~Figueroa, J.~Bochmann, G.~Rempe, An elementary quantum network of single
  atoms in optical cavities, Nature 484~(7393) (2012) 195--200.
\newblock \href {https://doi.org/10.1038/nature11023}
  {\path{doi:10.1038/nature11023}}.

\bibitem{Yilmaz2010}
S.~T. Y{\i}lmaz, P.~Fallahi, A.~Imamo{\u{g}}lu, Quantum-dot-spin single-photon
  interface, Physical Review Letters 105~(3) (jul 2010).
\newblock \href {https://doi.org/10.1103/physrevlett.105.033601}
  {\path{doi:10.1103/physrevlett.105.033601}}.

\bibitem{Usmani2012}
I.~Usmani, C.~Clausen, F.~Bussi{\`{e}}res, N.~Sangouard, M.~Afzelius, N.~Gisin,
  Heralded quantum entanglement between two crystals, Nature Photonics 6~(4)
  (2012) 234--237.
\newblock \href {https://doi.org/10.1038/nphoton.2012.34}
  {\path{doi:10.1038/nphoton.2012.34}}.

\bibitem{Bernien2013}
H.~Bernien, B.~Hensen, W.~Pfaff, G.~Koolstra, M.~S. Blok, L.~Robledo, T.~H.
  Taminiau, M.~Markham, D.~J. Twitchen, L.~Childress, R.~Hanson, Heralded
  entanglement between solid-state qubits separated by three metres, Nature
  497~(7447) (2013) 86--90.
\newblock \href {https://doi.org/10.1038/nature12016}
  {\path{doi:10.1038/nature12016}}.

\bibitem{Narla2016}
A.~Narla, S.~Shankar, M.~Hatridge, Z.~Leghtas, K.~Sliwa, E.~Zalys-Geller,
  S.~Mundhada, W.~Pfaff, L.~Frunzio, R.~Schoelkopf, M.~Devoret, Robust
  concurrent remote entanglement between two superconducting qubits, Physical
  Review X 6~(3) (sep 2016).
\newblock \href {https://doi.org/10.1103/physrevx.6.031036}
  {\path{doi:10.1103/physrevx.6.031036}}.

\bibitem{Riedinger2016}
R.~Riedinger, S.~Hong, R.~A. Norte, J.~A. Slater, J.~Shang, A.~G. Krause,
  V.~Anant, M.~Aspelmeyer, S.~Gröblacher, Non-classical correlations between
  single photons and phonons from a mechanical oscillator, Nature 530~(7590)
  (2016) 313--316.
\newblock \href {https://doi.org/10.1038/nature16536}
  {\path{doi:10.1038/nature16536}}.

\bibitem{Riedinger2018}
R.~Riedinger, A.~Wallucks, I.~Marinkovi{\'{c}}, C.~Löschnauer, M.~Aspelmeyer,
  S.~Hong, S.~Gröblacher, Remote quantum entanglement between two
  micromechanical oscillators, Nature 556~(7702) (2018) 473--477.
\newblock \href {https://doi.org/10.1038/s41586-018-0036-z}
  {\path{doi:10.1038/s41586-018-0036-z}}.

\bibitem{Braunstein2005}
S.~L. Braunstein, Squeezing as an irreducible resource, Physical Review A
  71~(5) (may 2005).
\newblock \href {https://doi.org/10.1103/physreva.71.055801}
  {\path{doi:10.1103/physreva.71.055801}}.

\bibitem{Makino2016}
K.~Makino, Y.~Hashimoto, J.~ichi Yoshikawa, H.~Ohdan, T.~Toyama, P.~van Loock,
  A.~Furusawa, Synchronization of optical photons for quantum information
  processing, Science Advances 2~(5) (may 2016).
\newblock \href {https://doi.org/10.1126/sciadv.1501772}
  {\path{doi:10.1126/sciadv.1501772}}.

\bibitem{Gao2018}
Y.~Y. Gao, B.~J. Lester, Y.~Zhang, C.~Wang, S.~Rosenblum, L.~Frunzio, L.~Jiang,
  S.~Girvin, R.~J. Schoelkopf, Programmable interference between two microwave
  quantum memories, Physical Review X 8~(2) (jun 2018).
\newblock \href {https://doi.org/10.1103/physrevx.8.021073}
  {\path{doi:10.1103/physrevx.8.021073}}.

\bibitem{Ding2017}
S.~Ding, G.~Maslennikov, R.~Hablützel, H.~Loh, D.~Matsukevich, Quantum
  parametric oscillator with trapped ions, Physical Review Letters 119~(15)
  (oct 2017).
\newblock \href {https://doi.org/10.1103/physrevlett.119.150404}
  {\path{doi:10.1103/physrevlett.119.150404}}.

\bibitem{Schimpf2021}
C.~Schimpf, M.~Reindl, F.~B. Basset, K.~D. Jöns, R.~Trotta, A.~Rastelli,
  Quantum dots as potential sources of strongly entangled photons: Perspectives
  and challenges for applications in quantum networks, Applied Physics Letters
  118~(10) (2021) 100502.
\newblock \href {https://doi.org/10.1063/5.0038729}
  {\path{doi:10.1063/5.0038729}}.

\bibitem{Zapletal2017}
P.~Zapletal, R.~Filip, Multi-copy quantifiers for single-photon states,
  Scientific Reports 7~(1) (may 2017).
\newblock \href {https://doi.org/10.1038/s41598-017-01333-y}
  {\path{doi:10.1038/s41598-017-01333-y}}.

\bibitem{Gessner2019}
M.~Gessner, A.~Smerzi, L.~Pezz{\`{e}}, Metrological nonlinear squeezing
  parameter, Physical Review Letters 122~(9) (mar 2019).
\newblock \href {https://doi.org/10.1103/physrevlett.122.090503}
  {\path{doi:10.1103/physrevlett.122.090503}}.

\bibitem{Wolf2019}
F.~Wolf, C.~Shi, J.~C. Heip, M.~Gessner, L.~Pezz{\`{e}}, A.~Smerzi, M.~Schulte,
  K.~Hammerer, P.~O. Schmidt, Motional fock states for quantum-enhanced
  amplitude and phase measurements with trapped ions, Nature Communications
  10~(1) (jul 2019).
\newblock \href {https://doi.org/10.1038/s41467-019-10576-4}
  {\path{doi:10.1038/s41467-019-10576-4}}.

\bibitem{Niset2009}
J.~Niset, J.~Fiur{\'{a}}{\v{s}}ek, N.~J. Cerf, No-go theorem for gaussian
  quantum error correction, Physical Review Letters 102~(12) (mar 2009).
\newblock \href {https://doi.org/10.1103/physrevlett.102.120501}
  {\path{doi:10.1103/physrevlett.102.120501}}.

\bibitem{Michael2016}
M.~H. Michael, M.~Silveri, R.~Brierley, V.~V. Albert, J.~Salmilehto, L.~Jiang,
  S.~Girvin, New class of quantum error-correcting codes for a bosonic mode,
  Physical Review X 6~(3) (jul 2016).
\newblock \href {https://doi.org/10.1103/physrevx.6.031006}
  {\path{doi:10.1103/physrevx.6.031006}}.

\bibitem{CampagneIbarcq2020}
P.~Campagne-Ibarcq, A.~Eickbusch, S.~Touzard, E.~Zalys-Geller, N.~E. Frattini,
  V.~V. Sivak, P.~Reinhold, S.~Puri, S.~Shankar, R.~J. Schoelkopf, L.~Frunzio,
  M.~Mirrahimi, M.~H. Devoret, Quantum error correction of a qubit encoded in
  grid states of an oscillator, Nature 584~(7821) (2020) 368--372.
\newblock \href {https://doi.org/10.1038/s41586-020-2603-3}
  {\path{doi:10.1038/s41586-020-2603-3}}.

\bibitem{Chabaud2020}
U.~Chabaud, D.~Markham, F.~Grosshans, Stellar representation of non-gaussian
  quantum states, Physical Review Letters 124~(6) (feb 2020).
\newblock \href {https://doi.org/10.1103/physrevlett.124.063605}
  {\path{doi:10.1103/physrevlett.124.063605}}.

\bibitem{Chabaud2021}
U.~Chabaud, G.~Roeland, M.~Walschaers, F.~Grosshans, V.~Parigi, D.~Markham,
  N.~Treps, Certification of non-gaussian states with operational measurements,
  {PRX} Quantum 2~(2) (jun 2021).
\newblock \href {https://doi.org/10.1103/prxquantum.2.020333}
  {\path{doi:10.1103/prxquantum.2.020333}}.

\bibitem{Leibfried2003}
D.~Leibfried, R.~Blatt, C.~Monroe, D.~Wineland, Quantum dynamics of single
  trapped ions, Reviews of Modern Physics 75~(1) (2003) 281--324.
\newblock \href {https://doi.org/10.1103/revmodphys.75.281}
  {\path{doi:10.1103/revmodphys.75.281}}.

\bibitem{Leibfried1996}
D.~Leibfried, D.~M. Meekhof, B.~E. King, C.~Monroe, W.~M. Itano, D.~J.
  Wineland, Experimental determination of the motional quantum state of a
  trapped atom, Physical Review Letters 77~(21) (1996) 4281--4285.
\newblock \href {https://doi.org/10.1103/physrevlett.77.4281}
  {\path{doi:10.1103/physrevlett.77.4281}}.

\bibitem{Roos1999}
C.~Roos, T.~Zeiger, H.~Rohde, H.~C. Nägerl, J.~Eschner, D.~Leibfried,
  F.~Schmidt-Kaler, R.~Blatt, Quantum state engineering on an optical
  transition and decoherence in a paul trap, Physical Review Letters 83~(23)
  (1999) 4713--4716.
\newblock \href {https://doi.org/10.1103/physrevlett.83.4713}
  {\path{doi:10.1103/physrevlett.83.4713}}.

\bibitem{McCormick2019}
K.~C. McCormick, J.~Keller, S.~C. Burd, D.~J. Wineland, A.~C. Wilson,
  D.~Leibfried, Quantum-enhanced sensing of a single-ion mechanical oscillator,
  Nature 572~(7767) (2019) 86--90.
\newblock \href {https://doi.org/10.1038/s41586-019-1421-y}
  {\path{doi:10.1038/s41586-019-1421-y}}.

\bibitem{Wan2014}
Y.~Wan, F.~Gebert, J.~B. Wübbena, N.~Scharnhorst, S.~Amairi, I.~D. Leroux,
  B.~Hemmerling, N.~Lörch, K.~Hammerer, P.~O. Schmidt, Precision spectroscopy
  by photon-recoil signal amplification, Nature Communications 5~(1) (jan
  2014).
\newblock \href {https://doi.org/10.1038/ncomms4096}
  {\path{doi:10.1038/ncomms4096}}.

\bibitem{Bhatti2015}
D.~Bhatti, J.~von Zanthier, G.~S. Agarwal, Superbunching and nonclassicality as
  new hallmarks of superradiance, Scientific Reports 5~(1) (dec 2015).
\newblock \href {https://doi.org/10.1038/srep17335}
  {\path{doi:10.1038/srep17335}}.

\bibitem{Kuhn2015}
A.~Kuhn, Cavity induced interfacing of atoms and light, in: Engineering the
  Atom-Photon Interaction, Springer International Publishing, 2015, pp. 3--38.
\newblock \href {https://doi.org/10.1007/978-3-319-19231-4_1}
  {\path{doi:10.1007/978-3-319-19231-4_1}}.

\bibitem{Wallucks2020}
A.~Wallucks, I.~Marinkovi{\'{c}}, B.~Hensen, R.~Stockill, S.~Gröblacher, A
  quantum memory at telecom wavelengths, Nature Physics 16~(7) (2020) 772--777.
\newblock \href {https://doi.org/10.1038/s41567-020-0891-z}
  {\path{doi:10.1038/s41567-020-0891-z}}.

\bibitem{Enzian2021}
G.~Enzian, J.~Price, L.~Freisem, J.~Nunn, J.~Janousek, B.~Buchler, P.~Lam,
  M.~Vanner, Single-phonon addition and subtraction to a mechanical thermal
  state, Physical Review Letters 126~(3) (jan 2021).
\newblock \href {https://doi.org/10.1103/physrevlett.126.033601}
  {\path{doi:10.1103/physrevlett.126.033601}}.

\bibitem{Chu2018}
Y.~Chu, P.~Kharel, T.~Yoon, L.~Frunzio, P.~T. Rakich, R.~J. Schoelkopf,
  Creation and control of multi-phonon fock states in a bulk acoustic-wave
  resonator, Nature 563~(7733) (2018) 666--670.
\newblock \href {https://doi.org/10.1038/s41586-018-0717-7}
  {\path{doi:10.1038/s41586-018-0717-7}}.

\bibitem{Gely2019}
M.~F. Gely, M.~Kounalakis, C.~Dickel, J.~Dalle, R.~Vatr{\'{e}}, B.~Baker, M.~D.
  Jenkins, G.~A. Steele, Observation and stabilization of photonic fock states
  in a hot radio-frequency resonator, Science 363~(6431) (2019) 1072--1075.
\newblock \href {https://doi.org/10.1126/science.aaw3101}
  {\path{doi:10.1126/science.aaw3101}}.

\bibitem{Axline2018}
C.~J. Axline, L.~D. Burkhart, W.~Pfaff, M.~Zhang, K.~Chou, P.~Campagne-Ibarcq,
  P.~Reinhold, L.~Frunzio, S.~M. Girvin, L.~Jiang, M.~H. Devoret, R.~J.
  Schoelkopf, On-demand quantum state transfer and entanglement between remote
  microwave cavity memories, Nature Physics 14~(7) (2018) 705--710.
\newblock \href {https://doi.org/10.1038/s41567-018-0115-y}
  {\path{doi:10.1038/s41567-018-0115-y}}.

\bibitem{Hu2019}
L.~Hu, Y.~Ma, W.~Cai, X.~Mu, Y.~Xu, W.~Wang, Y.~Wu, H.~Wang, Y.~P. Song, C.-L.
  Zou, S.~M. Girvin, L.-M. Duan, L.~Sun, Quantum error correction and universal
  gate set operation on a binomial bosonic logical qubit, Nature Physics 15~(5)
  (2019) 503--508.
\newblock \href {https://doi.org/10.1038/s41567-018-0414-3}
  {\path{doi:10.1038/s41567-018-0414-3}}.

\bibitem{Magnard2020}
P.~Magnard, S.~Storz, P.~Kurpiers, J.~Schär, F.~Marxer, J.~Lütolf, T.~Walter,
  J.-C. Besse, M.~Gabureac, K.~Reuer, A.~Akin, B.~Royer, A.~Blais, A.~Wallraff,
  Microwave quantum link between superconducting circuits housed in spatially
  separated cryogenic systems, Physical Review Letters 125~(26) (dec 2020).
\newblock \href {https://doi.org/10.1103/physrevlett.125.260502}
  {\path{doi:10.1103/physrevlett.125.260502}}.

\end{thebibliography}



\end{document}